\documentclass{aa}
\usepackage{graphicx}
\usepackage{grffile}
\usepackage{txfonts}
\usepackage{natbib}
\usepackage{hyperref}
\hypersetup{colorlinks=true, citecolor=blue}
\usepackage{graphicx}
\usepackage{rotating}
\usepackage{listings}
\usepackage{amsmath}
\usepackage{placeins}

\begin{document} 

   \title{New-generation dust emission templates for star-forming galaxies}
   \author{Médéric Boquien\inst{1,3} \and Samir Salim\inst{2,3}}
   \institute{Centro de Astronomía (CITEVA), Universidad de Antofagasta, Avenida Angamos 601, Antofagasta, Chile \email{mederic.boquien@uantof.cl}
              \and
              Department of Astronomy, Indiana University, Bloomington, IN 47405, USA
              \and
              Both authors have contributed equally to this article.}

   \date{}

  \abstract
  {The infrared (IR) emission of dust heated by stars provides critical information for galaxy evolution studies. Unfortunately, observations are often limited to the mid-IR, making templates a necessity. Previously published templates were based on small samples of luminous galaxies, not necessarily representative of normal star-forming galaxies.}     
   {We construct new-generation dust templates, including instrument-specific relations and software tools that facilitate the estimation of the total IR (TIR) luminosity and obscured+unobscured SFR based on one or several fluxes up to $z=4$. For the first time the templates include a dependence on both TIR luminosity and the specific SFR (sSFR), thereby increasing their reliability and utility for a wide range of galaxies. We also provide formulae for calculating TIR luminosities and SFR from JWST F2100W observations at $0<z\lesssim 2$.}
  {Our templates are based on 2584 normal star-forming galaxies spanning a wide range of stellar mass and sSFR, including sSFRs typical at higher redshifts. IR spectra and properties are obtained using \textsc{cigale} and the physically motivated \cite{draine2007a} dust models. The photometry from the GALEX-SDSS-WISE Legacy Catalog is supplemented with 2MASS and \textit{Herschel}-ATLAS --- up to 19 bands from FUV to 500~$\mu$m.}
    {The shape of the dust spectrum varies with TIR luminosity, but also independently with sSFR. Remarkably precise estimates of the dust luminosity are possible with a single band over the rest-frame 12--17~$\mu$m and 55--130~$\mu$m. We validate single-band estimates on diverse populations, including local LIRGs, and find no significant systematic errors. Using two or more bands simultaneously yields unbiased estimation of the TIR luminosity even of star-forming dwarfs.}
   {We obtain fresh insights regarding the interplay between monochromatic IR luminosities, spectral shapes and physical properties, and construct new templates and estimators of the dust luminosity and SFR. We provide software for generating templates and estimating these quantities based on 1--4 bands from WISE, JWST, \textit{Spitzer}, and \textit{Herschel}, up to $z=4$.}


   \maketitle

\section{Introduction}
The rest-frame ultraviolet (UV) emission of galaxies is one of the more direct probes of star formation in galaxies. However, its interpretation as star formation rate (SFR) depends on a number of factors, some of which may not be well constrained: star formation history (SFH), stellar metallicity, stellar atmospheres,  stellar tracks, and perhaps most importantly, dust attenuation.

Even though it represents typically less than 1\% of the mass of the interstellar medium \citep[e.g.,][]{remy2014a}, dust can have a considerable effect on the emerging UV radiation of galaxies, both dimming and reddening its emission. The SFR derived from the observed UV luminosity, without any dust corrections, is often referred to as unobscured star formation.

The reliability of the UV emission as a SFR estimator therefore critically depends on the ability to accurately correct for dust attenuation. One approach for achieving this goal is to relate the observed UV color ($\beta$) to the infrared (IR) excess (IRX), where the latter is closely related to the intrinsic UV attenuation \citep[e.g.,][]{meurer1999a}. Larger UV attenuation translates into a redder UV color so that the difference between the intrinsic and observed UV colors should ideally trace the attenuation. However, a key issue with this approach is that the relation between the observed UV color and the attenuation is strongly dependent on the attenuation curve \citep{boquien2009a, salim2019a}, which is often poorly known and yet spans a wide range of shapes and varies significantly from galaxy to galaxy \citep{salim2020a}. While, the determination of the total SFR using UV/optical Spectral Energy Distribution (SED) modeling \citep[e.g.,][]{conroy2013a} is a more sophisticated approach than the IRX--$\beta$ technique, it is based on the same fundamental principle (comparison of observed and intrinsic colors), and is therefore subject to similar systematic and random errors arising from poorly constrained or inappropriately assumed attenuation curves.

Stellar luminosity absorbed by dust grains is reprocessed and re-emitted at longer wavelengths, mainly in the mid-infrared (MIR, $\lambda\lesssim40$~$\mu$m) and far-infrared (FIR,  $\lambda\gtrsim40$~$\mu$m) domains. Dust emission is thus complementary to the UV emission emerging from galaxies as a tracer of star formation \citep[e.g.,][]{daddi2007a,elbaz2007a}. In heavily obscured galaxies, in which only a small fraction of UV photons can escape, and therefore a reliable correction of the UV or UV/optical-based SFR is particularly challenging, the dust emission traces nearly the entirety of star formation \citep{kennicutt1998a}. Overall, the IR emission provides the most reliable ``correction'' for the unobscured SFR determined from the UV/optical emission. Furthermore, as recognized previously \citep[e.g.,][]{inoue2002a}, and further demonstrated in this paper, the IR emission can be used as a reasonably good indicator of the total SFR of normal star-forming galaxies in its own right, even without explicitly accounting for the unobscured star formation.

Both the monochromatic ($\lambda\times L_\lambda\left(\lambda\right)$) and the total (bolometric) infrared luminosity ($L_{TIR}$) have been employed as SFR estimators \citep[e.g.,][]{calzetti2007a}. In this paper we will, among other objectives, tackle the question of which wavelength range is the best SFR indicator. Regardless of the answer to that question, the bolometric dust emission ($L_{TIR}$) is a fundamental galaxy property that we wish to be able to determine as reliably as possible, and its measurement represents the principal objective of this work.

Measuring the bolometric dust emission in a galaxy is a difficult and complex affair. The underlying reason for this difficulty is that finely sampled SED (or spectra) that would cover the entire IR range from a few microns to $\sim 1$~mm simply do not exist. Even the relatively well sampled SED, which are by no means common, still require interpolations and extrapolations in order to produce continuous IR spectra. These interpolations and extrapolations require theoretical modeling of dust emission, which is in itself a rather complex task.

The complexity of dust modeling arises in no small part from the intricate emission spectrum of dust, which depends on the nature of the grains (composition and size distribution), as well as the local physical conditions, in particular the shape and intensity of the incident radiation field. The most prominent feature of the dust emission spectrum consists in a gray body component peaking at around 100~$\mu$m, which modern models describe as a combination of the emission of large dust grains over a broad range of temperatures, reflecting the temperature distribution throughout a galaxy. At shorter wavelengths, the emission is dominated by Polycyclic Aromatic Hydrocarbons (PAH) that exhibit prominent bands that appear to be strongly dependent on metallicity. PAH emission, as well as the emission from very small grains that dominates at wavelengths between PAH and the peak, comes from a stochastic heating process. 

Over the past decades, dust emission models have increased in sophistication and complexity, improving
our ability to reproduce and interpret the observations \citep[e.g.,][]{desert1990a, dale2001a, draine2007a,dacunha2008a,compiegne2011a,draine2014a, jones2017a}. However, because of their relative complexity and ensuing flexibility, such models are of little help for deriving the total IR luminosity in cases when the observations are confined to only a small wavelength range, or even a single band, i.e., when huge extrapolations are required. Such is the situation for hundreds of thousands of galaxies observed by the Wide-field Infrared Survey Explorer \citep[WISE,][]{wright2010a}, as will also be the case of the galaxies that will be studied through the eye of the James Webb Space Telescope \citep[JWST,][]{gardner2006a}. In both cases the observations only extend up to $\sim25$~$ \mu$m. Fortunately, making inferences about the bolometric emission based on a very limited wavelength coverage is nevertheless possible, but one has to resort to using the templates.

Templates are important because they narrow down the vast multi-dimensional parameter space of theoretical spectra into a family that is parametrized on a single parameter, which is calibrated based on a sample of galaxies with good SED coverage. Parameterization should ideally be on a parameter that best correlates with the changes in the shape of the SED. The dust emission templates approach was pioneered by \cite{chary2001a}, who produced a set of templates dependent on the total infrared luminosity, which in effect provides constraints on the shape of the dust emission spectrum. If the galaxies from which the templates were constructed are representative of the galaxies to which they are applied, unbiased estimates of $L_{TIR}$ will be possible from limited data. Because in this case the parameterization is on the extensive quantity (i.e., it directly scales with the ``extent'' of the galaxy), even a single flux point (band) is sufficient to estimate the total dust luminosity.

However, these and other widely used templates (e.g., \citealt{dale2002a} and \citealt{rieke2009a}) were based on galaxies selected from relatively shallow surveys carried out by the Infrared Astronomical Satellite mission \citep[IRAS,][]{neugebauer1984a} and the Infrared Space Observatory \citep[ISO,][]{kessler1996a}. Such a selection will inevitably favor high-luminosity galaxies that are atypically luminous for their stellar or dust mass, possibly because of the more efficient star formation resulting from mergers that are common among these galaxies. As a result, they will have a warmer SED than normal star-forming galaxies of similar $L_{TIR}$, which could lead to offsets in the estimation of dust luminosities of normal galaxies, especially based on observations beyond the peak, even at low redshift \citep{lin2016a}. Furthermore, the relatively small sample sizes ($\sim$100 galaxies) that underlied these efforts precluded the investigation of more complex parameterizations. Such parameterizations may extend the validity of templates over a broader range of galaxies and redshifts.

The upcoming launch of JWST highlights the need for new templates that provide an un-biased view of galaxies at higher redshifts. In the absence of adequate detailed templates, the limited spectral coverage of JWST may negatively affect our ability to estimate the bolometric dust emission to trace star formation. Existing templates from local luminous samples are not necessarily representative of similarly luminous galaxies at higher redshifts \citep{safarzadeh2016a}. Indeed, when applied to higher redshifts, some of the existing templates are suspected to produce systematic offsets in $L_{TIR}$ derived from MIR monochromatic luminosities \citep{lin2016a}. Efforts to produce IR templates using the actual high-redshift galaxies have made significant progress in recent years, primarily as the result of the stacking of \textit{Herschel} observations. For now, the focus has been mostly on producing average templates in various redshift bins \citep{magdis2012a}, which cannot capture the diversity of dust emission properties present at a given redshift. The stacking approach was refined in \citet{schreiber2018a}, where the templates depend on the redshift but also on the SFR relative to the main  sequence. However, lacking a parameterization on an extensive quantity, such templates cannot be utilized when observations exist in only a single band.

To address these concerns with the existing templates based on local galaxies and to provide a more comprehensive physically motivated basis for interpreting IR observations in general, in this work we exploit a significantly larger sample of local galaxies from MIR (WISE) and FIR (\textit{Herschel}-ATLAS) surveys, and with photometry from the far UV to the sub-millimetric regime, to gain insight into the drivers of IR SED shape and the ability of different monochromatic luminosities to constrain dust parameters. Based on that analysis, we construct new single and two-parameter templates, where the role of the second parameter is to account for the range of normal star forming galaxies at different redshifts, thus making the new templates specifically well suited for JWST. Furthermore, in this work we highlight the use of explicit relations as a more practical alternative to discrete templates. Our templates and relations are constructed using some new approaches that are physically motivated and homogeneously applied to all the data. 

We introduce the multi-wavelength data set we build on in Sect.~\ref{sec:data}. In Sect.~\ref{sec:method} we describe the method we have developed for constructing and parameterizing the new templates. We present our results in Sect.~\ref{sec:results1} and \ref{sec:results2} and discuss them in Sect.~\ref{sec:discussion} before concluding in Sect.~\ref{sec:conclusion}. Throughout this paper we assume a WMAP7 cosmology \citep[$H_0=70.4$~km~s$^{-1}$~Mpc$^{-1}$, $\Omega_c=0.226$,][]{komatsu2011a}.

\section{Data and sample\label{sec:data}}
\subsection{\textit{Herschel}-ATLAS}
The sample for the construction of dust emission templates, or, more broadly, the relations that allow us to estimate the TIR luminosity and other physical properties from IR observations, should have the following characteristics: 1) good wavelength sampling across the full IR range in order to probe the main dust emission features, 2) deep IR observations in order to provide reliable and highly complete photometric measurements, 3) representative sampling of star-forming galaxies, including high-mass galaxies, to take into account variations in dust emission across galaxy types, 4) extensive UV and optical data to constrain the stellar masses and dust-corrected SFR, 5) ability to distinguish and remove AGN, which can contaminate the IR emission, and, finally 6) relatively large size, in order to adequately sample the parameter space and reduce statistical uncertainties. These conditions are fulfilled for a sample selected from the \textit{Herschel}-ATLAS survey \citep[H-ATLAS,][]{valiante2016a,maddox2018a}, combined with multi-wavelengths data from SDSS \citep[Sloan Digital Sky Survey,][]{york2000a}, GALEX \citep[Galaxy Evolution Explorer,][]{martin2005a}, and WISE \citep[Wide-field Infrared Explorer,][]{wright2010a}, which form the basis for the SDSS-GALEX-WISE Legacy Catalog \citep[GSWLC,][]{salim2016a,salim2018a}.

\subsection{Data}
H-ATLAS is the largest uniform imaging survey carried out by the \textit{Herschel Space Observatory}, covering 660 sq.~deg. It observed the sky in five FIR/sub-millimeter bands, using the PACS \citep[100~$\mu$m and 160~$\mu$m,][]{poglitsch2010a} and SPIRE \citep[250~$\mu$m, 350~$\mu$m, and 500~$\mu$m,][]{griffin2010a} cameras. H-ATLAS data release 1 \citep[DR1,][]{valiante2016a} consists of three equatorial fields with a total area of 160 sq.~deg covering the Galaxy and Mass Assembly \citep[GAMA,][]{driver2011a} survey, whereas two fields covering a total of 500 sq.~deg around the north and the south Galactic poles were released as DR2 \citep{maddox2018a}.

The SPIRE photometry of the H-ATLAS catalog is based on detections in the 250~$\mu$m image, which has the highest density of sources in any \textit{Herschel} band \citep{maddox2018a}. This photometry is carried out by successively measuring and subtracting the sources in the order of their brightness, separately in each filter. Most SDSS galaxies are unresolved in SPIRE images, the beam size ranges from 18\arcsec\ at 250~$\mu$m to 35\arcsec\ at 500~$\mu$m, so the default Point Spread Function (PSF) photometry is adequate for them. For extended sources (based on their optical size), which are found among $z<0.1$ galaxies, aperture photometry is also performed. The flux we adopt (reported in the catalog as ``best'') is whichever of the aperture or PSF flux is the brightest. Because the PACS beam is smaller (7\arcsec\ at 100~$\mu$m and 11\arcsec\ at 160~$\mu$m), only aperture photometry is performed in these bands, centered on the position of the optical counterpart. The H-ATLAS catalogs provide matched counterparts in SDSS, which were identified following the methodology of \cite{bourne2016a}. We add in quadrature a flux calibration uncertainty of 7\% for PACS and 4\% for SPIRE to the flux uncertainties reported in the H-ATLAS catalogs. 

For the MIR photometry (12~$\mu$m and 22~$\mu$m), we use the all-sky survey from WISE, processed by the unWISE project \citep{lang2016a}. They performed forced photometry using SDSS positions and galaxy profiles, thus obtaining the fluxes that are more accurate compared to the standard pipeline photometry, which treats all sources as unresolved. Forced photometry is important in particular for sources that are resolved and/or blended in WISE images. We add in quadrature a 2\% flux calibration uncertainty to the formal uncertainties reported in the unWISE catalog.

For deriving the stellar masses and SFR we use near-IR (NIR), optical, and UV photometry. NIR photometry in three bands (J, H, and K$_\mathrm{s}$) comes from the 2MASS  \citep[Two Micron All Sky Survey,][]{skrutskie2006a} Extended Source Catalog, as described in \cite{salim2016a}. Optical photometry in five bands ($u$, $g$, $r$, $i$, and $z$) comes from SDSS DR10. Far- (FUV) and near-UV (NUV) photometry is obtained from the GALEX data release GR6/7, with the various flux corrections that are discussed in \cite{salim2016a}, which also described the matching algorithm to SDSS. We use GALEX data regardless of the depth of UV observations (GSWLC-X2 dataset), covering 90\% of SDSS spectroscopic targets. 

To be included in GSWLC, and therefore in our current sample, in addition to being covered by GALEX, SDSS spectroscopic targets need to have an SDSS Petrosian magnitude in the $r$ band brighter than 18 and a redshift within $0.01<z<0.3$. In other words, they need to be similar to galaxies selected for a statistically complete SDSS Main Galaxy Survey \citep{strauss2002a}.

\subsection{Sample}
The H-ATLAS catalogs\footnote{\url{https://www.h-atlas.org/public-data/download}} (DR1 v1.2 and DR2 v1.4) contain a total of 10676 GSWLC-X2 galaxies over $\sim$400 sq.\ deg. From this initial sample we remove 127 AGN with broad lines (based on their SDSS spectroscopic classification), 3759 galaxies classified as AGN based on the BPT emission line diagram\footnote{Emission lines used in this paper come from the MPA/JHU SDSS DR7 catalog, \url{https://wwwmpa.mpa-garching.mpg.de/SDSS/DR7}.} and a further 29 galaxies not contained in the MPA/JHU catalog. The removal of AGN is essential, as they can significantly contribute to the IR spectrum, in particular in the WISE bands. From the remaining 6761 galaxies we remove 1786 galaxies with negative fluxes in the W3 and/or W4 bands. Furthermore, of the remaining 4975 galaxies we remove 2386 galaxies with a signal-to-noise ratio lower than 2 at 100~$\mu$m and/or 160~$\mu$m, and, finally, another 5 galaxies with $\mathrm{sSFR<10^{-12}}$~yr$^{-1}$. Our final sample consists of 2584 galaxies. Its median redshift is 0.08 with a 90 percentile redshift range between 0.02 and 0.17. Our sample is an order of magnitude larger than the samples used in most previous efforts to build dust emission templates\footnote{When used as adjectives, we will consider ``IR'' and ``dust-emission'' as interchangeable.}.

\subsection{Characteristics\label{ssec:char}}
\begin{figure}[!htbp]
 \includegraphics[width=\columnwidth]{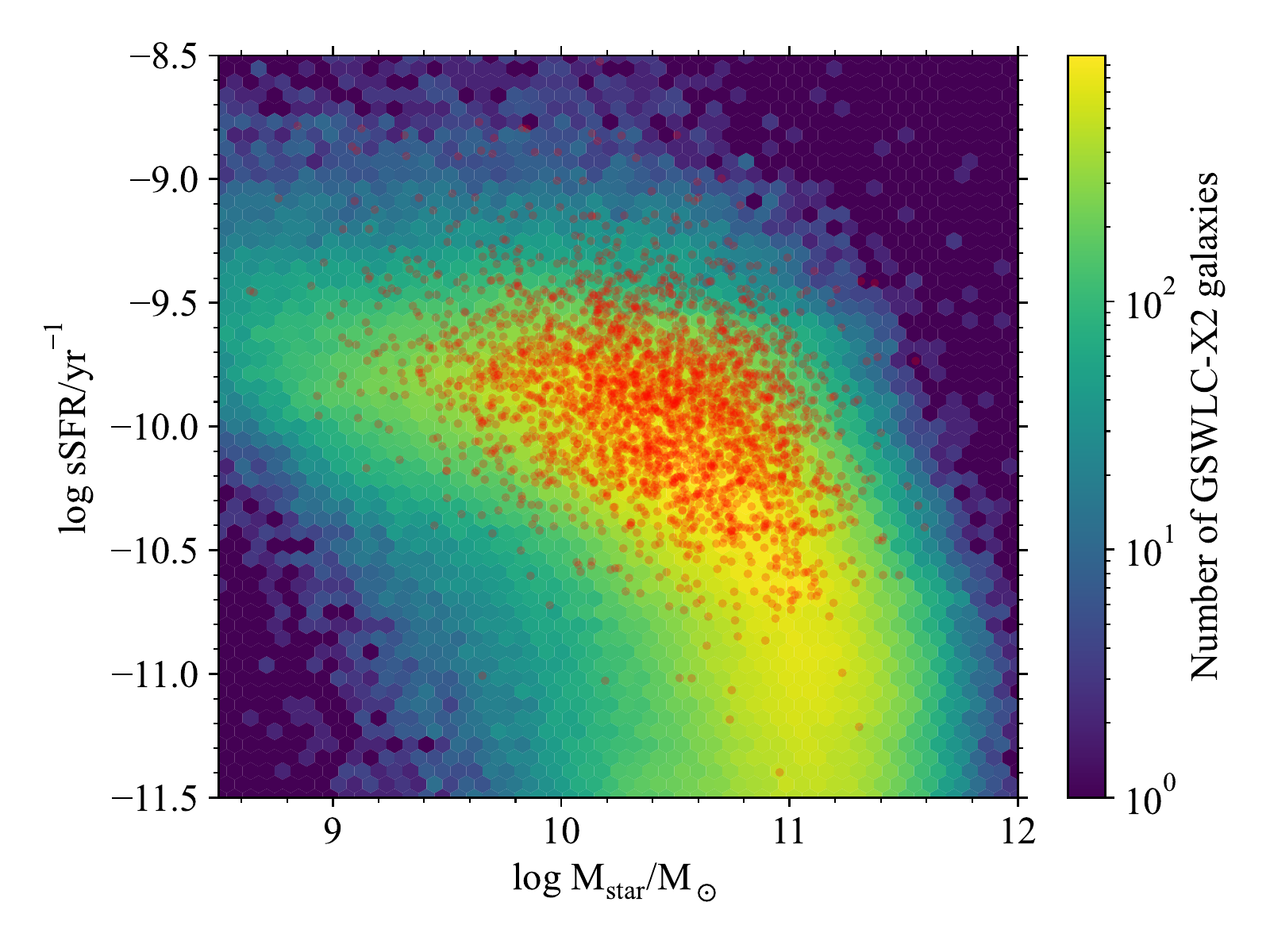}
 \caption{Distribution of the selected sample in the sSFR--$M_{star}$ plane (red circles) compared to the full GSWLC-X2 catalog (hexagons, with the density indicated by the color bar).\label{fig:sSFR-Mstar}}
\end{figure}

To estimate the physical properties of our sample, we have carried out SED modeling as described in detail in Sect.~\ref{ssec:method}. In Fig.~\ref{fig:sSFR-Mstar} we present a comparison of our sample with respect to all GSWLC-X2 galaxies in the sSFR--$M_{star}$ plane. 
We see that our sample follows well the main sequence of star-forming galaxies.

\begin{figure*}[!htbp]
 \includegraphics[width=0.33\textwidth]{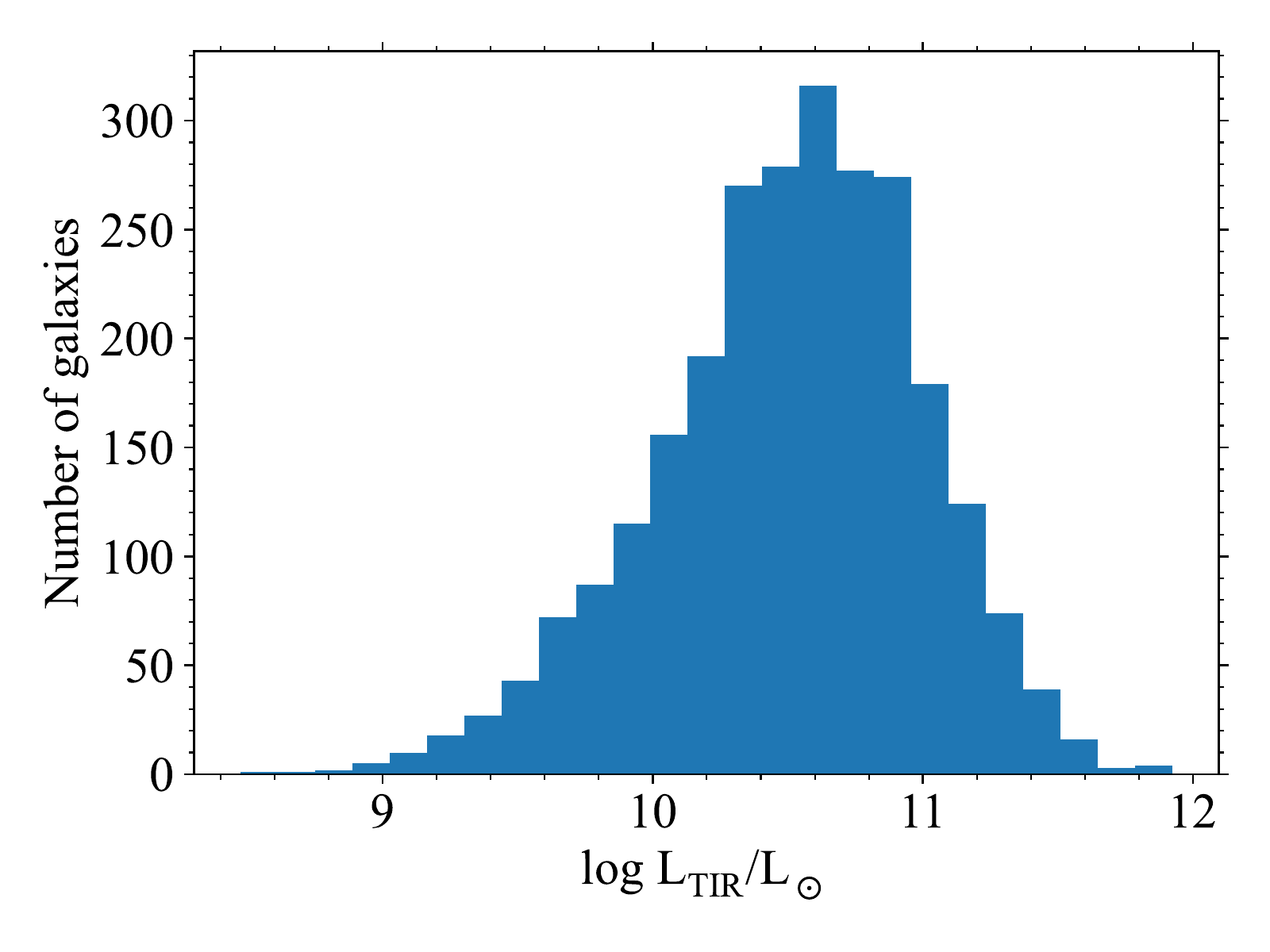}
 \includegraphics[width=0.33\textwidth]{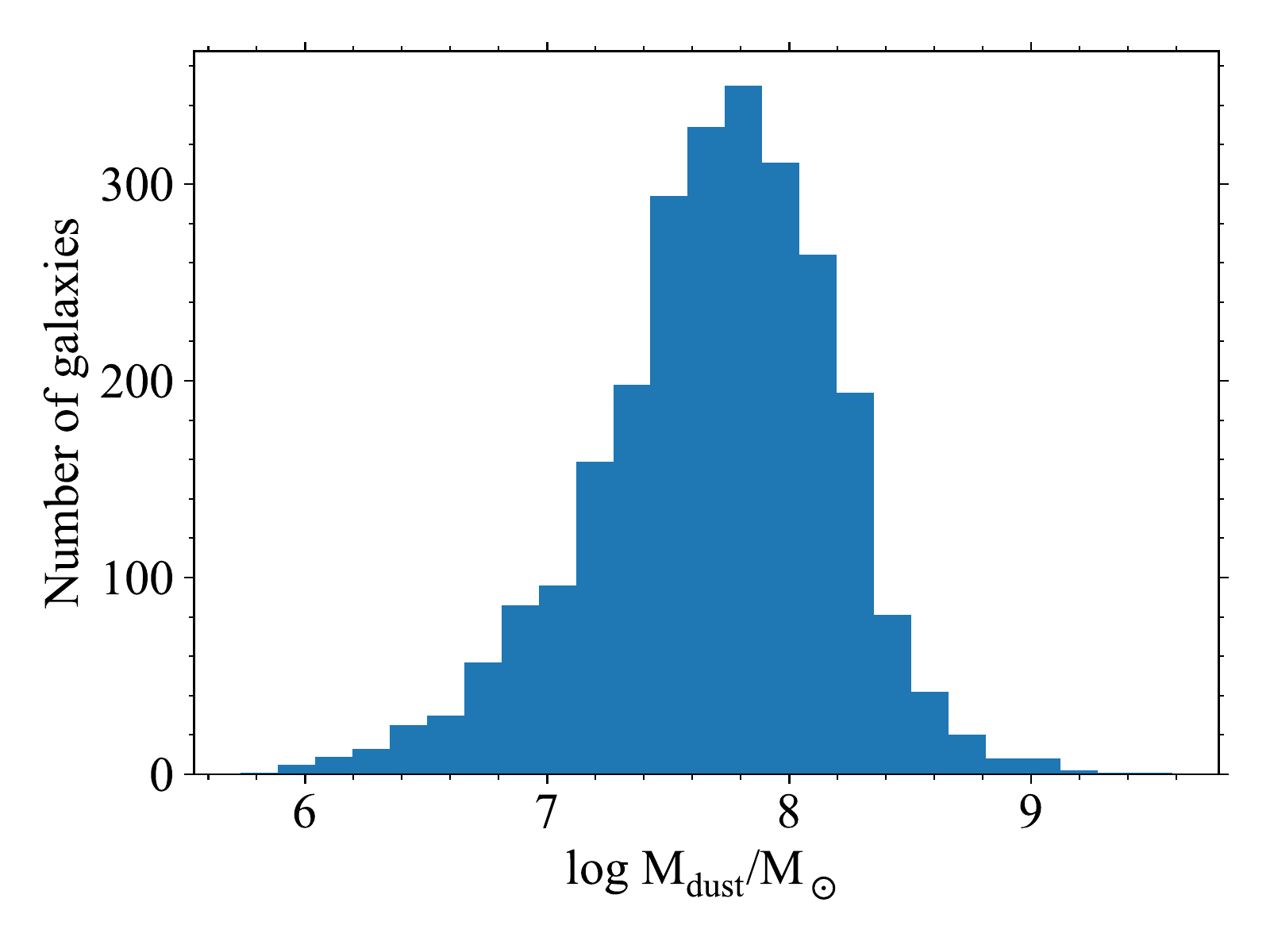}
 \includegraphics[width=0.33\textwidth]{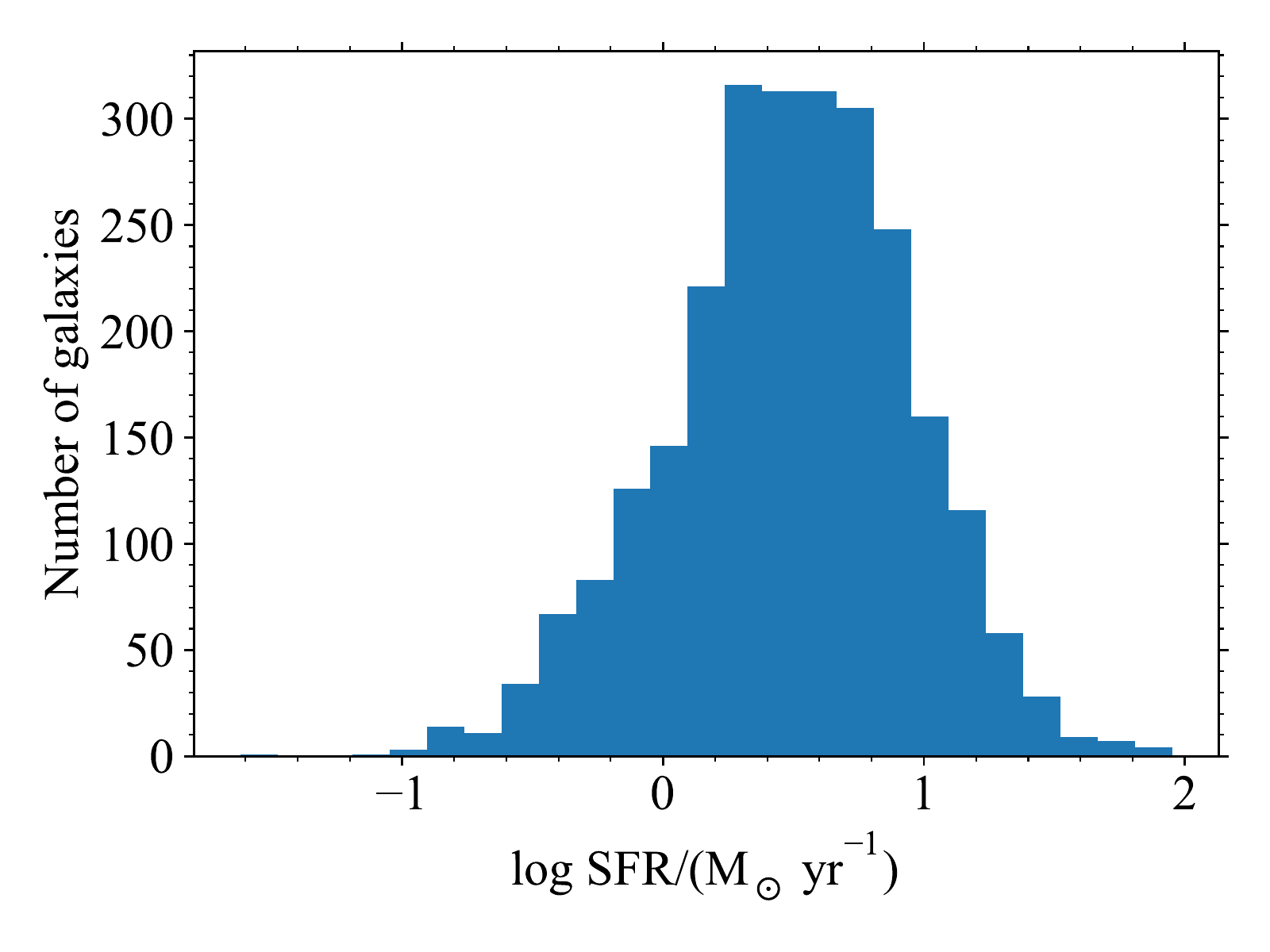}\\
 \includegraphics[width=0.33\textwidth]{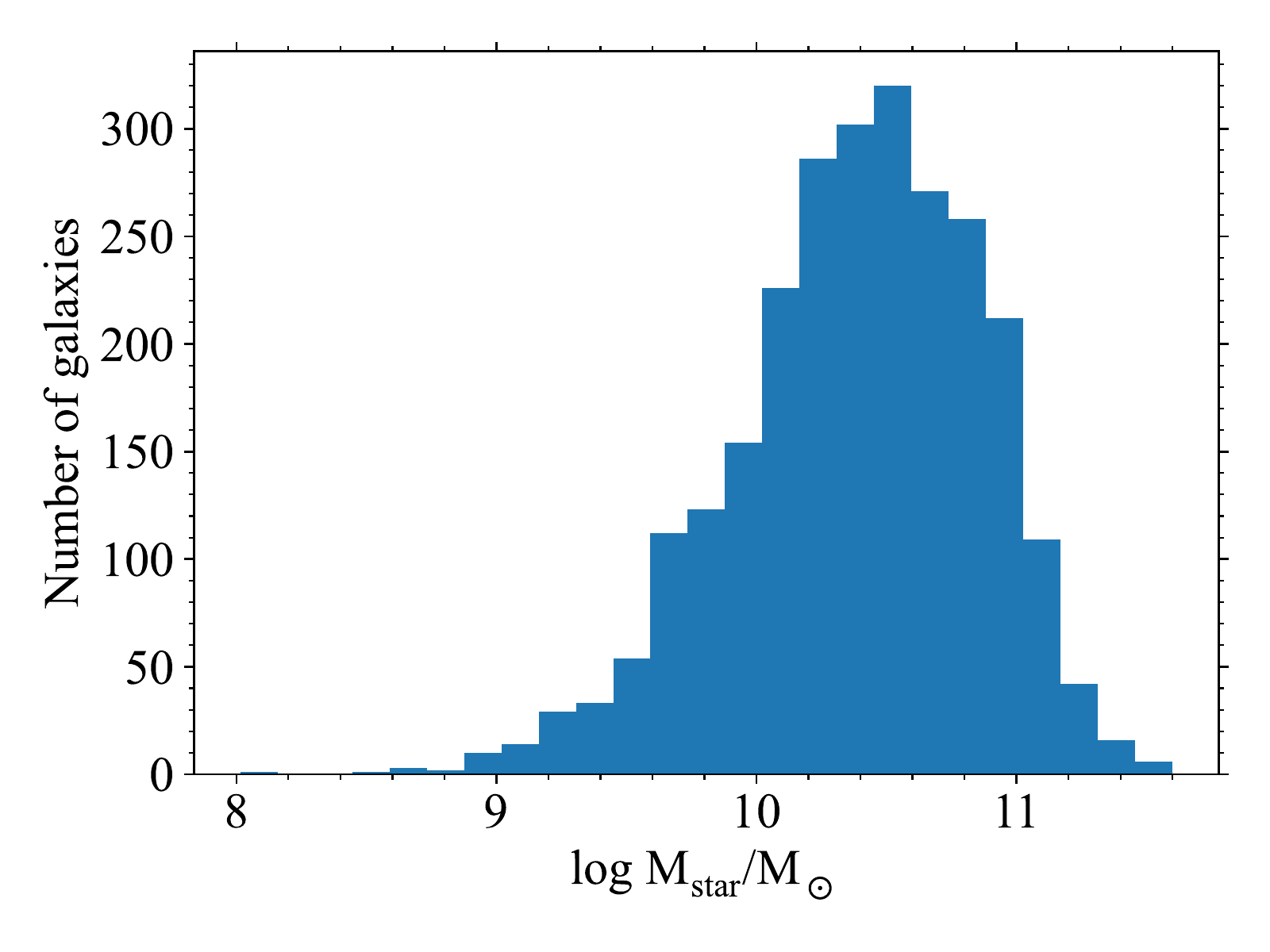}
 \includegraphics[width=0.33\textwidth]{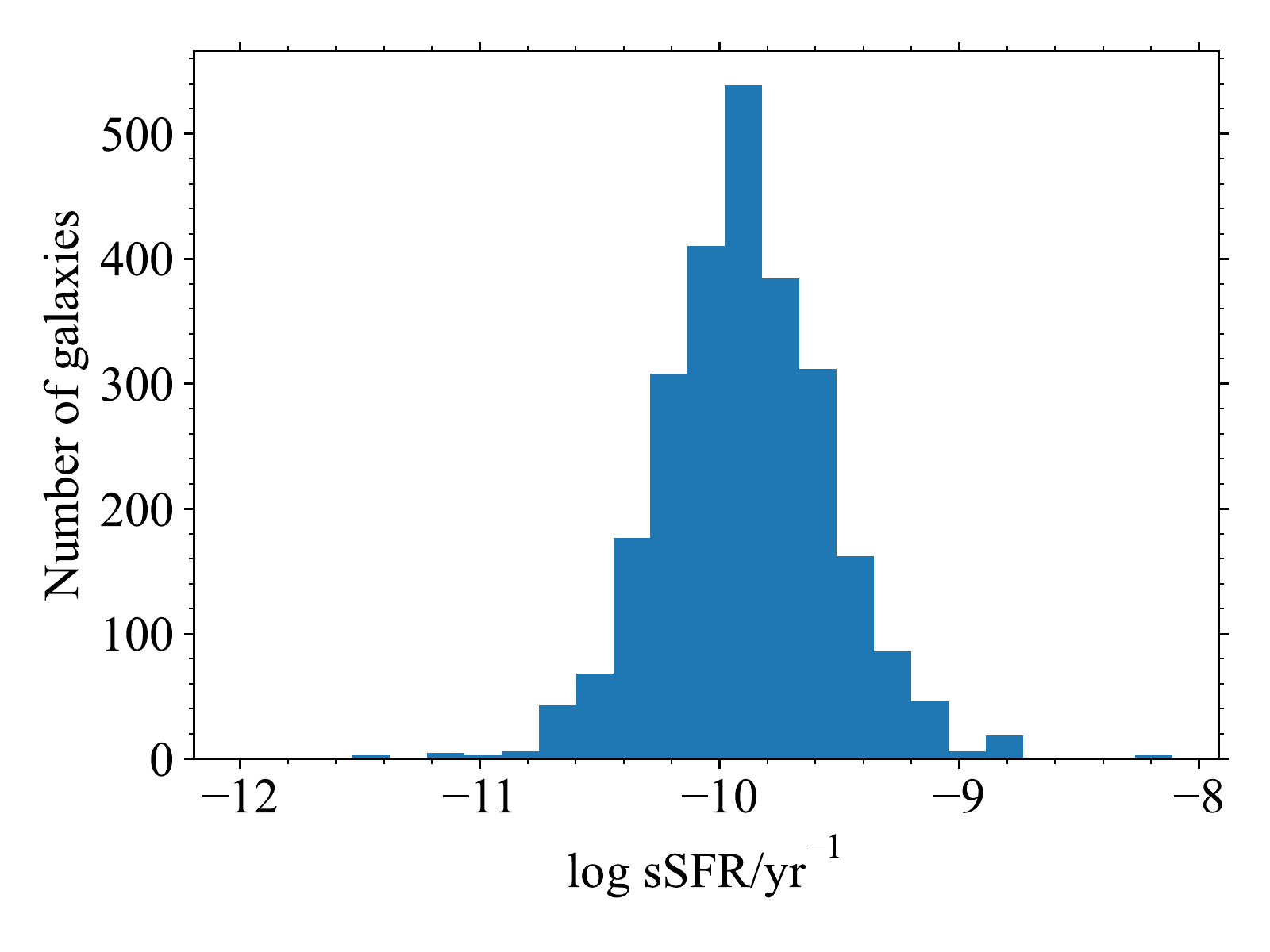}
 \includegraphics[width=0.33\textwidth]{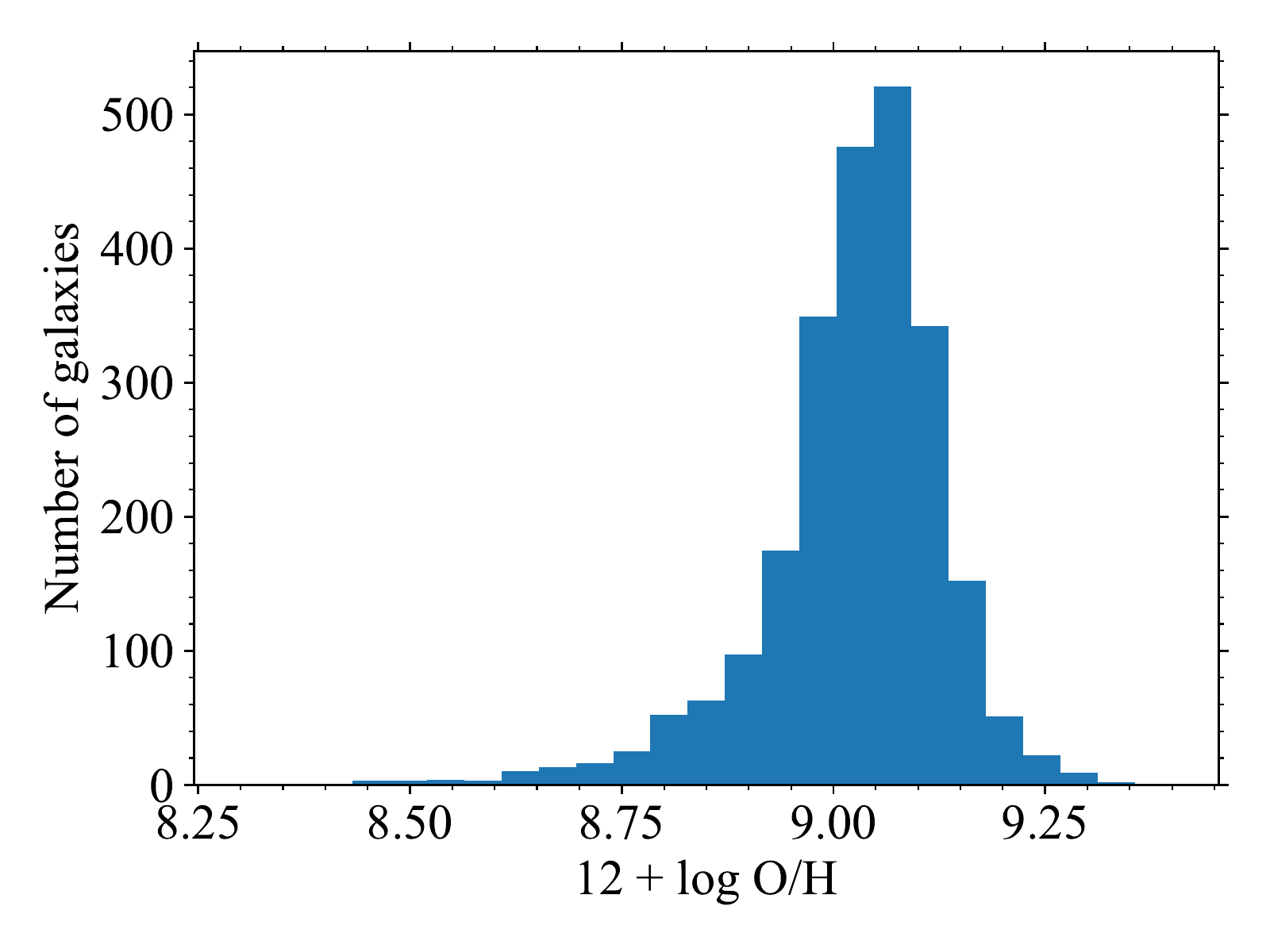}
 \caption{From the top-left to the bottom-right, distribution of $L_{TIR}$, $M_{dust}$, SFR, $M_{star}$, sSFR, and the oxygen abundance. For the latter the sample is restricted to 2391 objects rather than 2584 for the other physical properties as the oxygen abundance could not be computed for all the objects.\label{fig:sample-properties}}
\end{figure*}

In Fig.~\ref{fig:sample-properties} we present the distributions of the physical properties of the sample measured: SFR averaged over the last 100~Myr, stellar mass ($M_{star}$), total infrared luminosity ($L_{TIR}$, corresponding to the integral of the dust emission between 8~$\mu$m and 1~mm), dust mass ($M_{dust}$), specific SFR (sSFR), and oxygen abundance.
Overall, the sample covers a wide range of non-dwarf, normal star-forming galaxies, including LIRGs. Furthermore, it includes galaxies with a large range of sSFR at a given mass, up to $\log \mathrm{sSFR}\sim-9$. It does not include more extreme, but relatively rare galaxies such as ULIRGs ($\log L_{TIR}>12$), which can have a markedly different dust emission spectrum compared to more typical star-forming galaxies \citep[e.g.,][]{rieke2009a}. In Sect.~\ref{ssec:external} we discuss the cases in which the relations that we will derive in Sect.~\ref{sec:results2} can be used for galaxies that fall outside of the range of properties of our sample.

\section{Method\label{sec:method}}
The central aspect of this work is the construction of relations between one or more IR fluxes and general galaxy properties. These relations will serve as estimators of the bolometric dust luminosity and will also the allow the generation of dust emission templates parameterized on the TIR luminosity\footnote{We will throughout the paper refer to the bolometric dust luminosity and the TIR luminosity interchangeably. The exact definition of the TIR luminosity in relation to the dust luminosity is given in Sect.~\ref{ssec:char}.} and/or other physical properties. To carry out these tasks it is first necessary to estimate both the physical properties (stellar mass, dust mass, SFR, sSFR, etc.) of each galaxy in our sample and their associated dust emission spectrum.

\subsection{Estimating the physical properties and spectra of galaxies\label{ssec:method}}
One of the most physically motivated techniques to estimate the physical properties of galaxies is via the modeling of their electromagnetic emission using synthesis population models. Since the first models of \cite{tinsley1972a}, numerous codes have been developed over the years with increasing sophistication and power. The latest generation of codes can model galaxies from the FUV to the FIR, including dust in absorption and in emission in a consistent way, a key ingredient to break degeneracies and estimate the physical properties more reliably \citep{burgarella2005a,dacunha2008a,noll2009a,leja2017a,carnall2018a,boquien2019a,bowman2020a,robotham2020a}.

For this work we adopt \textsc{cigale} \citep{burgarella2005a,noll2009a,boquien2019a}\footnote{\url{https://cigale.lam.fr}} version 2020.0, a highly versatile code that has been successfully used for modeling galaxies and study a broad range of questions at different redshifts \citep[e.g.,][]{buat2011a, buat2012a, buat2018a, buat2019a, burgarella2011a, burgarella2020a, boquien2012a, boquien2014a, boquien2016a, ciesla2015a, ciesla2016a, ciesla2017a, ciesla2018a, ciesla2020a, salim2016a, salim2018a, salim2019a, hunt2019a, franco2020a, dale2020a, mountrichas2021a}. Under its standard operating mode, \textsc{cigale} generates and fits a large number of multi-wavelength spectro-photometric models to observations, and estimates the physical properties and the related uncertainties from their probability distribution function. Here, however, we aim not just at measuring the general physical properties of galaxies, but also at deriving the best-fitting dust emission spectrum for each galaxy to be used as a ``ground truth'' for constructing the aforementioned relations or templates. Both of these aims could in principle be achieved with \textsc{cigale} by simultaneously modeling and fitting the UV/optical/NIR photometry (i.e., the stellar populations) together with the MIR/FIR photometry (dust emission). However, if we wish to use the  dust emission models required to fit the IR emission with sufficiently high resolution in model parameters (the model grid), we would end up with an excessively large number of combined (stellar populations and dust) models (in our case 1 trillion). We have thus decided to adopt and adapt (Sect.~\ref{sssec:correction}) the two-step approach developed in \cite{salim2018a}. In a nutshell, we first model only the emission of the dust to estimate the physical properties that are largely independent from the stellar population modeling (e.g., dust mass and luminosity), as well as to obtain the dust emission spectrum. We require a detailed dust emission spectrum as a basis for the construction of relations and templates that are continuous in wavelength, and therefore can be used with any instrument and at any redshift. In the second step, we fit models of the stellar populations to UV/optical/NIR photometry and estimate different physical properties (SFR, stellar mass, etc.). Importantly, we do so while also including in the fitting the dust luminosity ($L_{dust}$) determined in the first step as a critical information to break the age-attenuation degeneracy. Finally, with the knowledge of the stellar spectra, we correct for the stellar contamination in the MIR bands, and repeat these two steps.

\subsubsection{Dust modeling\label{sssec:dustmodeling}}
We model the dust emission using the \cite{draine2014a} update of the \cite{draine2007a} models. We base our choice on the fact that these models are physically motivated while providing enough flexibility to accurately fit our 7-band IR photometry. The \cite{draine2007a} models are known to adequately reproduce well-sampled IR SED \citep[e.g., ][]{ciesla2014a, hunt2019a} and have been successful at reproducing the emission of galaxies \citep[e.g.,][for the KINGFISH sample that  contains diverse galaxies in the nearby universe]{aniano2012a,aniano2020a}, while showing great flexibility to simultaneously reproduce the emission of the PAH, and the warm and cold dust components. It is thus unlikely that there is a major issue with this underlying model. The main caveat may be in the 25~$\mu$m to 60~$\mu$m range as we will see in Sect.~\ref{ssec:comparison1}, with little data being available in this range to test and constrain the models in the first place. Each model contains four free parameters and is based on a combination of two components. One, illuminated by an incident radiation field of intensity $U_{min}$, represents the diffuse dust across the galaxy. The other component models the dust in star-forming regions, with an illumination intensity following a power law $dM_{dust}/dU\propto U^{-\alpha}$, where $M_{dust}$ is the dust mass, $U$ the incident radiation field intensity, and $\alpha$ is the adjustable power law index. A coefficient $\gamma$ sets the mass fraction of the component associated with star-forming regions. We use the entire parameter space provided by the model grid: $U_{\min}$ ranges from 0.1 to 50, $\alpha$ ranges from 1 to 3, and $q_{PAH}$, the mass fraction of PAH, ranges from 0.47\% to 7.32\%. We sample $\gamma$ from $10^{-2.5}$ to $10^{-0.3}$ in 15 logarithmically-spaced steps, a range over which it has the most pronounced effect on the shape of the spectra. The final grid consists of 124740 models at each redshift. This modeling is performed with the \textsc{dl2014} module in \textsc{cigale}.

We model the dust emission based on WISE and \textit{Herschel} data. In the Herschel SPIRE bands, we sometimes only have upper limits. In order to exploit this information we also include these upper limits in the modeling, with the computation of the goodness-of-fit described in detail in \cite{boquien2019a}. Since the modeling is performed at the observed redshifts known from SDSS spectroscopy, no K-correction is required. The output of dust modeling includes estimates of $L_{dust}$ and $M_{dust}$, their uncertainties, as well as the best-fitting spectra made of 1001 flux densities at a constant separation in log space of 0.004~dex from 1~$\mu$m to 10~mm.

An inspection of the fits revealed that there was a relatively small, but statistically significant  systematic offset between the observations and the best-fit model at 160~$\mu$m, with the former being 7.3\% higher than the latter on average. The cause of this offset is unclear. Fitting the observations using the THEMIS models \citep{jones2017a}, we observe a similar offset, suggesting that it may not have a physical origin. Private communication with the members of the \textit{Herschel}-ATLAS team did not allow us to eliminate the possibility of an instrumental origin. In any case, deriving the suite of templates and estimators by adjusting the fluxes 7.3\% downward, we find that the differences are very small, typically within 0.01~dex for monochromatic estimators. Given the minute differences, we adopt the \textit{Herschel}-ATLAS fluxes as-is.

\subsubsection{Stellar population modeling\label{sssec:sspmodeling}}
With the dust luminosity in hand, we model the stellar populations with \textsc{cigale} to estimate other physical properties. The stellar emission is computed adopting the \cite{bruzual2003a} single stellar populations (\textsc{cigale} module \texttt{bc03}) following a \cite{chabrier2003a} initial mass function and a metallicity ranging from subsolar ($Z=0.004$) to supersolar ($Z=0.05$). The SFH is described by two decaying exponentials (\textsc{cigale} module \texttt{sfh2exp}), one modeling the general stellar populations and the other one the latest episode of star formation. We assume the galaxies to be 10~Gyr old. The exponential describing the general stellar populations has an $e$-folding time ranging from 850~Myr to 20~Gyr. The onset of the latest episode of star formation, which has a large timescale, occurred between 100~Myr and 5~Gyr before the time of observation and this episode represents between 0\% and 50\% of the total mass of star formed over the lifetime of the galaxy. Our models also include the nebular emission (\textsc{cigale} module \texttt{nebular}) to account for the contamination of broadband fluxes by gas ionized by massive stars. This comprises both emission lines and continuum emission (free-free, free-bound, and 2-photon processes). Finally, we attenuate the stellar and nebular emission using the \cite{noll2009a} generalization of the \cite{calzetti2000a} attenuation law (\textsc{cigale} module \texttt{dustatt\_calzleit}). In short, the diversity of attenuation curves \citep{salim2020a} is obtained by multiplying the \cite{calzetti2000a} curve by a power law of index $\delta$ ranging from $-1.4$ (steeper) to 0.4 (shallower) and by adding a bump at 217.5~nm in strengths ranging from zero to twice that of the Milky Way. We allow for a differential reddening of a factor 0.44 between stars younger and older than 10~Myr. The absolute reddening of the younger population goes from 0.0125 mag to 0.7 mag, with a particular emphasis on lower values. Overall this constitutes a grid of 8,156,736 models at each redshift.

For modeling the stellar populations we use the GALEX FUV and NUV bands, the SDSS $u$, $g$, $r$, $i$, and $z$ bands, and the 2MASS $J$, $H$, and $K\mathrm{_s}$ bands. In addition, we use $L_{dust}$, the bolometric dust luminosity, as estimated in Sect.~\ref{sssec:dustmodeling} as a further constraint.

For some of the analysis we will be using oxygen abundances, which have been determined from dust-corrected [N\textsc{ii}]6584/[O\textsc{ii}]3726,3729 emission lines ratio, using the calibration of \cite{kewley2002a}. This particular calibration has an advantage over other common calibrations in that it is less sensitive to the effects of the ionization parameter and that it has a unique mapping between the line ratio and the oxygen abundance.

\subsubsection{Correction of stellar contamination in MIR bands\label{sssec:correction}}
Because stellar emission contributes to MIR bands, we have expanded the 2-step approach of \cite{salim2018a} described in Sect.~\ref{ssec:method} into a 4-step strategy to reliably correct for this contamination. First, we fitted the observed fluxes from 12~$\mu$m to 500~$\mu$m, as presented in Sect.~\ref{sssec:dustmodeling}, in order to estimate the dust luminosity. With this constraint in hand, we fitted the stellar populations as described in Sect.~\ref{sssec:sspmodeling}, from which we derived the expected stellar fluxes for each galaxy in the WISE 12~$\mu$m and WISE 22~$\mu$m bands. Stellar contribution is typically small, around a few percent. The last two steps consisted in repeating steps 1 and 2 but with the estimated stellar fluxes subtracted from the WISE bands. The final difference in the derived physical properties is small and neglecting this correction would not have affected the results in any substantial way.

\subsection{A new approach to the construction and use of dust emission ``templates''\label{ssec:approach}}
Empirical dust emission templates allow the determination of dust luminosities and other dust properties in cases when the wavelength coverage of observations is sparse. Because IR spectroscopy is in general limited to small portions of the IR spectrum and boquiis not widely available, all efforts to construct empirical templates ultimately rely on some theoretical modeling (even if it is just a simple gray body) in order to produce dust emission spectra that are continuous in wavelength. Previous work in this area utilized various combinations of empirical MIR templates and theoretical FIR modeling to achieve this goal. Sophisticated theoretical dust emission models (or model grids), such as the \cite{draine2007a} models used in this work, have the flexibility required for reproducing the important variations in the spectral shape and the strength of the PAH bands seen in star-forming galaxies. Because the model grids are too unconstrained when the SED is not well sampled, and are impossible to use for estimating the dust luminosity in the case when only one IR flux point is available, the purpose of the empirical dust emission templates is to effectively narrow down that parameter space of spectral shapes to what is actually found in galaxies. Templates are typically a family of spectra, discretely dependent on some parameter. The usual choice for this single parameter is either the FIR color \cite[i.e., the dust temperature, ][]{dale2002a}, or the dust luminosity \citep{chary2001a}. The two quantities are correlated, but the use of dust luminosity has the advantage that, being an extensive quantity, it allows a template to be fitted and the dust luminosity to be estimated even when only one IR flux point is available. When additional fluxes are available, and are reasonably well separated in wavelength, the family of empirical templates can be used to fit the relative fluxes (i.e., the colors) instead of the absolute flux, conceptually mimicking the usual SED fitting process. In such case, no use is made of the dust luminosity attached to each individual template. We refer to these two ways of estimating the dust parameters from the templates as the traditional approach. 

A straightforward method for constructing the templates would be to combine different spectra in bins of a given physical property and average them in some way. This approach has several important drawbacks. The discretization of the templates that results from the binning is somewhat arbitrary, the number of objects in a bin can be highly variable causing non-uniform accuracy, and the stochastic nature of averaging small samples could lead to non-physical, or at least odd, spectral shapes under some circumstances. Templates have been made to avoid the latter issues by forcing monotonic relations between monochromatic and total luminosity \citep[e.g., ][]{chary2001a}, but they are still discrete. In light of these disadvantages, we build our non-discrete ``templates'' by producing functional \textit{relations} that connect monochromatic IR luminosities to any physical property of interest. The IR spectrum corresponding to any value of this physical property can then be obtained by the reciprocal relations. This technique has multiple advantages. First, it allows for the IR spectra to be defined over a continuum of one or more parameters rather than in more or less sparse and arbitrary discrete bins. Since the relations are fitted linearly (in logarithm space) to the entire sample, the ``template'' spectra iron out the noise from the stochastic diversity of galaxies in any given bin. 

Second, while one can export the newly derived IR spectra as discrete templates, and use them in the traditional way described above (essentially, fitting them to one or more bands), our approach makes this extra step unnecessary, because one can derive a desired physical property from the relations directly, as a function of one or more monochromatic luminosities, or even as a function of some additional parameter. In the case of a single flux, the process is conceptually equivalent to, and the results are identical to, the template fitting. When additional fluxes are available, we have found that the functional relations actually provide more accurate estimates of the parameter (specifically, TIR luminosity) than the fitting method. The principal disadvantage of the relations method compared to the fitting method is that the estimation of the uncertainties of the derived parameter (TIR luminosity) would require a Monte Carlo simulation with perturbation of the input fluxes. In what follows, we will generally discuss relations, reserving the term templates for the discrete set of SEDs.

In practical terms, the construction of relations starts by adopting, for each galaxy, the rest-frame dust emission spectrum and associated physical properties (both for the dust and the stellar populations) corresponding to the minimum $\chi^2$ (the best-fitting model). Then for each wavelength $\lambda$ we compute the relation between each physical property $p$ and the luminosity $\lambda L_\lambda\left(\lambda\right)$. As the flux density of the emission spectrum of galaxies does not necessarily vary linearly with the physical properties, we have elected to fit simple power laws. In effect, considering the logarithms of the respective quantities, the approach reduces to:

\begin{equation}
  \log \lambda L_\lambda\left(\lambda\right) = \alpha\left(\lambda\right)\times\log p + \beta\left(\lambda\right)\label{eqn:fit},
\end{equation}
with $\alpha\left(\lambda\right)$ and $\beta\left(\lambda\right)$ the wavelength-dependent coefficients determined from the fits. As we will see in Sect.~\ref{sec:results1}, $\alpha\left(\lambda\right)$ is a particularly interesting quantity as it indicates whether the luminosity at wavelength $\lambda$ changes linearly ($\alpha\left(\lambda\right)=1$) or non-linearly ($\alpha\left(\lambda\right)\neq 1$) with $p$. We will eventually adopt a more complex formulation by considering two parameters simultaneously, but for now we focus on a single-parameter dependence. Naturally, as we will see in Sect.~\ref{ssec:single}, the reciprocal of Eq.~\ref{eqn:fit} allows us to estimate a physical property $p$ from $\lambda L_\lambda\left(\lambda\right)$. We use the former formulation to study parameterized IR spectra (Sect.~\ref{sec:results1}) and the latter to provide estimations of the physical properties from one or more IR photometric bands (Sect.~\ref{sec:results2}).

\begin{figure*}
 \includegraphics[width=0.33\textwidth]{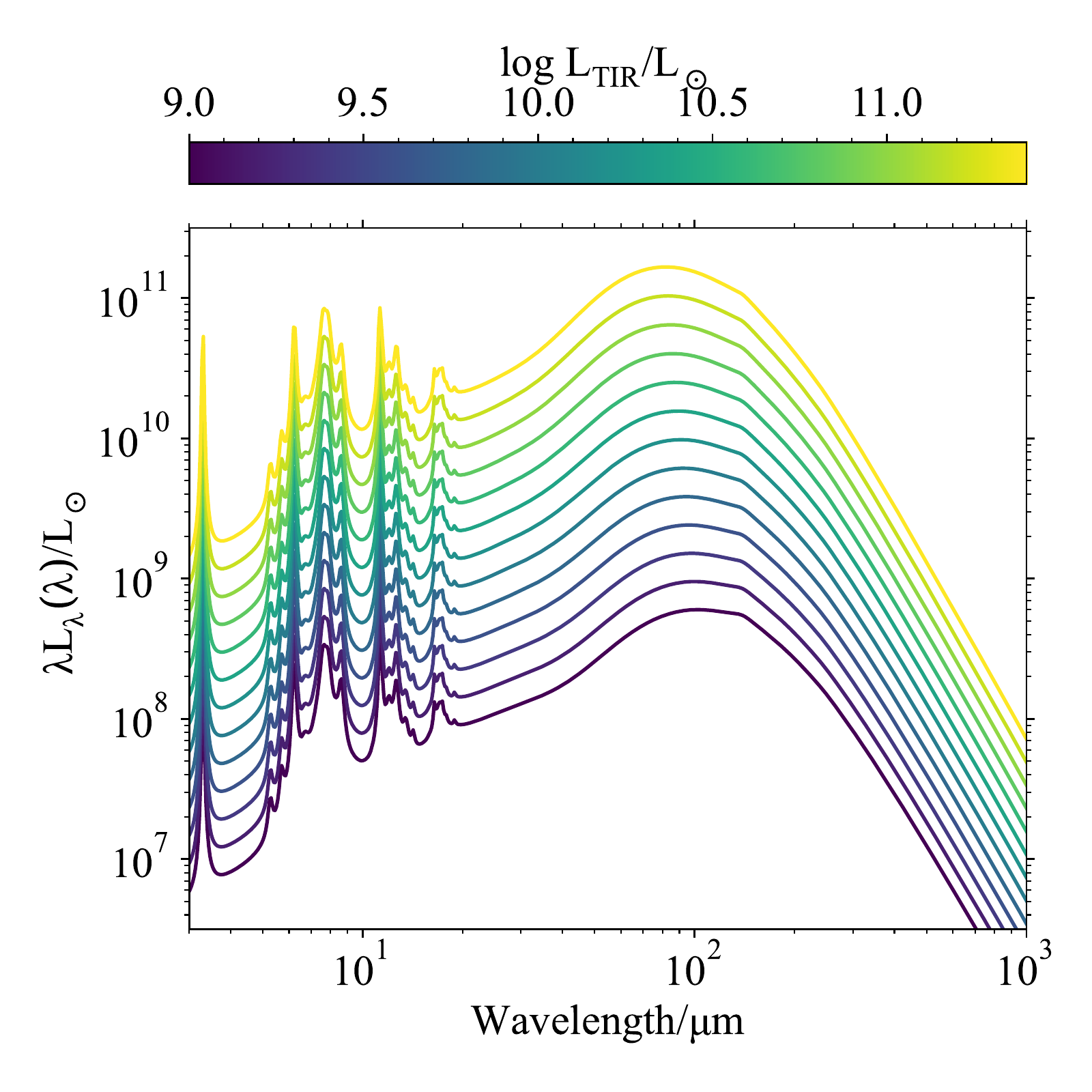}
 \includegraphics[width=0.33\textwidth]{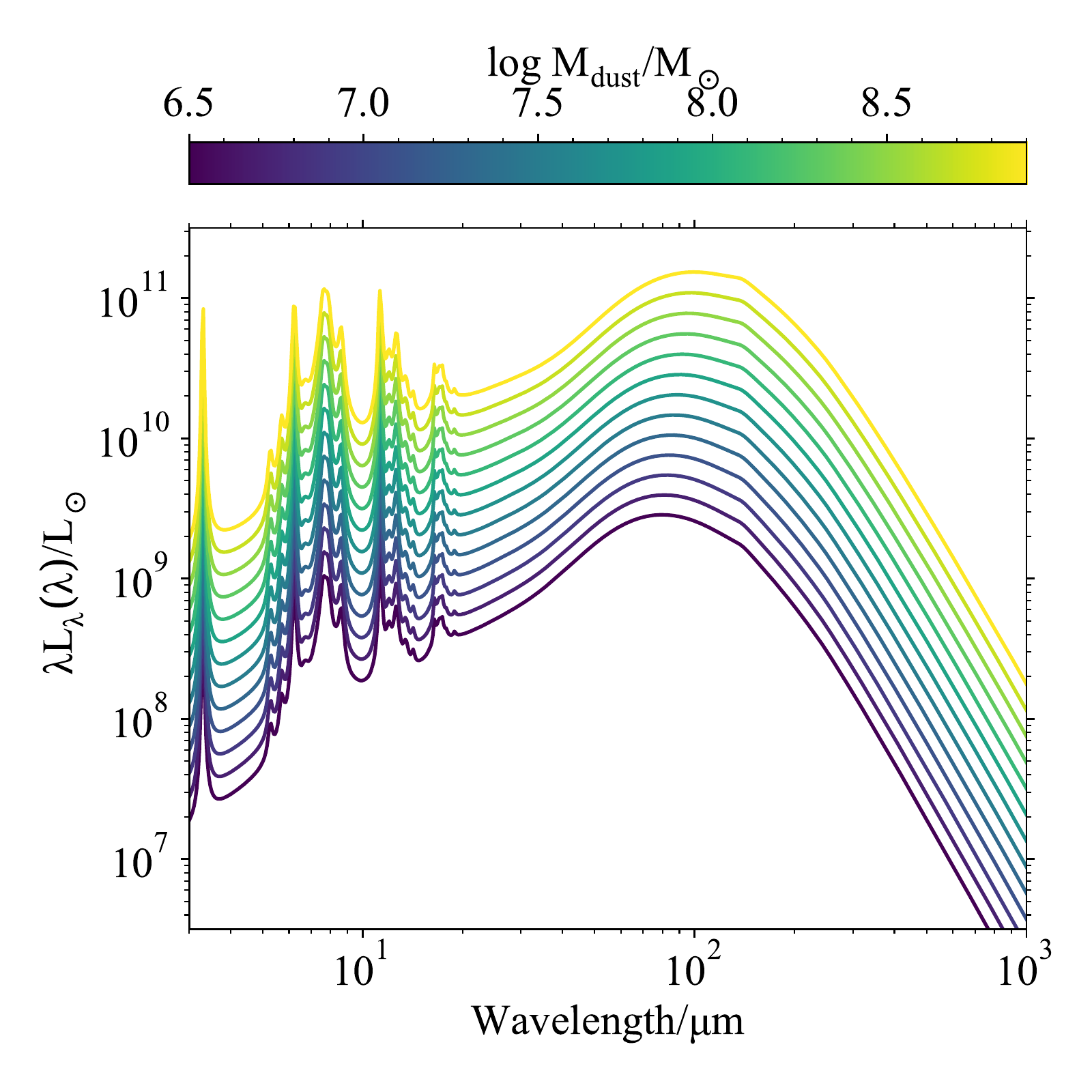}
 \includegraphics[width=0.33\textwidth]{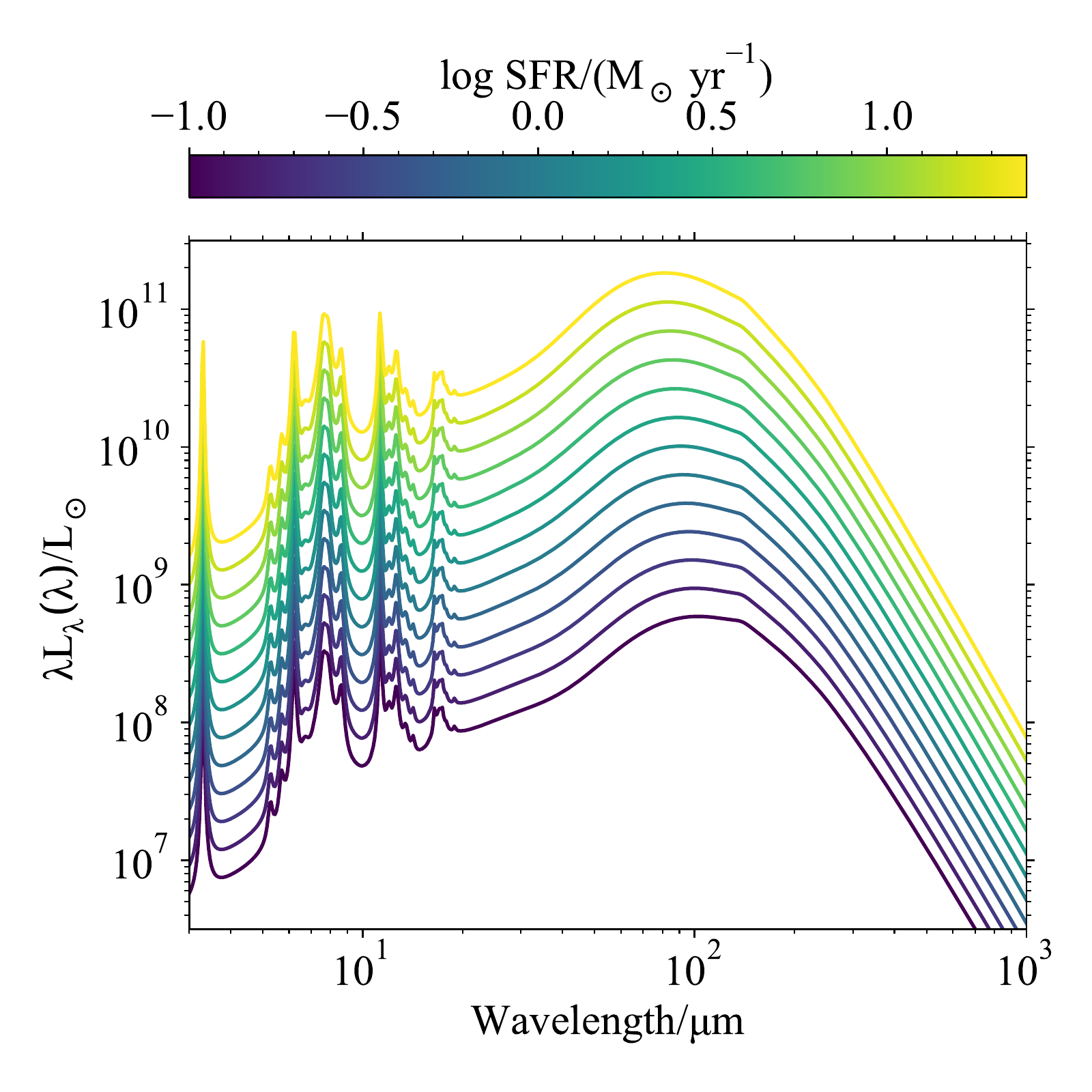}\\
 \includegraphics[width=0.33\textwidth]{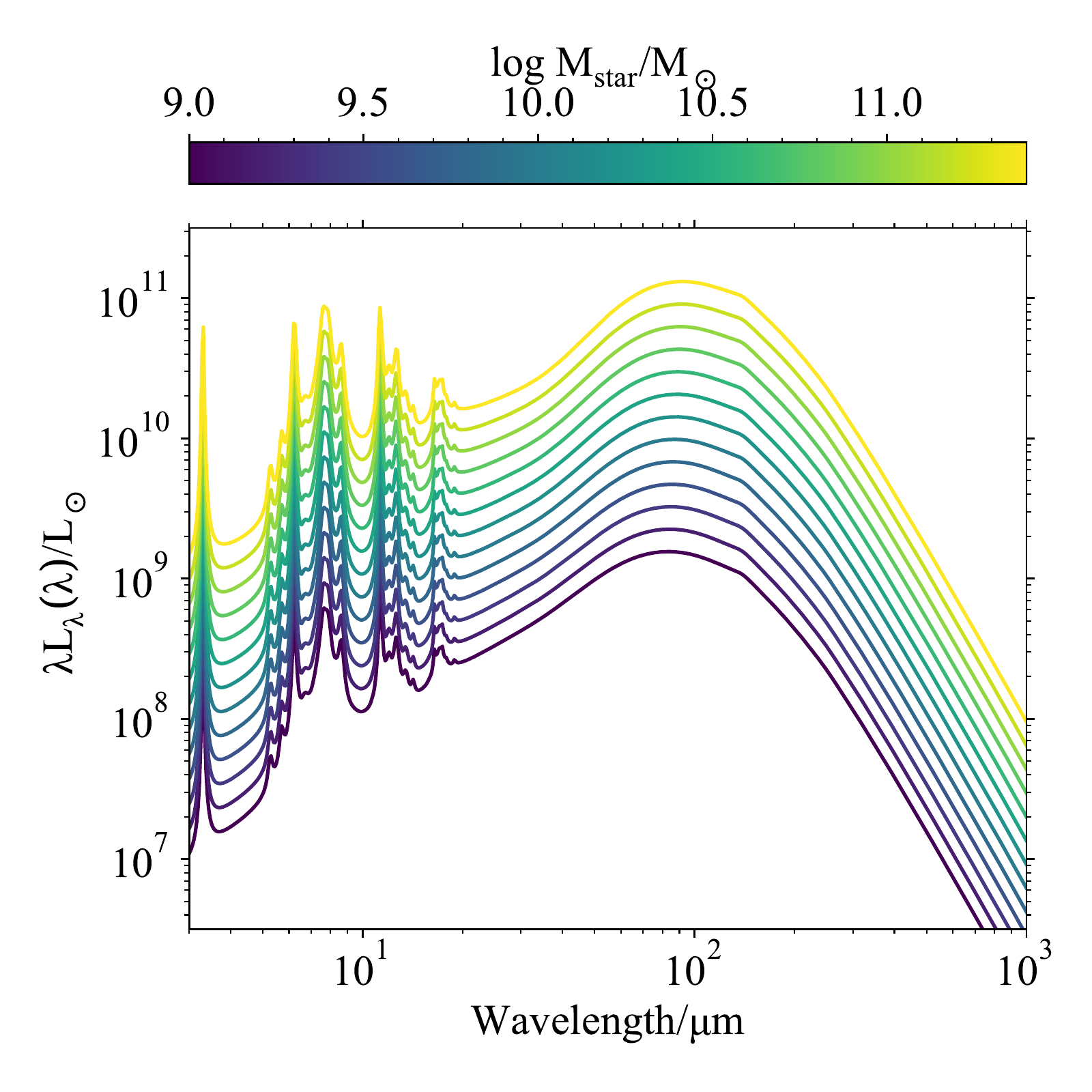}
 \includegraphics[width=0.33\textwidth]{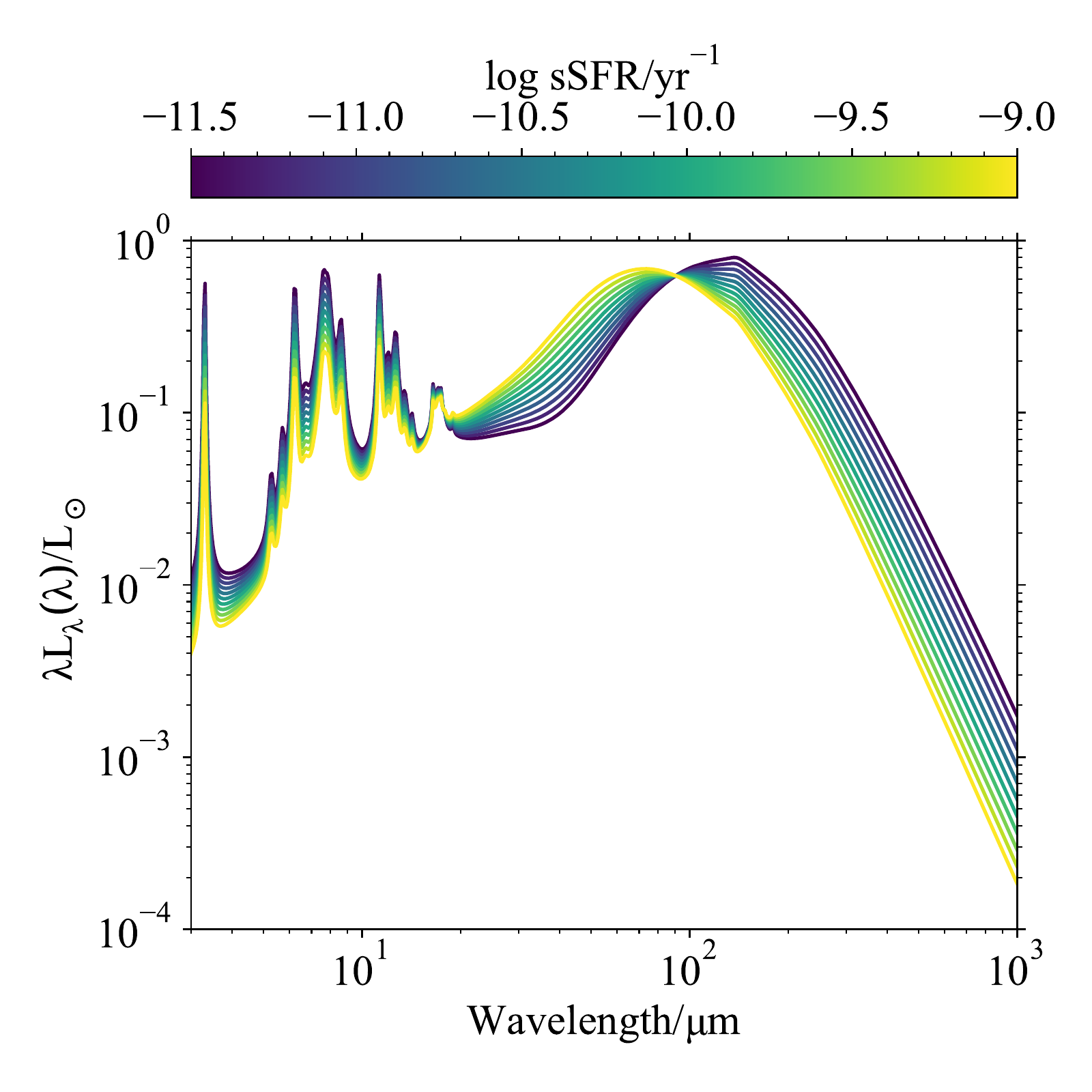}
 \includegraphics[width=0.33\textwidth]{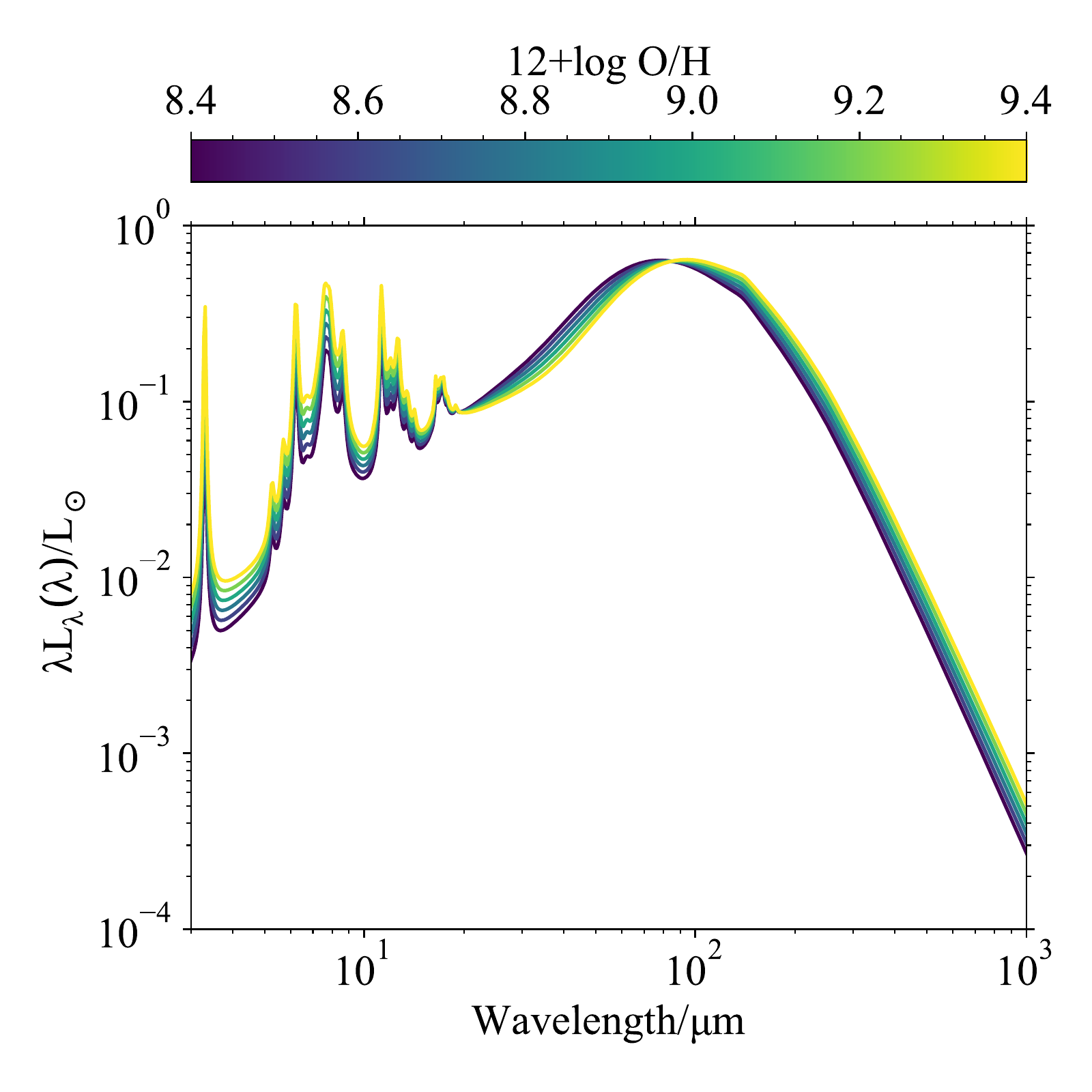}
 \caption{From the top-left to the bottom-right, dust emission spectra parameterized on $L_{TIR}$, $M_{dust}$, SFR, $M_{star}$, sSFR, and the oxygen abundance. The spectra are calibrated in absolute luminosity for the extensive properties, but they are normalized to $L_{TIR}$ for the intensive properties. The color of each spectrum follows the corresponding physical property as indicated by the color bar to the right of each panel.\label{fig:templates-1-prop}}
\end{figure*}

\section{Results: physical drivers of the diversity of dust emission spectra\label{sec:results1}}

We divide the presentation of our results into two sections. In the current section we explore how IR spectra depend on the physical properties, in particular, dust luminosity, SFR, sSFR, stellar mass, dust mass, and oxygen abundance (gas-phase metallicity). In addition to providing us with a physical insight into what drives the diversity of dust emission spectra, we also investigate how well can each of these properties be determined from different parts of the IR spectrum. Informed by this analysis, in the next section we discuss the estimation of two of these properties (dust luminosity and SFR) using one or more IR bands.

\subsection{IR spectra as a function of extensive and intensive properties\label{ssec:templates}}
In this section we explore relationships between the dust emission spectra and various physical properties, which can be divided into extensive and intensive properties. Extensive properties (SFR, $M_{star}$, $M_{dust}$, and $L_{TIR}$) involve absolute rather than relative fluxes, i.e., they directly scale with the ``extent'' of the galaxy. Intensive properties (sSFR and oxygen abundance), on the other hand involve normalized or relative fluxes. Even though it is inherently not possible to parameterize dust emission spectra on an intensive property, this becomes possible if they are first normalized by an extensive property. We therefore explore relationships between IR SED normalized by $L_{TIR}$ and two intensive physical properties.

We present in Fig.~\ref{fig:templates-1-prop} the dust emission spectra as a function of the six aforementioned physical properties.
The spectra parameterized on extensive physical properties show large absolute variations. First and foremost, their monochromatic luminosities (in log) scale very well with the physical properties, often in a nearly linear way, as we will see later. There are more subtle but nevertheless clear relative variations as well, in particular regarding the location of the peak of the modified black body emission, such that it moves to shorter (respectively longer) wavelengths with increasing $L_{TIR}$ and SFR (resp. $M_{dust}$ and $M_{star}$). The shift is intuitively explained in that higher levels of $L_{TIR}$ and SFR probably trace more intense radiation fields that heat the dust to warmer temperatures. Conversely, in dustier galaxies the energy is distributed over a larger number of dust grains (everything else being the same) and higher mass galaxies tend to have a softer, less intense radiation field than low mass galaxies, which are optically bluer. 

\begin{figure*}[!htbp]
 \includegraphics[width=\columnwidth]{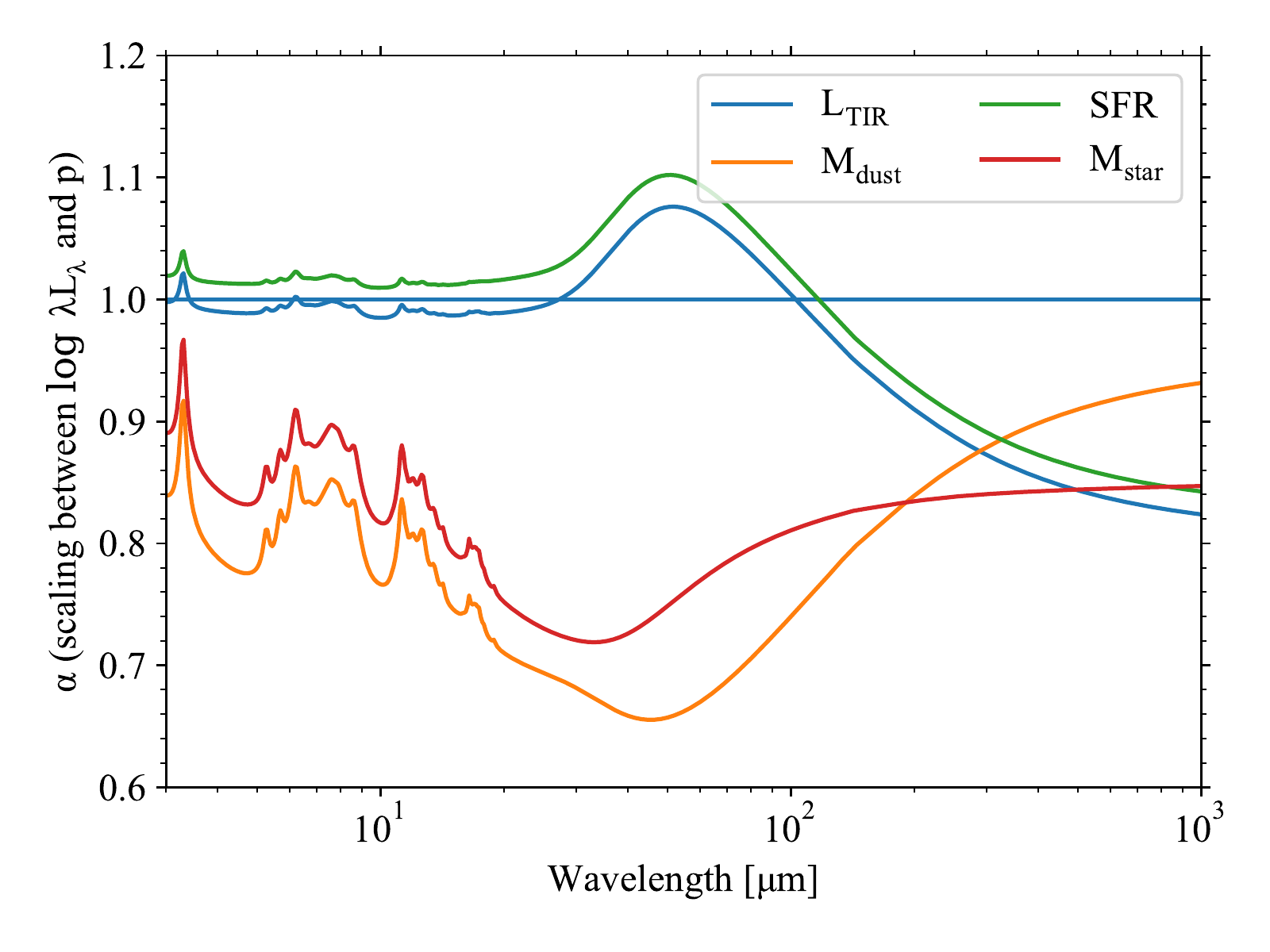}
 \includegraphics[width=\columnwidth]{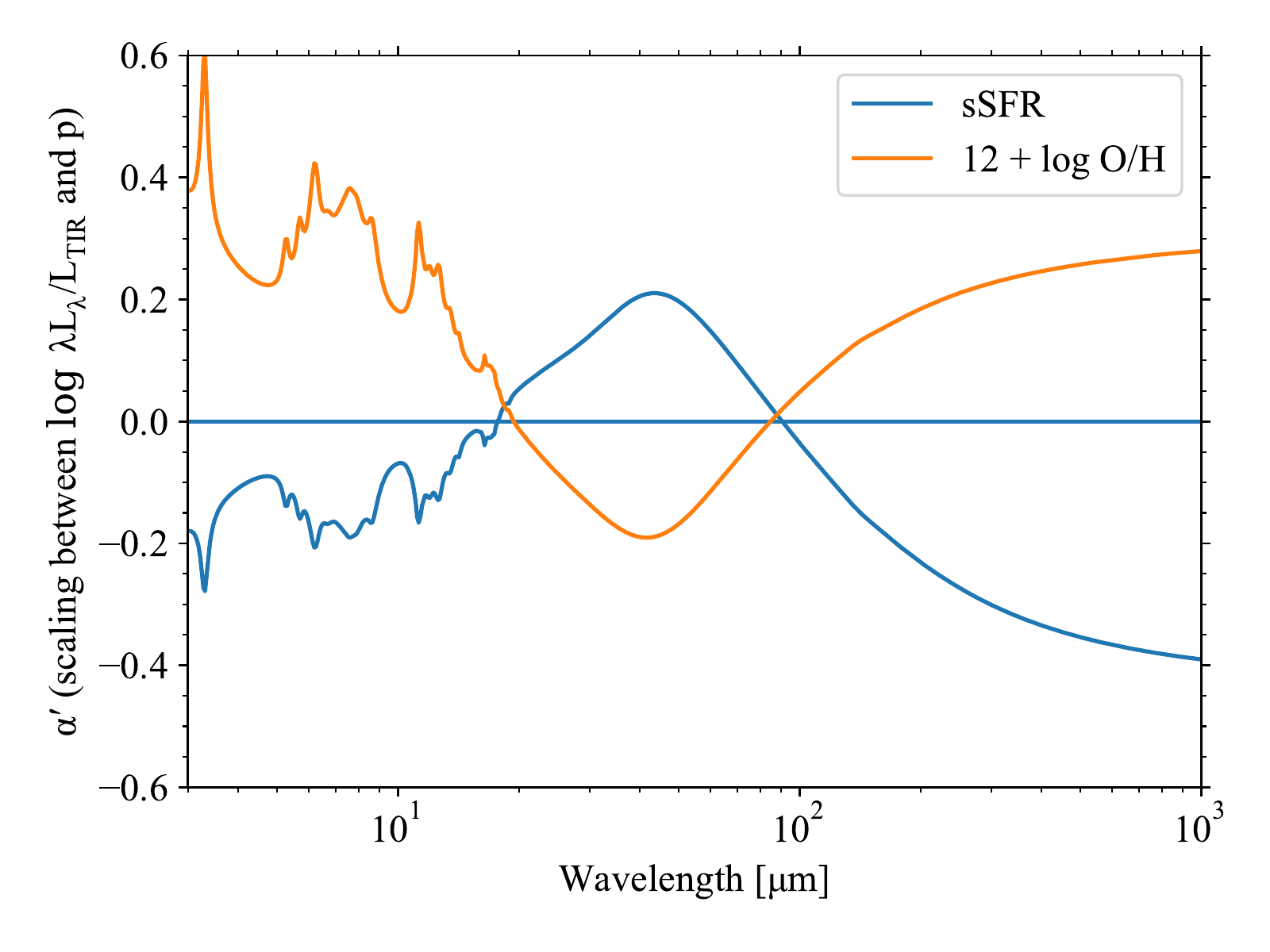}
 \caption{The scaling coefficient between the monochromatic luminosity and a physical property versus the wavelength. The scaling coefficient $\alpha$ for the extensive properties $L_{TIR}$ (blue), $M_{dust}$ (orange), SFR (green), and $M_{star}$ (red) is shown in the left panel. The scaling coefficient $\alpha^\prime$ for the intensive properties sSFR (blue) and the oxygen abundance (orange) is shown in the right panel. For the latter, $\mathrm{\lambda L_\lambda}$ has been normalized to $L_{TIR}$, i.e., it shows the residuals at fixed luminosity. The black horizontal lines indicate respectively the threshold between super-linearity and sub-linearity (left), and between correlation and anti-correlation (right). \label{fig:alpha}}
\end{figure*}

Even though the above description provides a satisfactory overall qualitative explanation, the physical properties are not independent from one another and there can be important variations due to secondary parameters. Such variations can actually be seen in the trends of the dust emission spectrum with intensive properties. For instance, a higher sSFR corresponds to a warmer dust temperature with a displacement of the peak towards shorter wavelengths, which is what we would intuitively expect. This strong trend with sSFR shows that neither SFR nor $M_{star}$ are sufficient by themselves to account for the full range of the dust emission spectral diversity, and indeed the trend with sSFR may be more fundamental \citep{dacunha2008a,nordon2012a,magnelli2014a}. Finally, the oxygen abundance has a more moderate impact on the variation of the spectrum. As expected, but still interesting that we see that in these ``monotonized'' spectra, the galaxies with the higher oxygen abundance have a stronger PAH emission and a slightly colder peak compared to lower metallicity galaxies. However, even at lower oxygen abundance PAH emission remains fairly important. The reason is that we have a dearth of very low metallicity objects in this sample. Indeed the MIR emission of low-metallicity galaxies tends to be more difficult to detect at the shallow depth of large surveys. With over 99.9\% of the sample galaxies having an oxygen abundance larger than 8.4, this is consistent with the results obtained for instance by \cite{engelbracht2005a,engelbracht2008a} with \textit{Spitzer}.

\subsection{How monochromatic IR luminosities scale with physical properties\label{ssec:scaling}}
To gain additional insight into the physical drivers of variations in the dust emission spectra, we show in Fig.~\ref{fig:alpha} how $\alpha$, the scaling between a monochromatic luminosity and a physical property (see Eq.~\ref{eqn:fit}), varies as a function of wavelength. For extensive quantities, a value of 1 discriminates between the super-linear (the monochromatic luminosity increases faster than the physical property) and the sub-linear (the monochromatic luminosity increases slower than the physical property) regimes.

Both $L_{TIR}$ and the SFR present a remarkably similar behavior over the full wavelength range, showing the strength of the dust emission to estimate the SFR of galaxies. Note that the SFR discussed in this paper is not what is sometimes referred to as the ``obscured'' SFR, but rather the true, total SFR. Interestingly, the scaling only varies slightly depending on whether the wavelength lies within a PAH line or is situated in the continuum. This constancy helps in cases when the redshift of the source and therefore where the filter is placed in rest frame is not precisely known.  At longer wavelengths, the relations become clearly super-linear for both properties, peaking close to 50~$\mu$m, before decreasing, becoming briefly linear again around 100~$\mu$m. This super-linear range is likely due to the progressively warmer modified-blackbody at higher $L_{TIR}$ and SFR, which causes a rapid increase of the emission as the peak shifts to shorter wavelengths. Finally, the scaling becomes increasingly sub-linear beyond 100~$\mu$m. At these wavelengths the emission progressively becomes dominated by the cold dust, which contributes less to $L_{TIR}$  and is only weakly related to the SFR,  as these physical properties more tightly relate to the emission of warmer dust emitting at shorter wavelengths.

Our results regarding the scaling between the monochromatic luminosity and  $L_{TIR}$ are qualitatively similar to those of \cite{rieke2009a}, derived for nearby, FIR selected galaxies. They find an $\alpha \approx 0.85$ sub-linearity at 8 and 12 $\mu$m and an $\alpha \approx 1.1$ super-linearity at 24 and 60 $\mu$m.

The relation based on $M_{star}$ shows a systematic sub-linearity, with $\alpha$ varying typically between $0.75$ and $0.95$. A sub-linearity is expected as higher mass galaxies tend to have progressively redder stellar populations. Therefore an increase in $M_{star}$ translates into a smaller increase of the energy absorbed and re-emitted by dust. Scaling is closer to linear in PAH lines, possibly due to the role of the interstellar radiation field, and therefore, older stars, in the excitation of these lines \citep[e.g., ][]{haas2002a}.

The case of $M_{dust}$ is more interesting in that $\alpha$ shows a stronger dependence on wavelength than $M_{star}$. Up to around 200~$\mu$m, $\alpha$ is lower for $M_{dust}$ than it is for $M_{star}$. Dust dominating the emission at these wavelengths represents only a minor fraction of $M_{dust}$. There is a change in the regime at longer wavelengths, with $\alpha$ progressively pushing above 0.9 in the sub-millimeter, reflecting the fact that the bulk of the dust is cold and dominates the emission in the Rayleigh-Jeans regime. This forms the basis for using a single sub-millimeter band as a measure of the dust mass, and by extension of the gas mass too \cite[e.g.,][]{dunne2001a, groves2015a, scoville2016a, millard2020a}.

In the case of intensive properties, we have normalized the spectra to $L_{TIR}$, which means that the scaling coefficient, which we denote $\alpha^\prime$ to differentiate it from $\alpha$ that is used in the case for extensive properties, probes variations of the shape of the emission spectra, and not their absolute normalization. Relations with $\alpha^\prime\left(\lambda\right)=0$, would indicate the normalized emission spectra are invariant at wavelength $\lambda$ with the considered physical property, i.e., no dependence between the normalized monochromatic luminosity and the intensive property. Fig.~\ref{fig:alpha} shows that the sSFR and the oxygen abundance are almost perfectly anti-correlated, with their $\alpha^\prime\left(\lambda\right)$ of opposite signs at almost all wavelengths. This behavior is probably the consequence of the anti-correlation between the oxygen abundance and sSFR at a fixed stellar mass, the extension of the mass--metallicity relation \citep{ellison2008a,salim2014a}. There are however two regions where $\alpha^\prime\left(\lambda\right)\simeq0$, slightly bellow $\sim20$~$\mu$m and around $\sim90$~$\mu$m. With the dust emission being largely independent from these physical properties at these wavelengths, the monochromatic dust luminosity and $L_{TIR}$ should be in linear relation with one another. This is consistent with what we found previously. Between these two wavelengths, $\alpha^\prime\left(\lambda\right)>0$ ($\alpha^\prime\left(\lambda\right)<0$) for the sSFR (oxygen abundance), which translates into the increase (decrease) of dust temperature with respect to the sSFR (oxygen abundance). Conversely, at shorter wavelengths we see the strengthening of PAH features with increasing metallicity and their slight decrease with the sSFR. At long wavelengths, the sub-millimeter emission is strongly dependent on metallicity, probably as a consequence of an increase of the gas-to-dust mass ratio, whereas it weakens at high sSFR as a larger fraction of the luminosity is emitted at shorter wavelengths.

\subsection{How monochromatic IR luminosities correlate with physical properties}
\begin{figure}[!htbp]
 \includegraphics[width=\columnwidth]{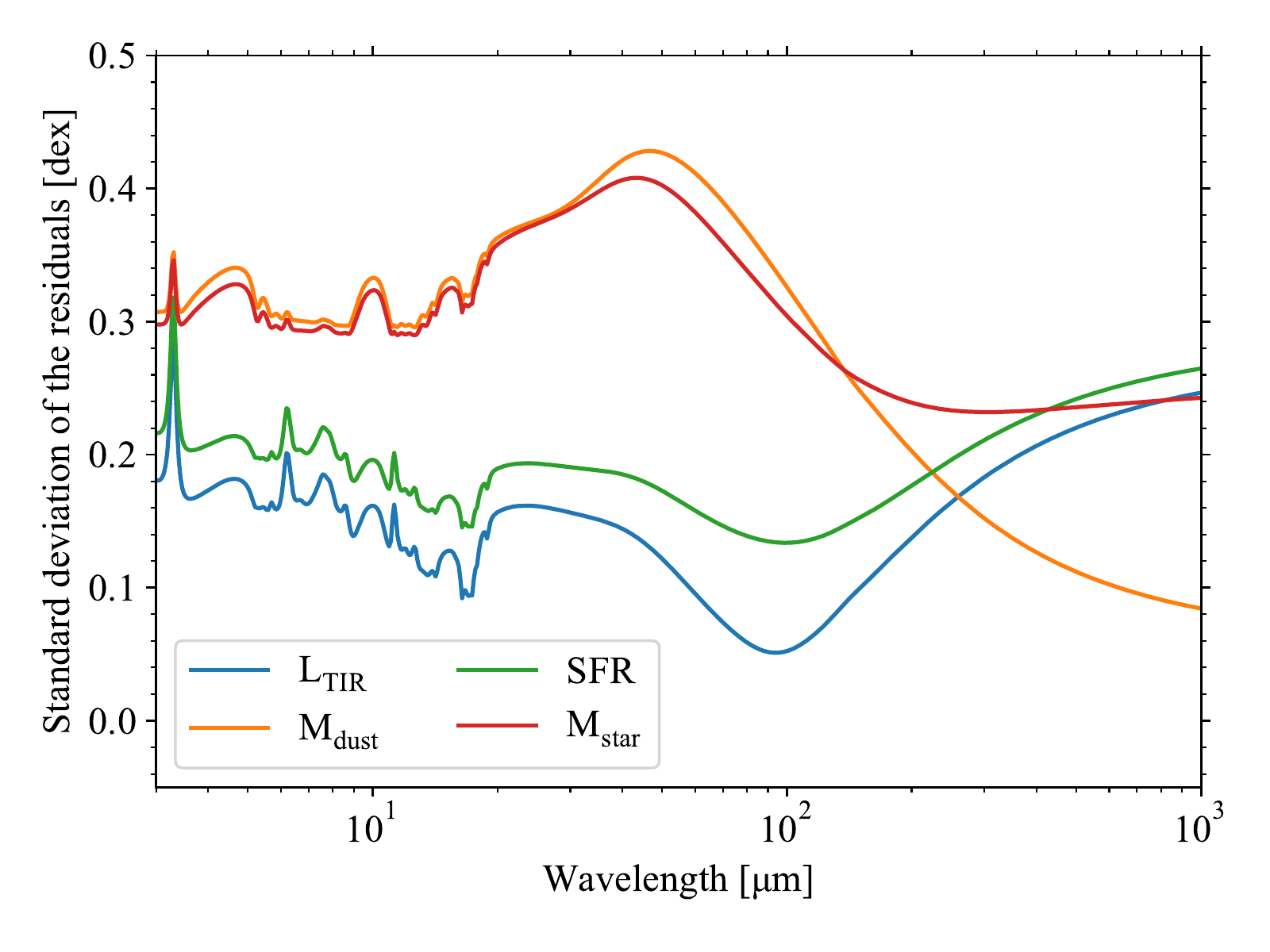}
 \caption{The standard deviation of the residuals around the relation between the monochromatic luminosity and a physical property (Eq.~\ref{eqn:fit}) versus the wavelength, for the extensive properties $L_{TIR}$ (blue), $M_{dust}$ (orange), SFR (green), and $M_{star}$ (red).\label{fig:scatter}}
\end{figure}

Besides the linearity, another important quantity to consider is the scatter around the relation, i.e., the degree of the correlation between monochromatic luminosities and some physical property. Indeed, a relation that is linear but presents a very large scatter may ultimately not be sufficiently reliable for individual objects and a non-linear relation but with a smaller scatter will be more desirable for parameter estimation. To investigate this aspect, we present in Fig.~\ref{fig:scatter} the scatter of the residuals around the relations versus the wavelength.

First, it appears that in addition to being strongly non-linear, the $M_{star}$ relation presents a high standard deviation that is typically larger than 0.3~dex. Interestingly, the scatter is smallest in the MIR and then at $\lambda>100~\mu$m, lending additional support to the notion that the two regions are governed by interstellar radiation field heating \citep{haas2002a}.  Up to $\sim100$~$\mu$m the $M_{dust}$ relation has a behavior that is very similar to that of $M_{star}$. However, for $M_{dust}$ the standard deviation keeps decreasing with increasing wavelength, reaching less than 0.1~dex at 1 mm, confirming once again the reliability of a single long wavelength band to estimate the dust mass in star-forming galaxies, and showing that the longer the wavelength the better the estimator of $M_{dust}$ it is.

$L_{TIR}$ shows by far the smallest standard deviation of all extensive quantities, with a minimum of slightly more than $\sim 0.05$~dex (corresponding to a relative scatter of $\sim13$\% in linear scale) at $\sim$95~$\mu$m. In the 20-40~$\mu$m range however the standard deviation is higher, from 0.15~dex to 0.16~dex. Even though on average the monochromatic luminosity and $L_{TIR}$ are in linear relation with one another in the MIR, the bulk of $L_{TIR}$ is emitted around the peak of the modified black body. Thus we can expect that variations in the dust emission emerging in the MIR would have a relatively smaller influence on $L_{TIR}$, leading to an increase of the standard deviation of the residuals in this spectral region. It is interesting to note that the region between 20~$\mu$m and 50~$\mu$m actually has a higher standard deviation than the region between 10~$\mu$m and 20~$\mu$m, invalidating a simplified notion that going to longer wavelengths towards the peak is always preferred. The reason for this may lie in the prominent role of stochastically heated very small grains, the emission of which peaks in this region \citep{desert1990a}. Only beyond 50~$\mu$m does the standard deviation diminish as one moves towards the peak of the emission spectrum, dominated by large grain emission. This consideration may be relevant when designing IR detectors. Also, it is worth pointing out that the region beyond $\sim$200 $\mu$m is actually inferior to MIR, potentially informing the use of ALMA vs.\ JWST. Altogether, simultaneously taking into account the linearity, the lack of dependence on sSFR, and the small standard deviation, suggests that, even though no wavelength is flawless, the emission of the dust around 90-100~$\mu$m and $L_{TIR}$ trace each other best.

Compared to the $L_{TIR}$ relation, the SFR relation shows a scatter higher by 0.03~dex to 0.05~dex in the MIR, a region for which we showed the relations are close to being perfectly linear. Overall, the best monochromatic luminosities for estimating SFR are the same that best determine $L_{TIR}$ ($\sim18~\mu$m and $\sim100~\mu$m). There has been some debate in the literature whether a monochromatic luminosity is a better tracer of SFR than $L_{TIR}$.

Since SFR is the total of UV/optically obscured and unobscured contributions, the larger scatter in comparison to $L_{TIR}$ may be due to the diversity of attenuation properties across the sample (e.g., amplitude of the attenuation, shape of the attenuation curve, differential reddening between different stellar populations) as well as the range of relative contributions of young and old stellar populations to dust heating. The limitations in the ability to estimate the total SFR from dust emission alone could be lifted by using hybrid SFR estimators, which combine IR and UV luminosities \citep[e.g.,][]{elbaz2007a,daddi2007a,hao2011a,liu2011a,kennicutt2009a,boquien2016a}, a topic that we will explore in detail in a future publication, or more generally by performing SED fitting utilizing the full UV through IR SED.

The overall analysis in the current and the preceding sections provides more detail to our current understanding of the physical drivers of the dust emission, confirming the close connection of the MIR and 80-100~$\mu$m spectral regions with both $L_{TIR}$ and the SFR. The former connection is important in particular for JWST, which will only be able to target the MIR for galaxies up to $z\sim2$. The analysis also highlights the sensitivity of dust parameter estimation on sSFR, which we will address shortly. Finally, we confirm that the sub-millimeter emission is a reassuringly reliable tracer of the dust mass.

\section{Results: estimation of TIR luminosity and SFR from IR photometry\label{sec:results2}}
\subsection{Parameterization of dust emission templates and relations\label{ssec:param}}
A parameterization on an extensive quantity has the added benefit that it makes it possible to derive $L_{TIR}$ using a single flux point or band. We point out that this extensive quantity does not need to be $L_{TIR}$. In particular, it may be useful to parameterize the templates on the total SFR, which, like $L_{TIR}$ displays a close relation to the shape of the IR spectrum (Fig.\ \ref{fig:templates-1-prop}), and which in our modeling is known from the UV-to-IR SED fitting. While the unobscured star formation is not directly observable in the IR, it is in some instances possible to estimate it reasonably well without the UV data. Furthermore, parameterizing directly on SFR has the advantage of not having to use a fixed conversion factor to translate $L_{TIR}$ into SFR. 

Previous studies have emphasized either the total luminosity \citep[e.g., ][]{chary2001a} or the sSFR \citep[e.g., ][]{dacunha2008a} as the principal drivers of the shape of the dust emission spectrum. What the right panel of Fig.~\ref{fig:alpha} shows is that, except for particular wavelengths, both are important. From the right panel of Fig.~\ref{fig:alpha} we see that the change in the sSFR of one dex at fixed $L_{TIR}$ results in the difference in the luminosity at 8 $\mu$m of nearly 0.2 dex. If the templates (or relations) only depended on $L_{TIR}$, the estimates of the total IR luminosity produced by them would be biased for galaxies with atypically high or atypically low levels of star formation, potentially leading to the systematics when our estimators that are constructed based on low redshift galaxies are applied to objects at higher redshifts where the average sSFR is higher. This motivates us to present our templates and relations parameterized on sSFR as well. In Fig.~\ref{fig:templates-1-prop} we see that the peak of the emission shifts from 120 to 70~$\mu$m as sSFR increases. This range encompasses the peaks of stacked SEDs going from $z\sim0$ to $z\sim2$ \citep{bethermin2015a}, giving us confidence that by incorporating the sSFR dependence we are effectively producing templates that are applicable to a wide range of redshifts. On the other hand, in the presence of multiple flux points, the color term implicitly accounts for the sSFR dependence, and including the sSFR explicitly in the relations is superfluous, as verified by the tests that we carried out. 

Two-parameter templates covering FIR (30--1000~$\mu$m) and based on spectra of simulated galaxies have been previously introduced by \cite{safarzadeh2016a}. However, their second parameter is $M_{dust}$, which cannot be well constrained in the absence of sub-millimeter data, which are often not available. \cite{magdis2012a} and \cite{safarzadeh2016a} stress the importance of $L_{TIR}/M_{dust}$ as a driver of the IR SED shape. Being qualitatively similar to $L_{TIR}/M_{dust}$, our sSFR dependence confers similar benefits, but is more accessible, requiring essentially only the knowledge of $M_{star}$ (Sect.~\ref{ssec:sSFR}).

To conclude, in this work we produce the following four types of average dust emission spectra (templates):

\begin{itemize}
 \item Templates parameterized on $L_{TIR}$.
 \item Templates parameterized on total (obscured plus unobscured) SFR.
 \item Templates parameterized on $L_{TIR}$ and sSFR simultaneously.
 \item Templates parameterized on total SFR and sSFR simultaneously.
\end{itemize}

As pointed out in Sect.~\ref{ssec:approach}, we argue that the estimation of the physical properties can be more easily carried out by direct, continuous relations rather than the templates. Therefore, we also construct the following relations for estimating $L_{TIR}$ and total SFR:

\begin{itemize}
 \item Relations for single-band flux measurements (Sect.~\ref{ssec:single} and \ref{ssec:single-Mdust}).
 \item sSFR-dependent relations for single-band flux measurements (Sect.~\ref{ssec:sSFR}).
 \item Relations for multiple-band flux measurements (Sect.~\ref{ssec:multi}).
\end{itemize}

Finally, we also provide a tool within the software package to generate any set of discrete templates that the user may find useful, specified by $L_{TIR}$ or SFR, with or without the additional dependence on sSFR (Sect.~\ref{ssec:software}).

\begin{figure*}
 \centering
 \includegraphics[width=\columnwidth]{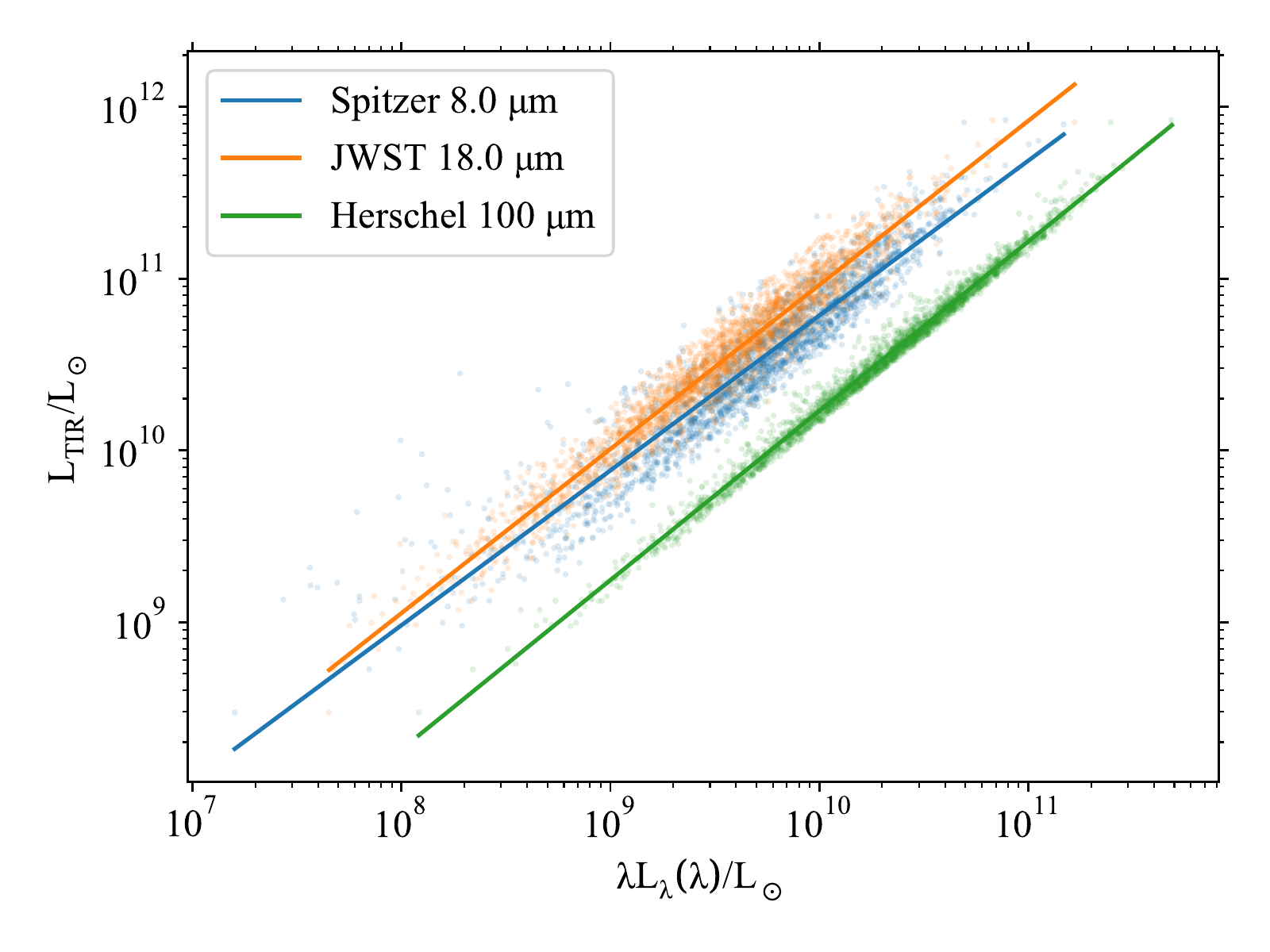}
 \includegraphics[width=\columnwidth]{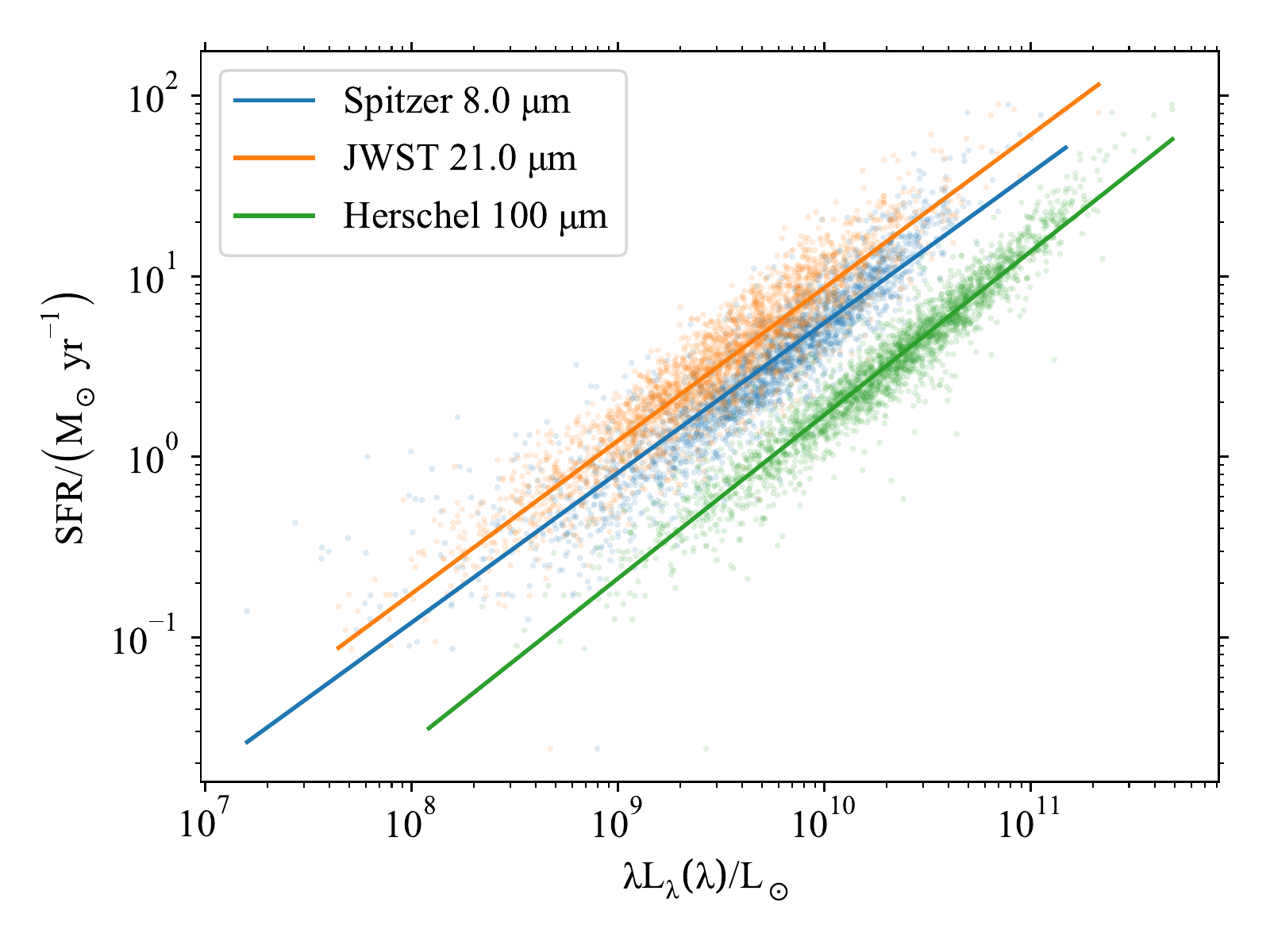}
 \caption{Estimation of $L_{TIR}$ (left) and the SFR (right) from the \textit{Spitzer} 8~$\mu$m (blue), JWST 18~$\mu$m (orange), and \textit{Herschel} 100~$\mu$m (green) bands. The dots represent our H-ATLAS sample and the solid lines the best-fit relation computed from Eq.~\ref{eqn:fit-LTIR-SFR}.\label{fig:LTIR-SFR-estimators}}
\end{figure*}

\subsection{Estimation of $L_{TIR}$ and the SFR from a single IR band\label{ssec:single}}
In this section we focus specifically on estimating $L_{TIR}$ and SFR from the emission in a single band and not taking into account the sSFR dependence. This is equivalent to using templates parameterized on $L_{TIR}$ or SFR. In our approach, the relations used for estimating $L_{TIR}$ and SFR are precomputed for particular bands. Specifically, we have selected a set of representative bands that are or will likely be extensively used for constraining the dust emission of galaxies: WISE 12~$\mu$m and 22~$\mu$m, \textit{Spitzer} 8.0~$\mu$m and 24~$\mu$m, \textit{Herschel} 70~$\mu$m, 100~$\mu$m, and 160~$\mu$m, and JWST 7.7~$\mu$m, 10.0~$\mu$m, 11.3~$\mu$m, 12.8~$\mu$m, 15.0~$\mu$m, 18.0~$\mu$m, 21.0~$\mu$m, and 25.5~$\mu$m. The fluxes in these bands are computed by integrating each best-fit template through the filter bandpasses. All the analysis in this subsection is performed assuming $z \approx 0$, but in the accompanying software package we provide estimators up to $z=4$ so that the estimation can be performed without an explicit K-correction. We present how to take variations of the sSFR into account in Sect.~\ref{ssec:sSFR}.

We determine the relation by inverting the dependent and independent variables of Eq.~\ref{eqn:fit} and repeating the fitting procedure:

\begin{equation}
  \log p = m\left(b\right)\times\log \lambda L_\lambda\left(b\right) + n\left(b\right)\label{eqn:fit-LTIR-SFR},
\end{equation}
with $p$ the physical property to be estimated ($L_{TIR}$ or the SFR), $\lambda L_\lambda\left(b\right)$ the luminosity in a photometric band $b$ of pivot wavelength $\lambda$, and $m\left(b\right)$ and $n\left(b\right)$ the coefficients obtained from the fit for that band. The luminosities are computed by integrating each spectrum through the corresponding bandpasses. 

As an illustration, we show in Fig.~\ref{fig:LTIR-SFR-estimators} the estimates of $L_{TIR}$ and the SFR based on Eq.~\ref{eqn:fit-LTIR-SFR} and three representative bands, \textit{Spitzer} 8~$\mu$m, JWST 18~$\mu$m, and \textit{Herschel} 100~$\mu$m. The data come from our sample used to derive the relations. As expected, all three bands correlate well with both $L_{TIR}$ and the SFR, however with visibly less scatter for the former, in agreement with analysis presented in Figure \ref{fig:scatter}. Some reduction in scatter, depending on the quality of the available sSFR estimates, is possible when using sSFR-dependent relations (Sect.~\ref{ssec:sSFR}). In Table~\ref{table:estimators-LTIR-SFR} we provide the coefficients $m$ and $n$ to estimate $L_{TIR}$ and the SFR for all the bands mentioned previously. These coefficients are valid only for redshifts close to zero. For use at other redshifts we direct the user to the provided software tool (Section \ref{ssec:software}).

\begin{table*}
 \centering
 \begin{tabular}{lcccccccc}
  \hline\hline
  Band&\multicolumn{4}{c}{$L_{TIR}$} & \multicolumn{4}{c}{SFR}\\
  {} &$m$&$n$&$R^2$&$\sigma$&$m$&$n$&$R^2$&$\sigma$\\\hline
\textit{Spitzer} 8.0~$\mu$m   & $0.9018$ & $1.7671$  & $0.8857$ & $0.1559$ & $0.8298$ & $-7.5567$ & $0.8149$ & $0.1825$\\
\textit{Spitzer} 24~$\mu$m    & $0.9078$ & $1.8731$  & $0.8915$ & $0.1523$ & $0.8403$ & $-7.5065$ & $0.8350$ & $0.1738$\\\hline
\textit{Herschel} 70~$\mu$m   & $0.9339$ & $0.9327$  & $0.9813$ & $0.0659$ & $0.8590$ & $-8.3202$ & $0.9155$ & $0.1289$\\
\textit{Herschel} 100~$\mu$m  & $0.9857$ & $0.3728$  & $0.9892$ & $0.0504$ & $0.9054$ & $-8.8225$ & $0.9209$ & $0.1250$\\
\textit{Herschel} 160~$\mu$m  & $1.0127$ & $0.3343$  & $0.9505$ & $0.1057$ & $0.9302$ & $-8.8580$ & $0.8797$ & $0.1514$\\
\textit{Herschel} 250~$\mu$m  & $0.9874$ & $1.0962$  & $0.8632$ & $0.1688$ & $0.9076$ & $-8.1650$ & $0.7884$ & $0.1930$\\
\textit{Herschel} 350~$\mu$m  & $0.9552$ & $1.9008$  & $0.7890$ & $0.2031$ & $0.8787$ & $-7.4309$ & $0.7111$ & $0.2187$\\
\textit{Herschel} 500~$\mu$m  & $0.9251$ & $2.7286$  & $0.7223$ & $0.2269$ & $0.8515$ & $-6.6736$ & $0.6417$ & $0.2372$\\\hline
JWST 7.7~$\mu$m      & $0.8925$ & $1.7981$  & $0.8755$ & $0.1619$ & $0.8210$ & $-7.5262$ & $0.8034$ & $0.1872$\\
JWST 10.0~$\mu$m     & $0.9250$ & $1.9536$  & $0.9035$ & $0.1444$ & $0.8534$ & $-7.4063$ & $0.8401$ & $0.1715$\\
JWST 11.3~$\mu$m     & $0.9242$ & $1.4463$  & $0.9104$ & $0.1395$ & $0.8508$ & $-7.8552$ & $0.8421$ & $0.1706$\\
JWST 12.8~$\mu$m     & $0.9526$ & $1.3800$  & $0.9395$ & $0.1163$ & $0.8778$ & $-7.9251$ & $0.8753$ & $0.1538$\\
JWST 15.0~$\mu$m     & $0.9598$ & $1.4862$  & $0.9450$ & $0.1111$ & $0.8861$ & $-7.8428$ & $0.8854$ & $0.1481$\\
JWST 18.0~$\mu$m     & $0.9556$ & $1.4071$  & $0.9419$ & $0.1141$ & $0.8830$ & $-7.9232$ & $0.8840$ & $0.1489$\\
JWST 21.0~$\mu$m     & $0.9153$ & $1.8410$  & $0.8963$ & $0.1492$ & $0.8470$ & $-7.5343$ & $0.8395$ & $0.1718$\\
JWST 25.5~$\mu$m     & $0.9053$ & $1.8678$  & $0.8914$ & $0.1523$ & $0.8380$ & $-7.5115$ & $0.8349$ & $0.1738$\\\hline
WISE 12.0~$\mu$m     & $0.9521$ & $1.4316$  & $0.9380$ & $0.1175$ & $0.8776$ & $-7.8795$ & $0.8743$ & $0.1543$\\
WISE 22.0~$\mu$m     & $0.9095$ & $1.8804$  & $0.8911$ & $0.1525$ & $0.8419$ & $-7.4995$ & $0.8344$ & $0.1741$\\\hline
 \end{tabular}
 \caption{Coefficients required to estimate $L_{TIR}$ in $\mathrm{L_\odot}$ and the SFR in $\mathrm{M_\odot~yr^{-1}}$ from a single \textit{Spitzer}, \textit{Herschel}, JWST, or WISE band (Eq.~\ref{eqn:fit-LTIR-SFR}) for $z\sim 0$ galaxies. These coefficients along with the variances and co-variances between $m$ and $n$ are available electronically at full numerical precision up to $z=4$.\label{table:estimators-LTIR-SFR}}
\end{table*}

We see that in line with our findings presented in Sect.~\ref{ssec:scaling}, the emission around 100~$\mu$m is one of the best tracers for both $L_{TIR}$ and the SFR. Surprisingly, most estimators are sublinear ($m<1$), even though the reverse relation is also sublinear ($\alpha<1$). Upon closer inspection, the reason for this counterintuitive behavior is because the variance for each band is larger than the covariance with $L_{TIR}$ or the SFR. We note that both in Sect.~\ref{ssec:scaling} and here we are using an ordinary least-square fit. In essence, in each case we know the dependent variables perfectly as we rely entirely on the best-fitting models for the emission spectra and the associated physical properties, and depending on the case, the objective is indeed to minimize the scatter either for the emission spectrum (Sect.~\ref{ssec:scaling}) or the physical properties (current section). It is important to keep this aspect in mind when considering the physical interpretations of our results.

\subsection{Estimation of $M_{dust}$ from a single IR band\label{ssec:single-Mdust}}
\begin{figure}
 \includegraphics[width=\columnwidth]{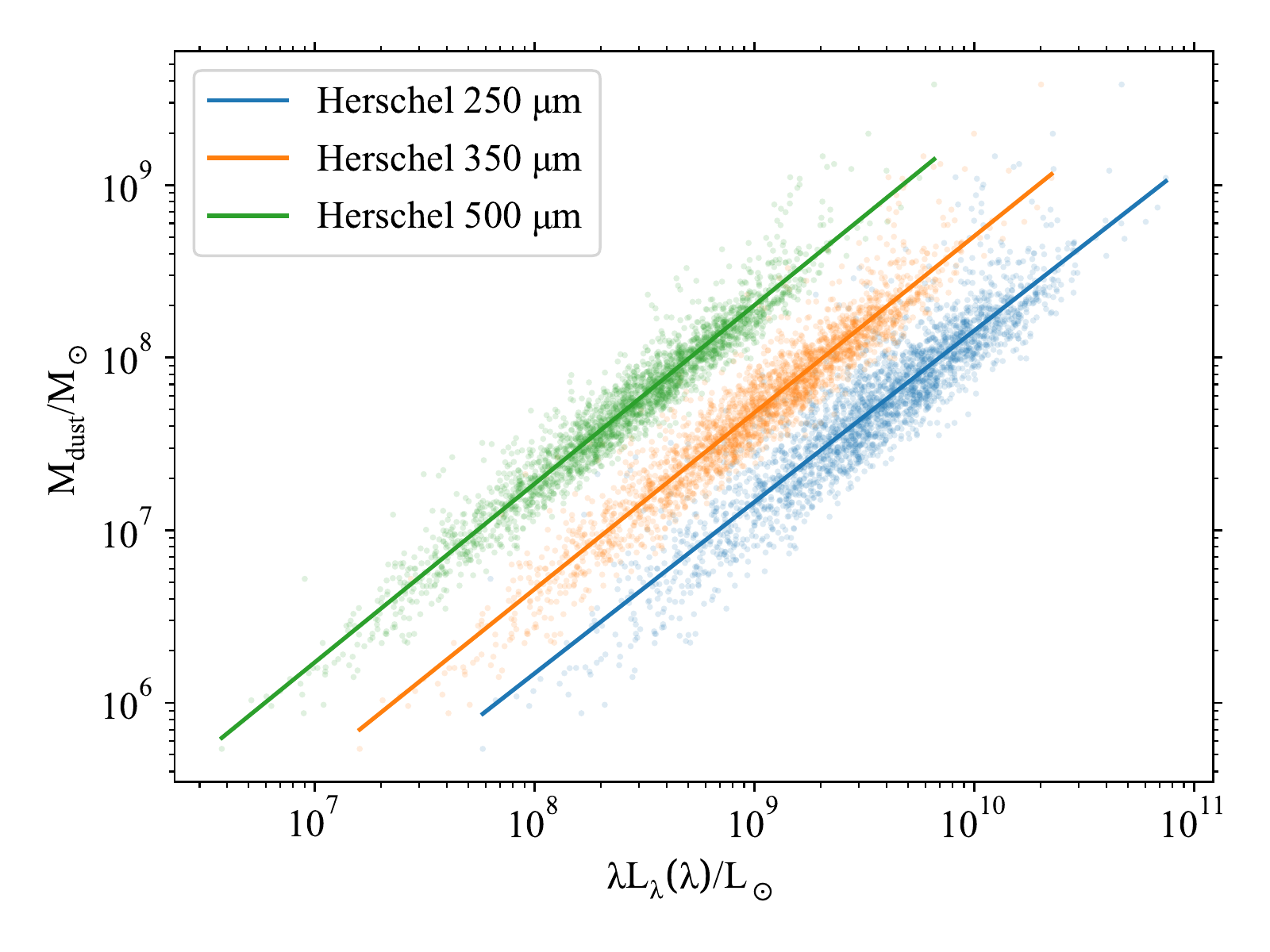}
 \caption{Estimation of $M_{dust}$ from the \textit{Herschel} 250~$\mu$m (blue), 350~$\mu$m (orange), and 500~$\mu$m (green) bands. The dots represent our H-ATLAS sample and the solid lines the best-fit relation computed from Eq.~\ref{eqn:fit-LTIR-SFR}.\label{fig:estimator-Mdust}}
\end{figure}

Another important physical property to estimate is $M_{dust}$, which can serve as a proxy for the total gas mass, while also providing constraints on the dust production and destruction processes. We saw in Sect.~\ref{sec:results1} that the estimation of $M_{dust}$ at longer wavelengths has an excellent potential, with $\alpha$ approaching 1, while the standard deviation of the residuals are progressively dropping with wavelength. We see in Fig.~\ref{fig:estimator-Mdust} that there is indeed a reasonably tight relation between $M_{dust}$ and the luminosities in the \textit{Herschel} SPIRE bands.
We do not consider here the \textit{Herschel} PACS bands because the shorter wavelengths are also sensitive to warmer dust and do not provide sufficiently precise results (Fig.~\ref{fig:scatter}). We give the coefficients for $M_{dust}$ estimation using Eq.~\ref{eqn:fit-LTIR-SFR} and \textit{Herschel} SPIRE bands at $z \sim 0$ in Table~\ref{table:estimators-Mdust}.
\begin{table}
 \centering
 \begin{tabular}{lcccc}
 \hline\hline
 Band&$m$&$n$&$R^2$&$\sigma$\\\hline
\textit{Herschel} 250~$\mu$m  & $0.9932$ & $-1.7761$  & $0.8281$ & $0.1904$\\
\textit{Herschel} 350~$\mu$m  & $1.0227$ & $-1.5235$  & $0.8979$ & $0.1513$\\
\textit{Herschel} 500~$\mu$m  & $1.0347$ & $-1.0090$  & $0.9345$ & $0.1232$\\\hline
 \end{tabular}
 \caption{Coefficients required to estimate $M_{dust}$ in $\mathrm{M_\odot}$ from Eq.~\ref{eqn:fit-LTIR-SFR} from \textit{Herschel} SPIRE bands for $z\sim 0$ galaxies. These coefficients along with the variances and co-variances between $m$ and $n$ are available electronically at full numerical precision up to $z=4$. \label{table:estimators-Mdust}}
\end{table}
All SPIRE bands are almost perfectly linear estimators of $M_{dust}$. The standard deviation of the residuals goes from $0.19$~dex at 250~$\mu$m down to $0.12$~dex at 500~$\mu$m, an improvement of slightly more than 30\%. One caveat to keep in mind using this estimator is that contrary to $L_{TIR}$ or the SFR, $M_{dust}$ is sensitive to the details of the underlying dust models, such as the emissivity index. Assumptions differing from that of the \cite{draine2007a} and \cite{draine2014a} models, including, for example, the choice to model carbonaceous grains as amorphous rather than graphite \citep{schreiber2018a}, may yield systematic differences in the resulting $M_{dust}$.

\subsection{Estimation of $L_{TIR}$ and the SFR from a single IR band and the sSFR\label{ssec:sSFR}}
Even though single-band estimators can provide us with good estimates for low-redshift samples on average, they may suffer from biases in some regions of the parameter space, in particular for galaxies that have atypically low or high sSFR at fixed $L_{TIR}$ (Sect.~\ref{ssec:param}), or for general population of galaxies at higher redshifts. For those cases it is recommended to use sSFR-dependent templates and relations.

Following the approach we used in Sect.~ \ref{ssec:single}, we have computed estimators as a function of one IR band and the sSFR:
\begin{equation}
  \log p = m\left(b\right)\times\log \lambda L_\lambda\left(b\right) + n\left(b\right) + s\times \log \mathrm{sSFR} \label{eqn:fit-LTIR-SFR-second-parameter}.
\end{equation}
We give the corresponding coefficients in Table~\ref{table:estimators-LTIR-SFR-second-parameter}.
\begin{table*}
 \centering
 \begin{tabular}{lcccccccccc}
  \hline\hline
  Band&\multicolumn{5}{c}{$L_{TIR}$} & \multicolumn{5}{c}{SFR}\\
  {} &$s$&$m$&$n$&$R^2$&$\sigma$&$s$&$m$&$n$&$R^2$&$\sigma$\\\hline
\textit{Spitzer} 8.0~$\mu$m   & $0.1827$ & $0.8898$ & $3.6919$ & $0.9082$ & $0.1411$ & $0.3453$ & $0.8073$ & $-3.9192$ & $0.9096$ & $0.1330$\\
\textit{Spitzer} 24~$\mu$m    & $-0.0595$ & $0.9196$ & $1.1719$ & $0.8937$ & $0.1508$ & $0.1335$ & $0.8139$ & $-5.9338$ & $0.8487$ & $0.1675$\\\hline
\textit{Herschel} 70~$\mu$m   & $-0.0667$ & $0.9466$ & $0.1415$ & $0.9837$ & $0.0616$ & $0.1284$ & $0.8344$ & $-6.7979$ & $0.9265$ & $0.1208$\\
\textit{Herschel} 100~$\mu$m  & $0.0401$ & $0.9799$ & $0.8291$ & $0.9901$ & $0.0483$ & $0.2199$ & $0.8737$ & $-6.3191$ & $0.9535$ & $0.0973$\\
\textit{Herschel} 160~$\mu$m  & $0.1691$ & $0.9997$ & $2.1397$ & $0.9676$ & $0.0862$ & $0.3332$ & $0.9045$ & $-5.2999$ & $0.9592$ & $0.0915$\\
\textit{Herschel} 250~$\mu$m  & $0.2669$ & $0.9822$ & $3.7884$ & $0.9124$ & $0.1381$ & $0.4214$ & $0.8995$ & $-3.9146$ & $0.9310$ & $0.1173$\\
\textit{Herschel} 350~$\mu$m  & $0.3146$ & $0.9587$ & $4.9849$ & $0.8650$ & $0.1679$ & $0.4652$ & $0.8838$ & $-2.8707$ & $0.9011$ & $0.1385$\\
\textit{Herschel} 500~$\mu$m  & $0.3458$ & $0.9360$ & $6.0610$ & $0.8226$ & $0.1889$ & $0.4942$ & $0.8670$ & $-1.9114$ & $0.8731$ & $0.1550$\\\hline
JWST 7.7~$\mu$m      & $0.1911$ & $0.8809$ & $3.8029$ & $0.9006$ & $0.1464$ & $0.3528$ & $0.7997$ & $-3.8246$ & $0.9039$ & $0.1367$\\
JWST 10.0~$\mu$m     & $0.1010$ & $0.9136$ & $3.0586$ & $0.9101$ & $0.1397$ & $0.2724$ & $0.8227$ & $-4.4248$ & $0.8975$ & $0.1408$\\
JWST 11.3~$\mu$m     & $0.1600$ & $0.9117$ & $3.1539$ & $0.9270$ & $0.1270$ & $0.3250$ & $0.8253$ & $-4.3876$ & $0.9228$ & $0.1236$\\
JWST 12.8~$\mu$m     & $0.1178$ & $0.9405$ & $2.6620$ & $0.9479$ & $0.1083$ & $0.2873$ & $0.8483$ & $-4.7976$ & $0.9354$ & $0.1138$\\
JWST 15.0~$\mu$m     & $0.0474$ & $0.9530$ & $2.0188$ & $0.9464$ & $0.1098$ & $0.2253$ & $0.8540$ & $-5.3105$ & $0.9218$ & $0.1244$\\
JWST 18.0~$\mu$m     & $0.0100$ & $0.9540$ & $1.5211$ & $0.9419$ & $0.1140$ & $0.1927$ & $0.8517$ & $-5.7166$ & $0.9106$ & $0.1323$\\
JWST 21.0~$\mu$m     & $-0.0329$ & $0.9213$ & $1.4580$ & $0.8970$ & $0.1487$ & $0.1561$ & $0.8181$ & $-5.7154$ & $0.8581$ & $0.1628$\\
JWST 25.5~$\mu$m     & $-0.0725$ & $0.9201$ & $1.0090$ & $0.8947$ & $0.1502$ & $0.1225$ & $0.8130$ & $-6.0602$ & $0.8464$ & $0.1685$\\\hline
WISE 12.0~$\mu$m     & $0.1138$ & $0.9402$ & $2.6725$ & $0.9460$ & $0.1102$ & $0.2838$ & $0.8478$ & $-4.7854$ & $0.9331$ & $0.1157$\\
WISE 22.0~$\mu$m     & $-0.0469$ & $0.9185$ & $1.3306$ & $0.8925$ & $0.1516$ & $0.1441$ & $0.8142$ & $-5.8096$ & $0.8504$ & $0.1666$\\\hline
 \end{tabular}
 \caption{Coefficients required to estimate $L_{TIR}$ in $\mathrm{L_\odot}$ and the SFR in $\mathrm{M_\odot~yr^{-1}}$ from a single \textit{Spitzer}, \textit{Herschel}, JWST, or WISE band and the sSFR (Eq.~\ref{eqn:fit-LTIR-SFR-second-parameter}) for $z\sim 0$ galaxies. These coefficients along with the variances and co-variances between $s$, $m$, and $n$ are available electronically at full numerical precision up to $z=4$.\label{table:estimators-LTIR-SFR-second-parameter}}
\end{table*}
Two quantities are of particular interest, the coefficient $s$, which scales log~sSFR, and $\sigma$, the standard deviation of the residuals. As for $\alpha^\prime$, a value of $s$ close to 0 indicates that the estimator is largely independent from sSFR. For $L_{TIR}$ this is in particular the case of the MIR bands that do not cover prominent PAH features, or the FIR around 100~$\mu$m, close to the peak of the emission. Conversely, bands with strong PAH features and long wavelength emission beyond the peak show the strongest dependence. The reason for the stronger dependence for bands overlapping prominent PAH features is not entirely clear as we would expect the PAH and $L_{TIR}$ emission to scale linearly with each other over a fairly large range of radiation field intensities \cite[see for instance Fig.~15 of][]{draine2007a}. However this could be an indirect effect of the PAH abundance, since in our sample galaxies with a higher oxygen abundance tend to have a lower sSFR, or the consequence of the role of older populations in heating both the PAH features and large grains (see Sect.~\ref{ssec:scaling}).

The standard deviations of the residuals in Table~\ref{table:estimators-LTIR-SFR-second-parameter} show systematically reduced values with respect to Table~\ref{table:estimators-LTIR-SFR}. Unsurprisingly the largest reductions correspond to the largest values of $s$. The relative improvement is greater for the estimation of SFR than $L_{TIR}$ because the proportionality between these two quantities is itself dependent on sSFR.  

The sSFR is a quantity that can be challenging to evaluate, ideally requiring SED modeling. The availability of the sSFR would also in many cases eliminate the need to separately estimate $L_{TIR}$ or the SFR. However, uncertainties on the sSFR are generally small compared to the full dynamical range of the sSFR, and the shape of the templates varies slowly and monotonically with the sSFR. This means that even an imprecise estimate of the sSFR (or one obtained indirectly using some population age estimate, e.g., the D4000 index or H$\alpha$ equivalent width) would still prove highly useful to reduce or even eliminate possible biases. As a matter of fact, as long as the stellar mass is available, it is possible to use the SFR estimators iteratively. For instance, the initial SFR could be estimated from the relations given in Table~\ref{table:estimators-LTIR-SFR}. Combined with $M_{star}$ it would yield an estimate of the sSFR. Then it would be possible to apply the relations provided in Table~\ref{table:estimators-LTIR-SFR-second-parameter}. Applying these simple steps appears to be sufficient to eliminate the dependency with the sSFR. As an example, for the \textit{Spitzer} 8~$\mu$m band, the $R^2$ coefficient of the difference between the exact $L_{TIR}$ from the model and the estimated $L_{TIR}$ from the estimator with respect to the sSFR goes from 0.1787 to 0.0197. Even though the determination of the SFR is not ideal as the relation itself shows one of the strongest dependencies on the sSFR among all the bands, the approximation remains sufficient to obtain satisfactory results to eliminate systematic biases. 

\subsection{Explicit estimation of SFR and $L_{TIR}$ from JWST 21 $\mu$m observed at various redshifts}
\begin{figure}
 \includegraphics[width=1.05\columnwidth]{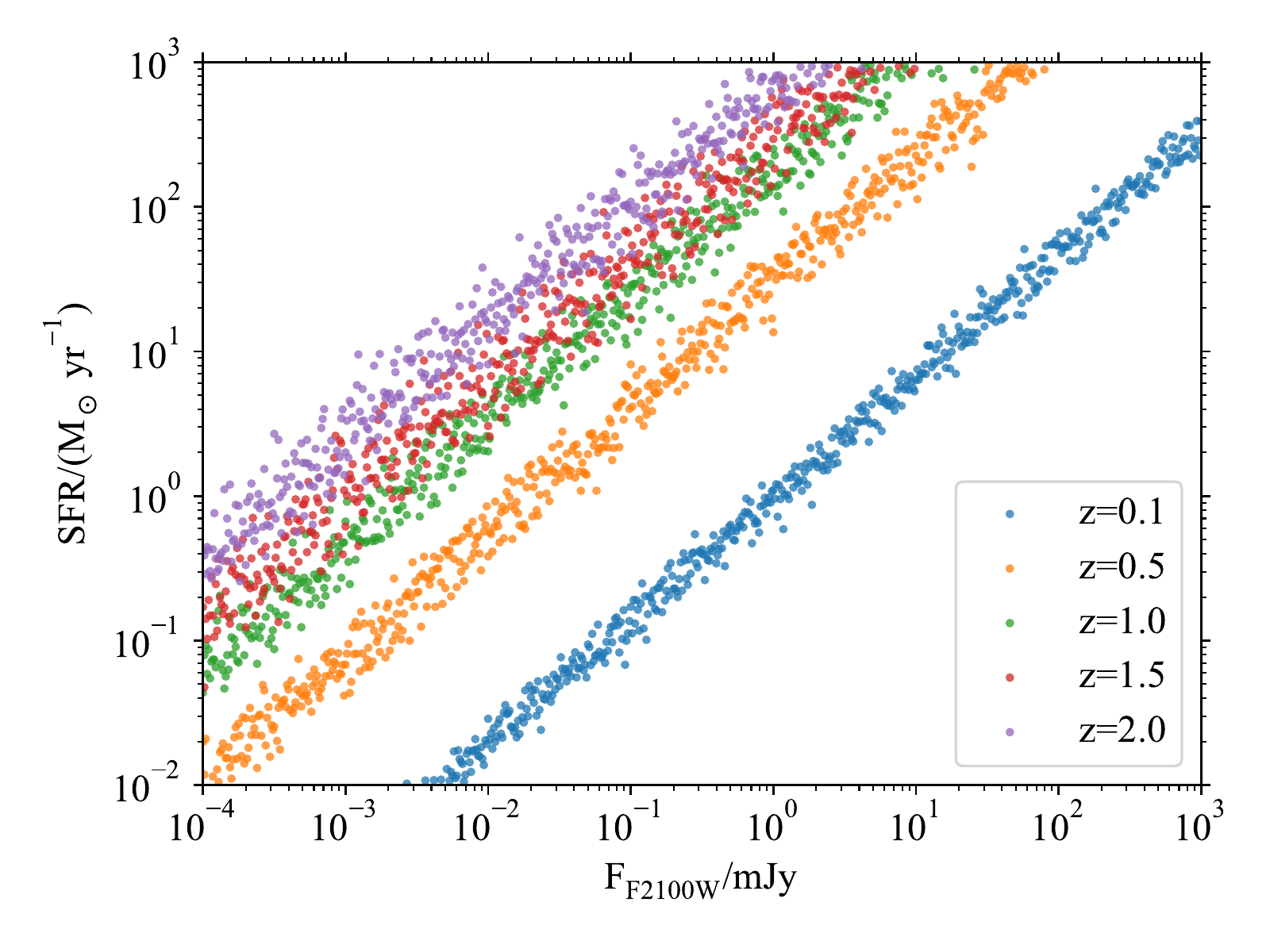}
 \caption{Estimated total (obscured plus unobscured) SFR as a function of the flux measured in JWST MIRI 21 $\mu$m filter (F2100W) and redshift, based on the relations that take into account the evolution of sSFR. The scatter results from assuming a 1-$\sigma$ scatter of 0.5 dex in sSFR of actively star-forming galaxies at each redshift. \label{fig:f2100w}}
\end{figure}

F2100W is the longest wavelength filter on JWST's Mid-Infrared Instrument \citep[MIRI,][]{rieke2015a} with very good sensitivity, and will be a workforce for extragalactic studies from the local universe up to the cosmic noon. The software tool that accompanies this paper allows the calculation of SFR and $L_{TIR}$ from F2100W flux for arbitrary redshift. Nevertheless, to facilitate quick calculation, in this section we provide simple, redshift-dependent formulas to derive SFR and $L_{TIR}$ from F2100W fluxes. We estimate these quantities at different redshifts for a range of fluxes using sSFR-dependent relations, where we assume the sSFR that corresponds to the main sequence at that redshift at $\log M_*=10.5$, according to \cite{speagle2014a}. In order to simulate the effect of the intrinsic width of the main sequence and the fact that the stellar mass may be different from $\log M_*=10.5$, we perturb sSFR by 0.5 dex 1-$\sigma$ Gaussian. Figure \ref{fig:f2100w} shows the dependence of SFR on the observed flux in five redshift bins. The scatter is the result of the assumed scatter in sSFR which makes it possible for the galaxies with the same flux and the same redshift to have different SFR or IR luminosities.

If stellar masses (and therefore sSFR) of galaxies are unknown, one can use the relations to estimate SFR and IR luminosity that assume sSFR evolution from \cite{speagle2014a}:
\begin{gather}
\log {\rm SFR}  = 0.83 \log F_{\rm F2100W}+2.23 \log z + 2.19, \qquad \sigma=0.19,\\
\log L_{TIR} = 0.92 \log F_{\rm F2100W}+2.19 \log z + 12.23, \qquad \sigma=0.14.
\end{gather}

If the stellar masses are known, SFR can be estimated using the above relation and the sSFR-dependent relations provided below can then be used to refine the result.
\begin{multline}
\log {\rm SFR} = 0.84 \log F_{\rm F2100W}+1.83 \log z +\ \\ 0.31 \log {\rm sSFR}+5.00,\qquad \sigma=0.10,
\end{multline}
\begin{multline}
\log L_{TIR} = 0.92 \log F_{\rm F2100W}+2.00 \log z +\ \\0.15 \log {\rm sSFR}+ 13.57,\qquad \sigma=0.12.
\end{multline}
SFR estimates especially benefit from the inclusion of the sSFR term because the relationship between SFR and $L_{TIR}$ is to first order dependent on the sSFR.

To get the full error on estimated quantities, the relative flux errors should be added in quadrature to the above uncertainties for the relationships. The relations should not be used above $z=2.2$ where the rest-frame wavelength covered by F2100W becomes rather short. Our estimates for the relation between F2100W and $L_{LIR}$ agree well with those of \citet{schreiber2018a}: at $\log L_{TIR}=11$ the difference is $<$0.1~dex at $z<0.5$ and $<$0.2~dex at $1<z<2$. The principal difference is that we do not assume a linear relation between monochromatic and total dust luminosities. Furthermore, we provide specific relations for SFR which are even more sublinear than the ones for $L_{TIR}$.

\subsection{Estimation of $L_{TIR}$ and the SFR from multiple IR bands\label{ssec:multi}}
Whenever possible, it is desirable to include information provided by several bands to determine $L_{TIR}$ or the SFR. The fundamental reason being that having more bands provides additional constraints on the shape of the emission spectrum. Following this idea, multiple derivations have been provided in the literature with excellent results using a diverse combination of bands from IRAS \citep{sanders1996a} to \textit{Spitzer} \citep{dale2002a,boquien2010a} and \textit{Herschel} \citep{boquien2011a,galametz2013a}. We will limit our relations to a maximum of 4 bands, yielding a total of 4047 band combinations. The fitting procedure is similar to that presented in Sect.~\ref{ssec:single} but extended to multiple bands:

\begin{equation}
  \log p = \sum_im_i\times\log \lambda_i L_\lambda\left(b_i\right) + n\label{eqn:fit-multiple-bands},
\end{equation}
with each scaling coefficient $m_i$ corresponding to band $b_i$, $i$ being the index of the band. Given the particularly large size of the resulting table, we provide these coefficients in an electronic form only.

We must mention, however, that this derivation is made with a small but important modification with respect to the single-band case. Previously it was not necessary to impose any bound on $m$ or $n$. When considering the case of Eq.~\ref{eqn:fit-multiple-bands}, a priori nothing would restrict $m_i$ to be negative, which could intuitively be understood as color terms. However, bands providing very similar information are degenerate. This is often the case of bands with close wavelengths, for instance JWST 21~$\mu$m and WISE 22~$\mu$m. Without imposed bounds, the two $m_i$ will be of opposite sign and have similar and very high absolute values (for instance $-50$ and $+50$). This means that in practical cases even a small amount of noise will be amplified to a considerable degree, strongly perturbing the estimate. To address this issue, we have imposed that all $m_i$ are bounded between 0 and 2. Analysis of these estimators with synthetic catalogs injected with noise shows that even though the residuals around the calibration sample are slightly higher when the $m_i$ coefficients are bound, they provide us with much more reliable estimates as they are more resilient to the photometric noise.

Throughout this section we primarily focused on relations for galaxies at $z\sim 0$. The behavior of the coefficients in the scaling relations at different redshifts in the case of a single band is presented in Appendix \ref{sec:appen_z}.

\subsection{Data products and software\label{ssec:software}}
Given the wealth of data contained in the relations we have presented, their practical use could be a challenge. To address this, we provide a number of products to the community on the  web site that hosts GSWLC \footnote{\url{https://salims.pages.iu.edu/bosa/}}, the sample upon which this study is based.

First, following the traditional approach, we provide a grid of templates parametrized on the physical properties as described in Sect.~\ref{ssec:param}, both as \textsc{ascii} and \textsc{fits} files. Both for flexibility and ease-of-use, we also provide a software tool to generate templates for any value of the physical properties and also for any combination of $L_{TIR}$ and sSFR as a second parameter. An interactive visualization tool of the templates based on these two parameters is also provided.

Furthermore, we provide a tool to estimate $L_{TIR}$ and SFR from an input table of fluxes and redshifts for any combination of up to four bands. For estimates based on single bands, a value that takes into account sSFR dependence is also provided. The fitting coefficients given in Table~\ref{table:estimators-LTIR-SFR} along with the corresponding covariance matrices for any combination of up to 4 bands from $z=0$ to $z=4$ with steps of $0.01$ are provided with this tool.

\section{Discussion\label{sec:discussion}}
\subsection{Comparison with previously published templates\label{ssec:comparison1}}
In this section we compare our templates to four previously published sets of templates: \cite{chary2001a}, \cite{dale2002a} (in particular the \cite{dale2014a} update), \cite{rieke2009a}, and \cite{smith2012b} templates updated to include \textit{Herschel} PACS bands of a larger sample than in the original paper\footnote{Private communication from D. J. B.\ Smith.}. Except for \cite{smith2012b}, the other templates were derived or constrained using samples, often heterogeneous, selected from shallow FIR surveys. Details of the construction of these four template sets are given in Appendix \ref{sec:appen_temp}. We limit our comparison to template sets that have been parameterized on an extensive quantity ($L_{TIR}$). 

A straightforward direct comparison between these templates is not always possible as $L_{TIR}$ may not have been computed over the exact same wavelength range, the templates are generally defined on a different discrete grid of $L_{TIR}$, or the templates may not have been parametrized on $L_{TIR}$ to begin with, as is the case of the \cite{dale2002a} templates. In order to provide a fair comparison we have therefore taken a number of steps to homogenize these template sets. First, we have parametrized the \cite{dale2002a} templates against $L_{TIR}$ by using the relation of \cite{marcillac2006a}, which links $L_{TIR}$ to the 60-to-100~$\mu$m ratio, in combination with Table~2 from \cite{dale2002a}, which links this ratio to $\alpha_{SF}$, the intrinsic parameter of the templates (not to be confused with the $\alpha(\lambda)$ coefficient used in the present article). Then we have recomputed the value of $L_{TIR}$ of each set of templates to correspond to the integral of the emission in the 8~$\mu$m to 1~mm wavelength range. Finally, we built an interpolator for each set of templates so that a spectrum could be computed for any $L_{TIR}$ within their range of validity, enabling different templates to be compared for the same values of $L_{TIR}$.

\begin{figure*}[!htbp]
 \centering
 \includegraphics[width=.495\textwidth]{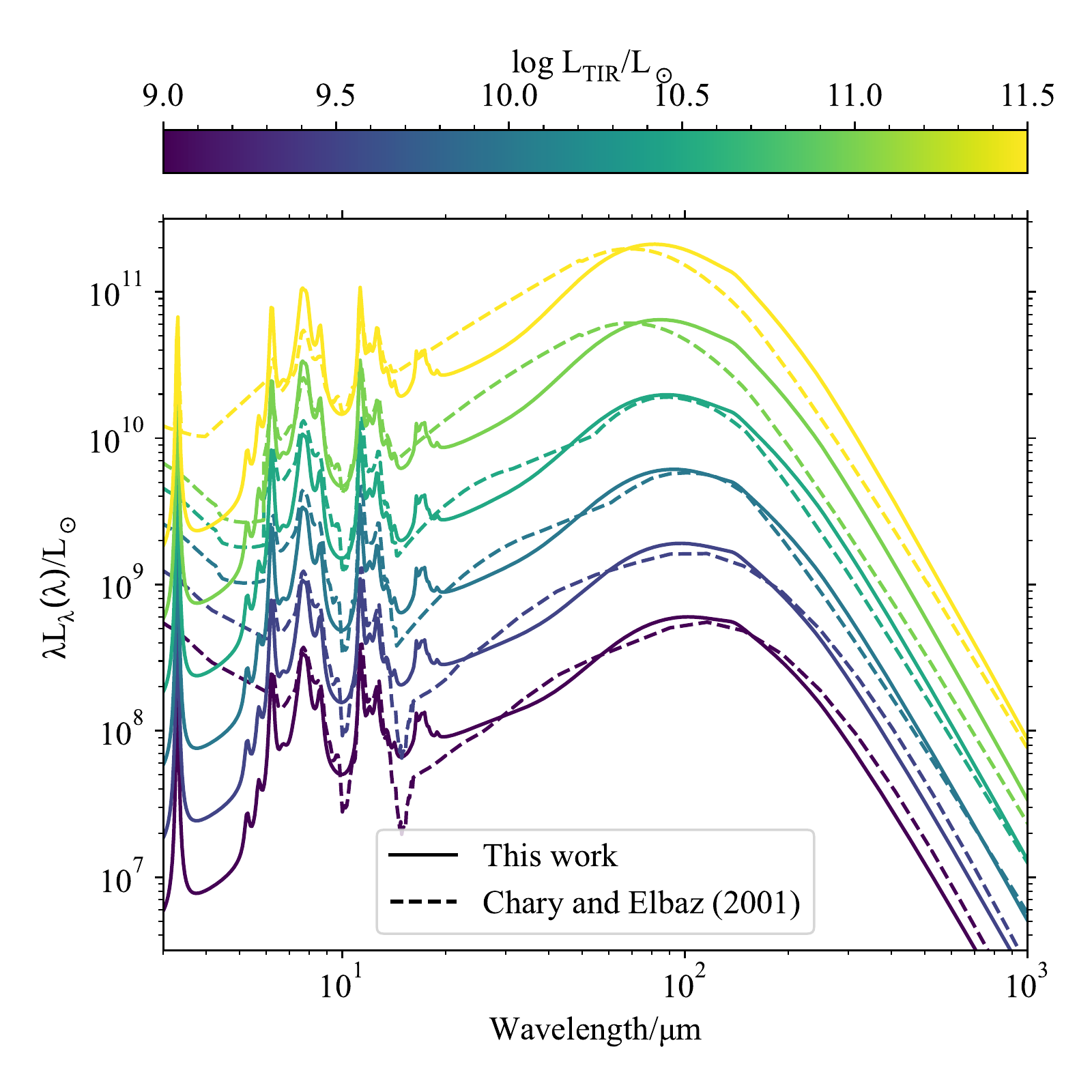}
 \includegraphics[width=.495\textwidth]{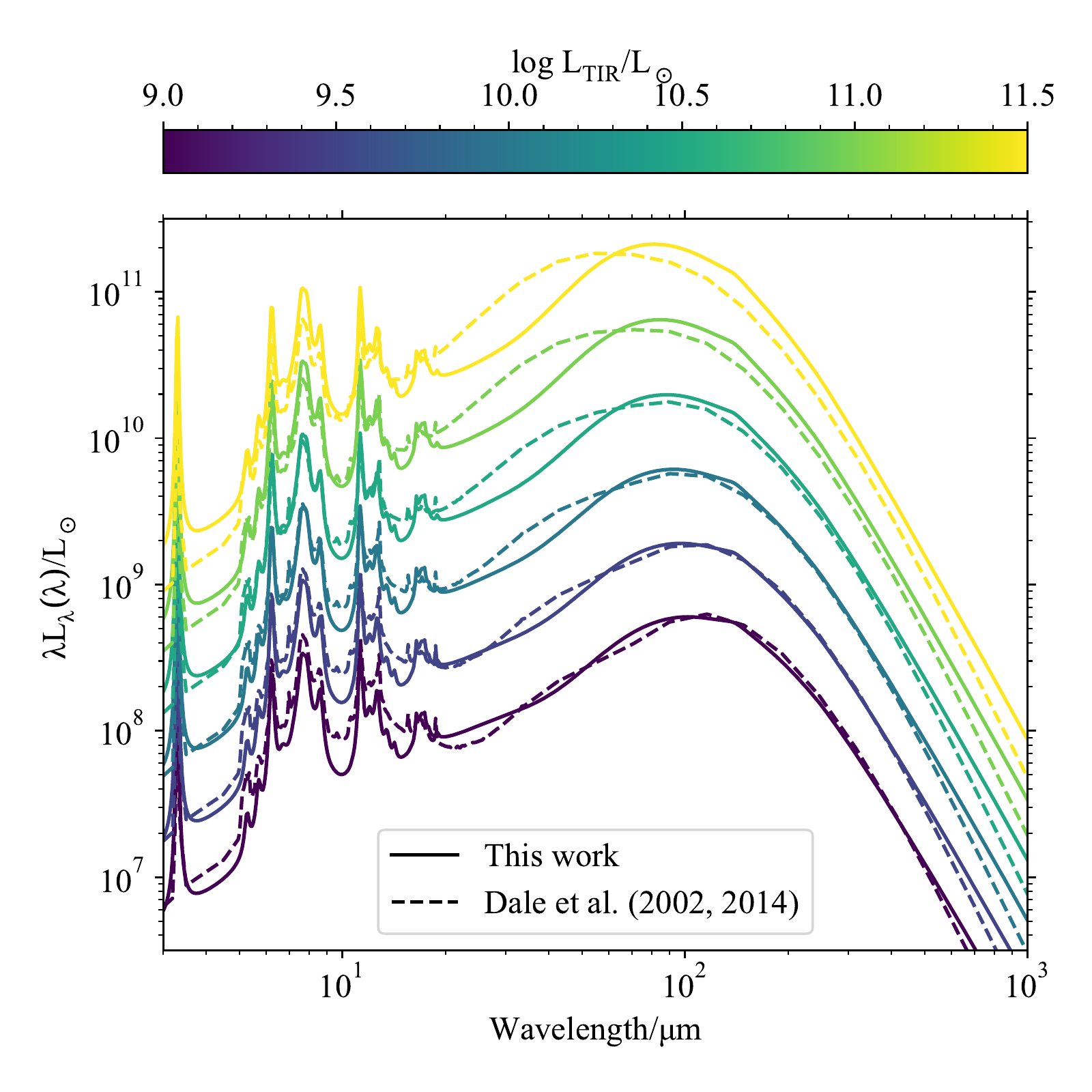}\\
 \includegraphics[width=.495\textwidth]{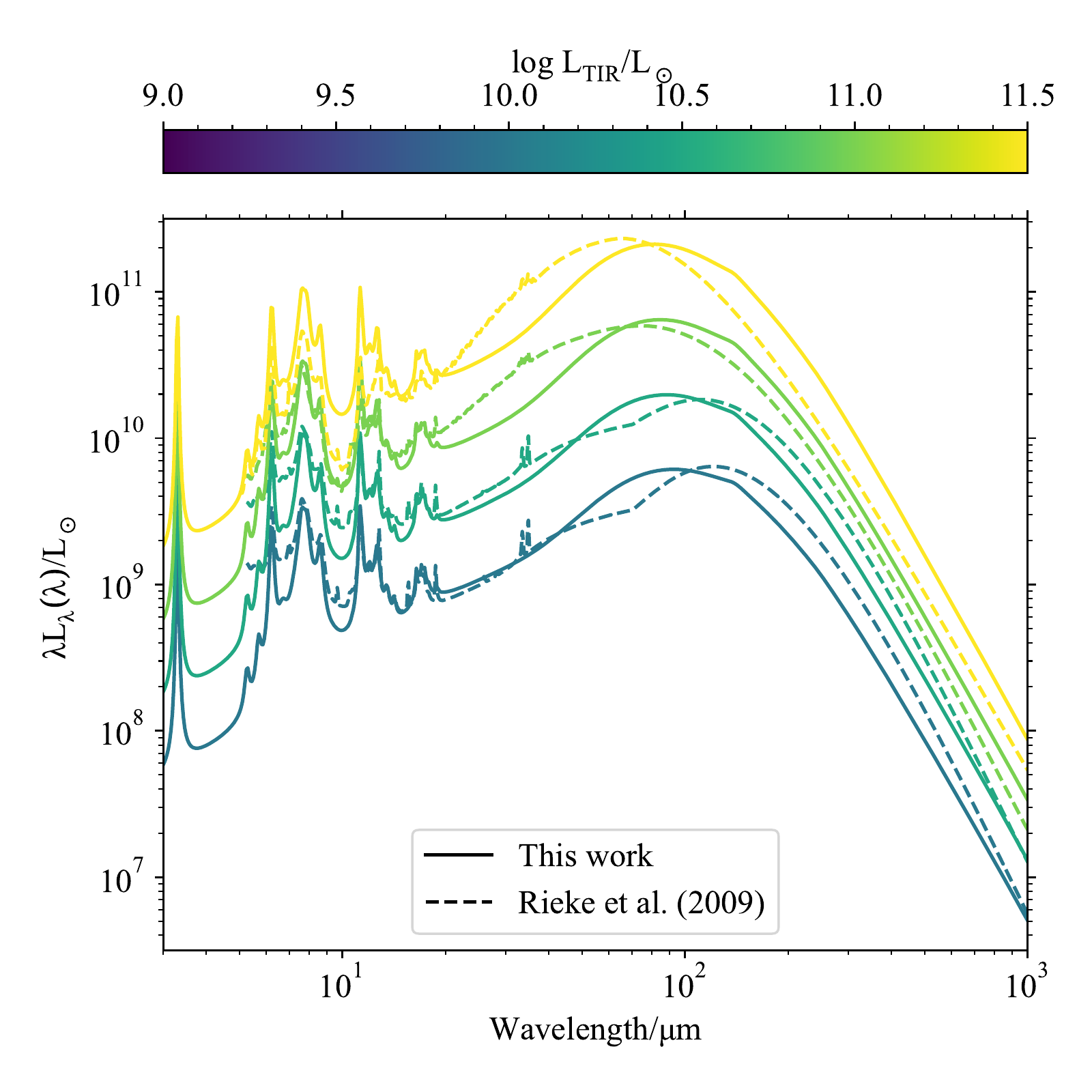}
 \includegraphics[width=.495\textwidth]{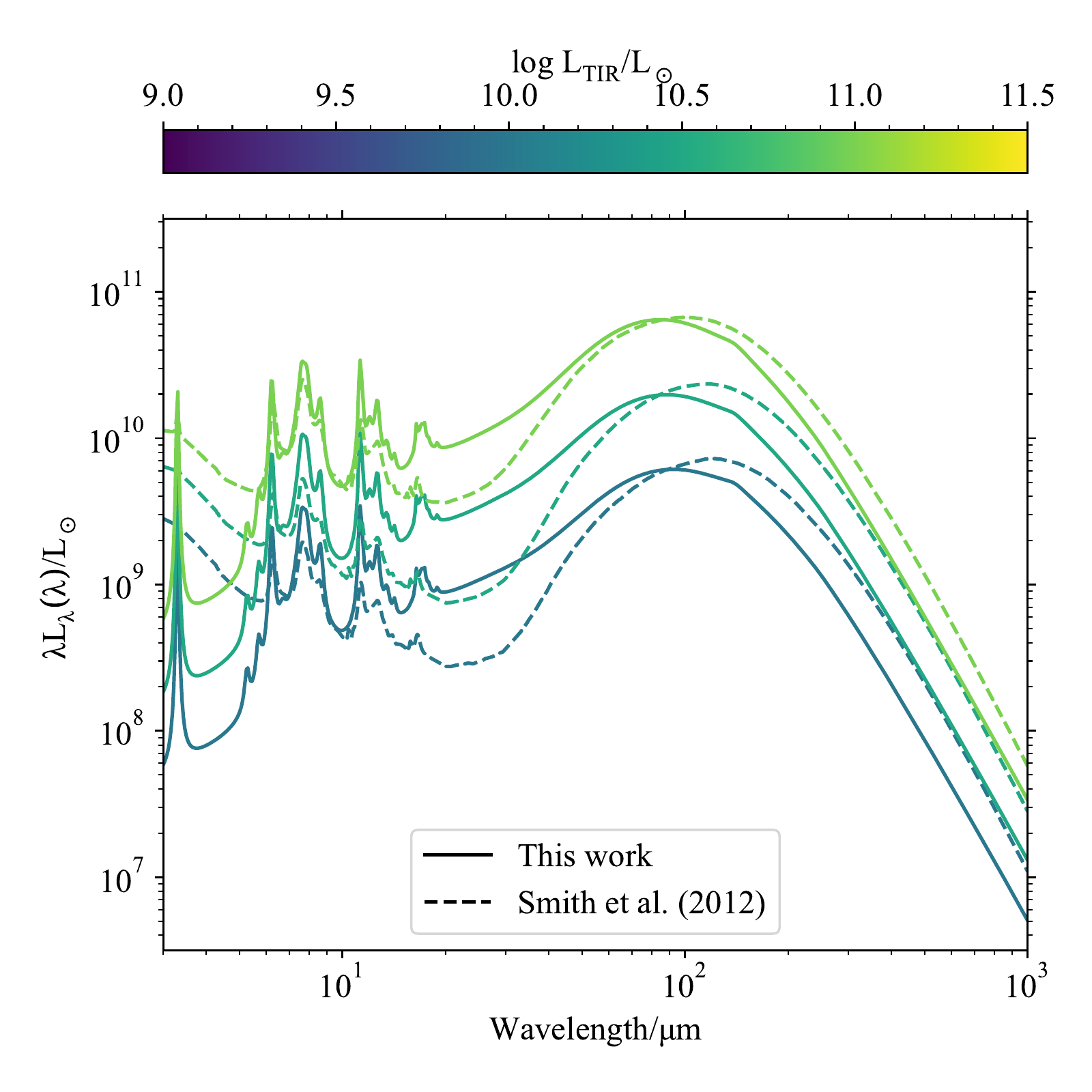}
 \caption{Comparison of the templates derived in this work with those from \cite{chary2001a} (top left), \cite{dale2002a,dale2014a} (top right), \cite{rieke2009a} (bottom left), and \cite{smith2012b} (bottom right). The solid lines correspond to the new templates and the dashed lines the literature templates. Each pair of lines of the same color corresponds to $\log~L_{TIR}/L_\odot$ of 9.0, 9.5, 10.0, 10.5, 11.0, and 11.5, following the bar to the right of each panel. Spectra outside of the definition range of a given set of templates have been omitted. The \cite{chary2001a} and \cite{smith2012b} templates include stellar populations, precluding any comparison at short wavelengths where such populations dominate.\label{fig:comparison-templates}}
\end{figure*}

We present in Fig.~\ref{fig:comparison-templates} the comparison of our $L_{TIR}$  dependent spectral templates to the previously published ones. Differences with respect to other templates, even if small at some wavelengths, are not random, but rather systematic and therefore require some discussion.
First of all, the PAH emission in our templates appears to be consistent with what was determined by \cite{chary2001a}, \cite{dale2002a,dale2014a}, and \cite{rieke2009a}. Moving to longer wavelengths, the different templates have clear discrepancies starting around 20~$\mu$m and extending up to about 70~$\mu$m, the region associated with the emission from very small grains \citep{desert1990a}. More specifically, templates other than \cite{smith2012b} tend to show higher emission levels and this excess is luminosity dependent, with the largest discrepancies found for the largest $L_{TIR}$. The origin of these differences is not certain. In the case of the \cite{dale2002a} templates, Fig.~10 from \cite{dale2001a} and Fig.~4 from \cite{dale2002a} suggest that there are systematic residuals between their observations and their models below 88~$\mu$m. Such residuals appear consistent with the difference we see here between their set of templates and our own work.

\begin{figure}[!htbp]
 \centering
 \includegraphics[width=.495\textwidth]{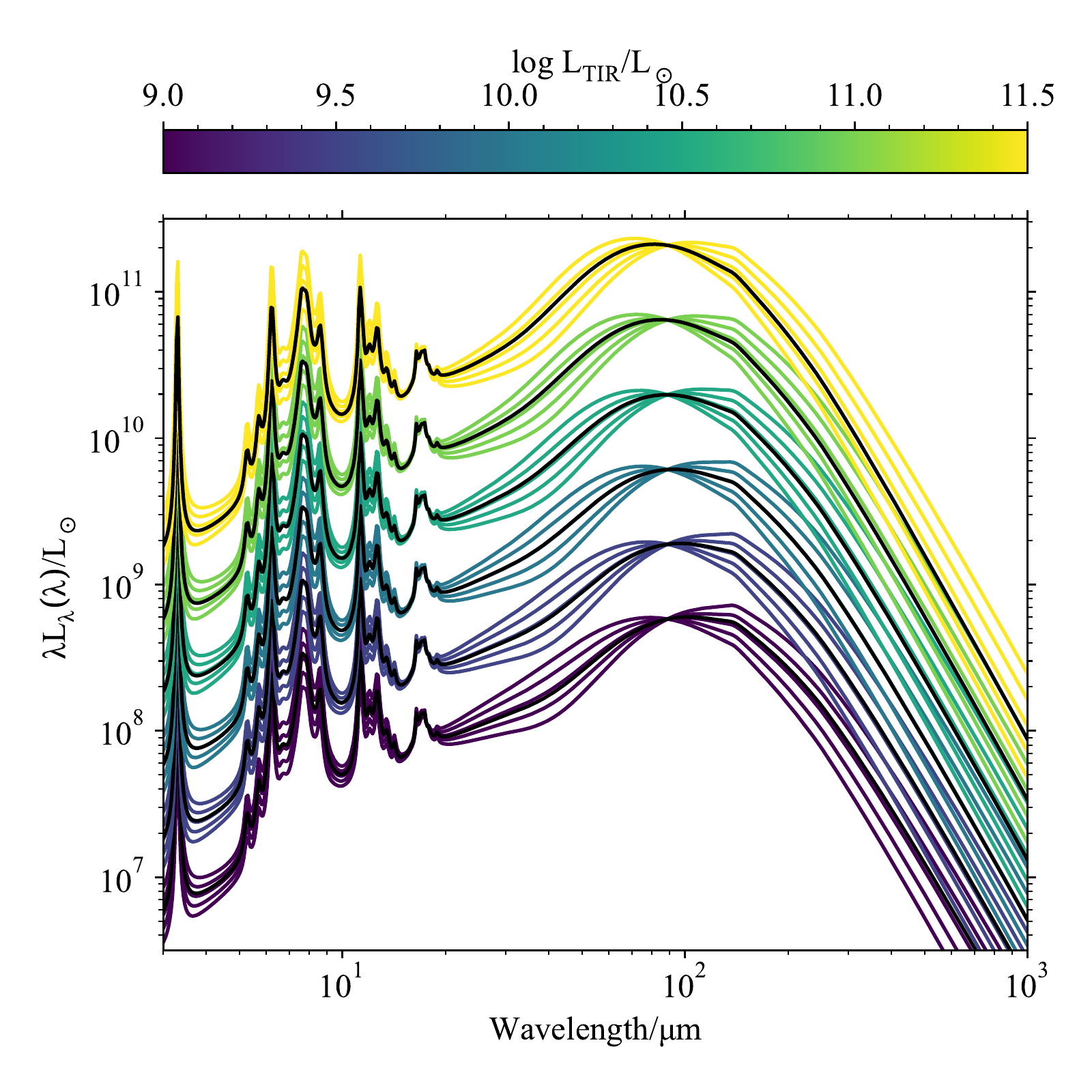}
 \caption{Comparison of the templates parametrized on $L_{TIR}$ at $\log~L_{TIR}/L_\odot$=9.0, 9.5, 10.0, 10.5, 11.0, and 11.5 (black lines), and the templates parametrized on both $L_{TIR}$ and sSFR for $\mathrm{sSFR/yr^{-1}}$=-11, -10.5, -10, -9.5, and -9. Templates corresponding to a high sSFR have a stronger emission below 90~$\mu$m. \label{fig:comparison-template-second}}
\end{figure}

More generally, relatively few normal galaxies have been observed in the spectral region between the mid and the far IR. IRAS and ISO had a gap between 25~$\mu$m and 60~$\mu$m. Likewise \textit{Spitzer} also had a gap between 24~$\mu$m and 70~$\mu$m in terms of broadband observations and \textit{Herschel} had no capability below 70~$\mu$m. The IRS spectrograph on-board \textit{Spitzer} had a coverage up to 40~$\mu$m and was used by \cite{rieke2009a}. However this was only done on a small number of objects. The LIRG and ULIRG templates may have a markedly different spectral shape compared to the galaxies in our sample and for galaxies with $\log~L_{TIR}/L_\odot<11$ their templates are based on four noise-free IRS spectra from \cite{smith2007b} parametrized on the ratio of the 12~$\mu$m and 25~$\mu$m bands. This may explain why we observe a significant deviation from $\log~L_{TIR}/L_\odot=11$. More generally, \cite{dale2009a} published IRS spectra of nuclear and extranuclear regions of SINGS galaxies \citep{kennicutt2003a}, showing that there is a clear diversity in the dust emission spectra in this wavelength range. Overall, it is likely that the relative lack of strong constraints makes models and templates uncertain over this domain. Furthermore, some templates show abrupt changes with $L_{TIR}$. This is the case in particular of the templates of \cite{rieke2009a}, where the peak of the emission shifts rapidly towards shorter wavelengths for $\log~L_{TIR}/L_\odot$ increasing from 10.5 to 11.5 (see in particular their Fig.~6). Conversely, in our templates the transition occurs at an almost constant rate of $\sim -11.5$~$\mu$m/dex in $L_{TIR}$. This abrupt change may be the result of different methodologies applied by \cite{rieke2009a} for galaxies above and below $\log L_{TIR}/L_\odot=11$. Namely, their high-luminosity templates use a modified black body to model the FIR regime, whereas low-luminosity templates are based on \cite{dale2002a} templates in that region (see Appendix \ref{sec:appen_temp} for details). More generally, what we see is that at fixed $L_{TIR}$, the templates of \cite{chary2001a}, \cite{dale2002a,dale2014a}, and \cite{rieke2009a} have a bluer (warmer) peak than our models, which may be the result of their basing of templates on FIR-selected samples from shallow surveys which would have preferentially detected galaxies with warmer dust. These elements points towards a strong influence of a small number of extreme objects at high sSFR for the most luminous objects in the templates of \cite{chary2001a}, \cite{dale2002a,dale2014a}, and \cite{rieke2009a}. Conversely, our large sample containing more moderate galaxies should not be affected by a small number of extreme objects. In order to test the influence of the sSFR at fixed $L_{TIR}$, we show in Fig.~\ref{fig:comparison-template-second} our templates parameterized on both $L_{TIR}$ and the sSFR.
We see that higher sSFR templates tend to have a stronger emission below 90~$\mu$m. This is qualitatively in agreement with the aforementioned literature templates and confirms the effect of a few extreme objects. The availability of our templates parametrized on both $L_{TIR}$ and sSFR make them highly flexible and applicable beyond just normal star-forming galaxies, as we will discuss in more detail in Sect.~\ref{ssec:external}.

\cite{smith2012b} templates, which are based on H-ATLAS, as for our sample, but without MIR photometry from WISE, show up to a $\sim0.5$ dex lower emission in the 20--40 $\mu$m range compared to our new templates, whereas they have a stronger emission in the Rayleigh-Jeans regime. The templates of \cite{smith2012b} are in stark contrast with the other sets of templates, which present a reasonably good agreement with each other at longer wavelengths. The large difference in the warm dust emission with respect to previous templates was already noted by \cite{smith2012b}. A possibility is that the lack of data below 100~$\mu$m for most galaxies in their sample coupled with modeling using MAGPHYS, did not allow for the spectra to be meaningfully constrained during the SED modeling process. MAGPHYS \citep{dacunha2008a} models the IR emission through a number of independent isothermal modified black bodies, whereas each \cite{draine2007a} model includes a specific temperature distribution (see Appendix \ref{sec:appen_temp}). It follows that even in the absence of observational constraints over a certain wavelength range, indirect constraints are provided through observations at other IR wavelengths. The discrepancy of \cite{smith2012b} with respect to other templates at longer wavelength was not readily apparent previously, but because here we compare spectra at identical $L_{TIR}$, a discrepancy in a given wavelength range has to be compensated with an opposite discrepancy in another range to maintain the respective luminosities equal. Ultimately, our modeling with \textsc{cigale} that combined the \cite{draine2007a} models and WISE bands at 12~$\mu$m and 22~$\mu$m, provides stronger constraints over this wavelength range.

Overall, we are confident that this new set of templates provides an excellent characterization of normal star-forming galaxies, and when sSFR dependent templates are considered, this characterization encompasses a full range of star-forming galaxies, from relatively quiescent to intensely star-forming, including those that are considered ``normal'' at higher redshift.

\subsection{Comparison with previously published relations\label{ssec:comparison2}}
In Sect.~\ref{ssec:comparison1} we compared different sets of templates to one another, without applying them to any observed data set. Here we focus on the KINGFISH sample of nearby star-forming galaxies, for which \cite{hunt2019a} derived $L_{TIR}$ with \textsc{cigale} and data from 32 bands from the FUV to 850~$\mu$m. We compare this reference $L_{TIR}$ to the estimates based on our own single-band relations and on single-band relations from \cite{galametz2013a}. For our estimates we do not take sSFR dependence into account because the KINGFISH sample and our H-ATLAS sample have a similar range of sSFR \citep{kennicutt2011a}. We adopt the photometric fluxes and distances of \cite{dale2012a,dale2017a}.

The agreement of the 70~$\mu$m and 100~$\mu$m estimates with the actual $L_{TIR}$ values is nearly perfect. The agreement for the 70~$\mu$m estimate is particularly important since we did not have observations probing this wavelength. The difference is small at 160~$\mu$m with perhaps a small offset of $\sim0.1$~dex for our estimator with respect to actual values at lower $L_{TIR}$. The main difference occurs for the \textit{Spitzer} 24~$\mu$m band, in particular at lower luminosities where $L_{TIR}$ from our estimators are in line with observations, whereas using the estimator of \cite{galametz2013a} gives results $\sim 0.2$~dex lower. An important point to keep in mind is that the definition ranges for these relations are different. Being based on a nearby sample, the \cite{galametz2013a} relations extend to fainter objects that are out of reach of large area surveys. This difference at 24~$\mu$m may suggest that there is an important change of regime in the MIR emission in fainter objects that is not apparent at longer wavelengths, which may generate a difference between the estimators at this wavelength.

If we now compare $L_{TIR}$ estimated using the relations given in Table~\ref{table:estimators-LTIR-SFR-second-parameter} to estimates using the previously published templates, we find the mean differences to lie typically within 0.1~dex. However, in the case of the templates of \cite{smith2012b} the mean difference reaches $\sim$0.25~dex at MIR wavelengths and goes even higher in some luminosity regimes. For instance, our estimates are consistent with those obtained with the templates of \cite{rieke2009a} in the JWST 10~$\mu$m filter up to $\log L_{TIR}/L_\odot=11$, but they abruptly deviate beyond this point. Multiple examples of such discrepancies in certain luminosity regimes can be found when comparing to different templates, translating some of their intrinsic characteristics probably originating from the techniques and the smaller samples used to build them. This emphasizes the importance of having estimators applicable over a broad range in $L_{TIR}$, as we will see in the next Section.

\subsection{Applicability of relations and templates to diverse galaxy populations\label{ssec:external}}
By construction, when applied to our full sample, the relations should recover the ``ground truth'' $L_{TIR}$ without a bias. We confirm that for various single-band and multi-band estimates the systematic offset is always smaller than 0.01 dex. This is, of course, the most ``favorable'' dataset to test for systematic errors. Analysis in Sect.~\ref{ssec:comparison2} demonstrated that the luminosities of KINGFISH galaxies are recovered reasonably well using single-band relations. A more stringent evaluation of the applicability of our templates and relations would be for the data sets that contain galaxies significantly different from the typical galaxies in our sample, either in terms of having $L_{TIR}$ outside of the range of our sample, or by having different $M_{\rm dust}$, even if $L_{TIR}$ (and even sSFR) are the same, as seems to be the case with local starbursts. For the results obtained based on single-band fluxes, we focus on sSFR-dependent relations as they produce more accurate results in all cases.

\subsubsection{Dusty galaxies in the green valley}
While the sample from which we derive the relations consists overwhelmingly of actively star-forming galaxies with $\log \mathrm{sSFR/yr^{-1}} > -11$ (Fig.~\ref{fig:sSFR-Mstar}), we have confirmed that the relations are also mostly applicable for non-AGN galaxies in our H-ATLAS sample that lie in the green valley region ($-12<\log \mathrm{sSFR/yr^{-1}}<-11$). Their $L_{TIR}$ estimated from single-band fluxes at 12, 24, and 100 $\mu$m, and utilizing sSFR-dependent relations, shows no bias with respect to the reference $L_{TIR}$ from full 7 IR fluxes \textsc{cigale} fits using the \cite{draine2014a} grids.

\subsubsection{Low-luminosity galaxies}

We next analyze the performance of our relations using a sample of very nearby galaxies, which therefore includes galaxies well below the IR luminosity range spanned by our sample from which the relations were constructed ($9 < \log L_{TIR}/L_\odot<12$). Specifically, we utilize \textit{Spitzer} observations in 8, 24, 70 and 160 $\mu$m bands from the Local Volume Legacy (LVL) survey \citep{dale2009a}. LVL contains 258 galaxies, of which 195 have detections in all four bands. Whereas the IR luminosities of these LVL galaxies span a very wide range ($6< \log L_{TIR}/L_\odot<11$), 3/4 are below our sample range ($\log L_{TIR}/L_\odot<9$). We derive four single-band estimates of $L_{TIR}$ based on different bands, various estimates including two bands, and one estimate based on all four bands. For the single bands, we estimate $L_{TIR}$ using  sSFR-dependent relations, where sSFR is based on the SFR estimated from the relation itself (value after the first iteration), and the stellar mass, taken from \cite{cook2014a}, is based on the 3.6~$\mu$m luminosity. At 70~$\mu$m and 160~$\mu$m we used PACS instead of MIPS bandpasses, which has no appreciable effect on the results. We compare our $L_{TIR}$ estimates to the reference values that we obtain by fitting all four bandpasses with full \cite{draine2014a} grid, with identical range of grid parameters as used to fit our H-ATLAS sample in Sect.~\ref{sssec:dustmodeling}. Our reference IR luminosities agree (0.01 dex offset, 0.04 dex standard deviation) with the ones given by \cite{dale2009a}, which they derive using a relation between MIPS bands and $L_{TIR}$ from \cite{dale2002a}. 

\begin{figure*}[!htbp]
  \centering
  \includegraphics[width=0.32\textwidth]{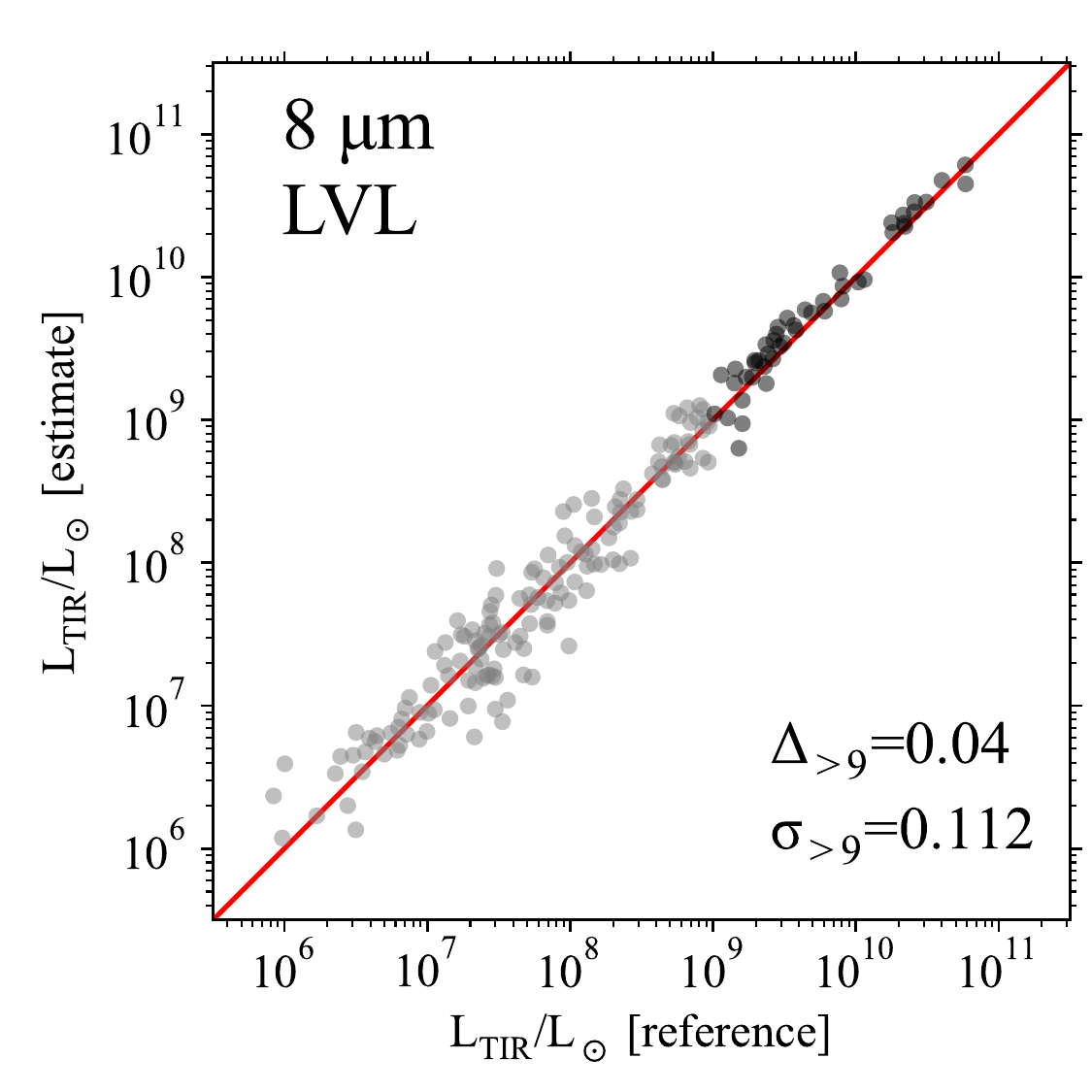}
  \includegraphics[width=0.32\textwidth]{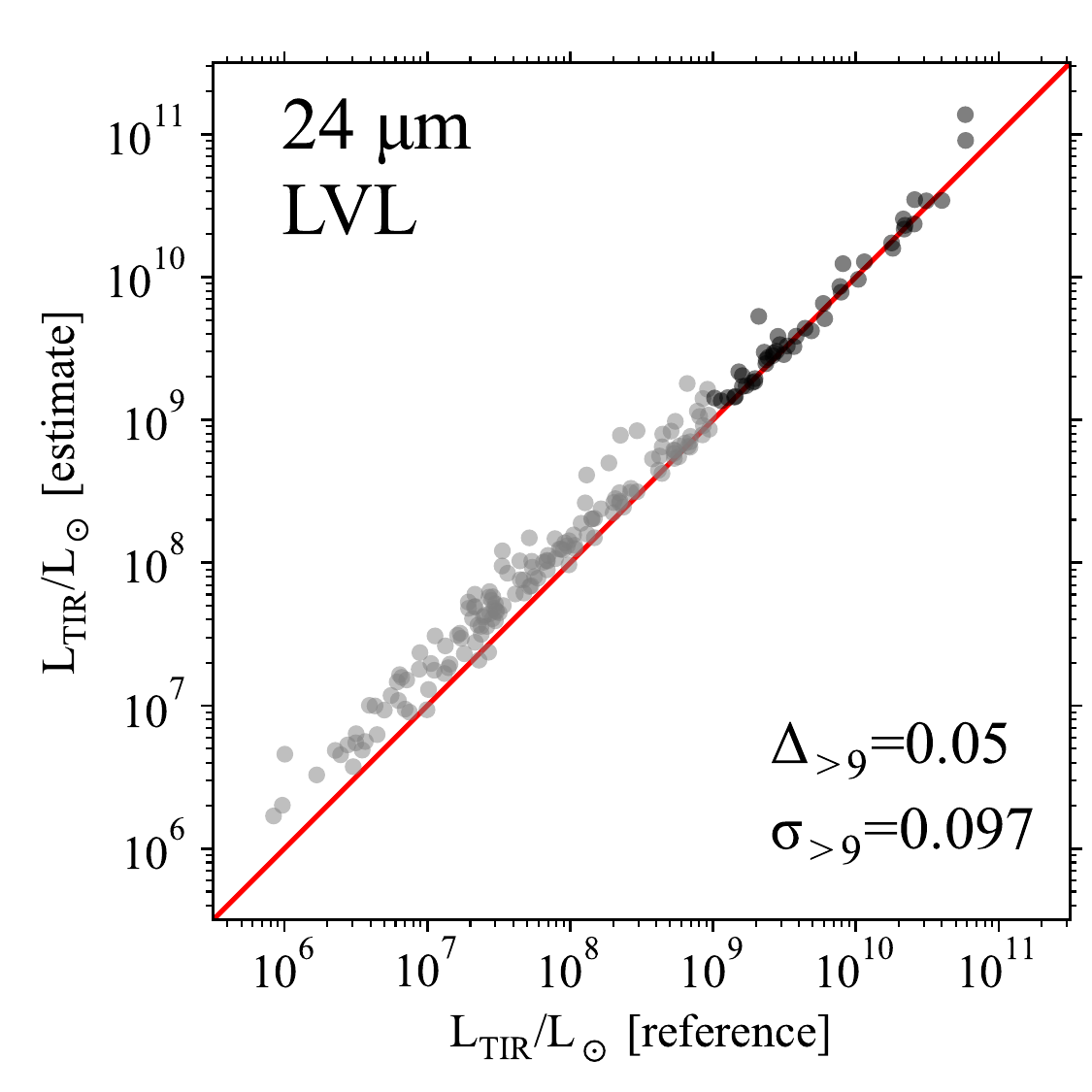}
  \includegraphics[width=0.32\textwidth]{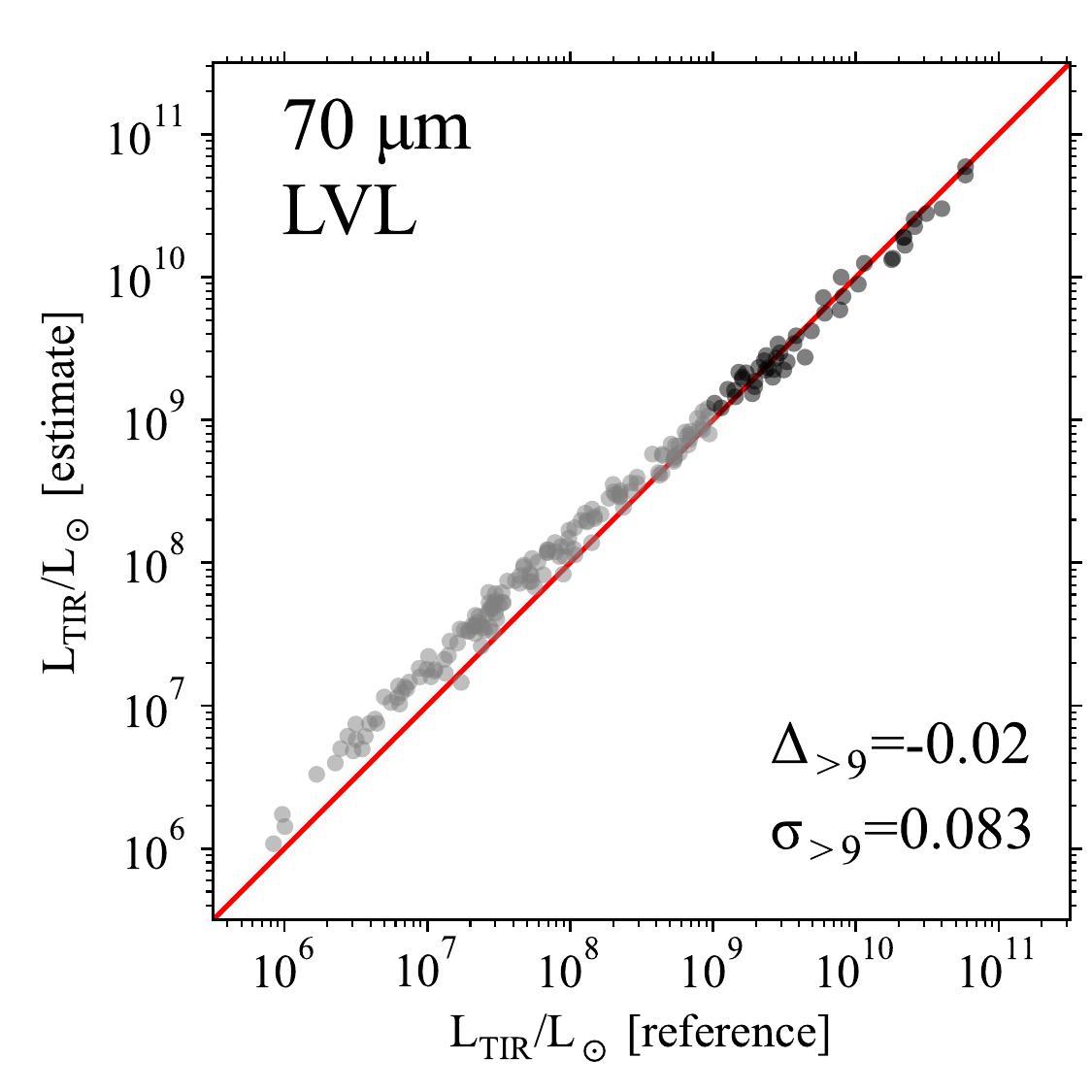}\\
  \includegraphics[width=0.32\textwidth]{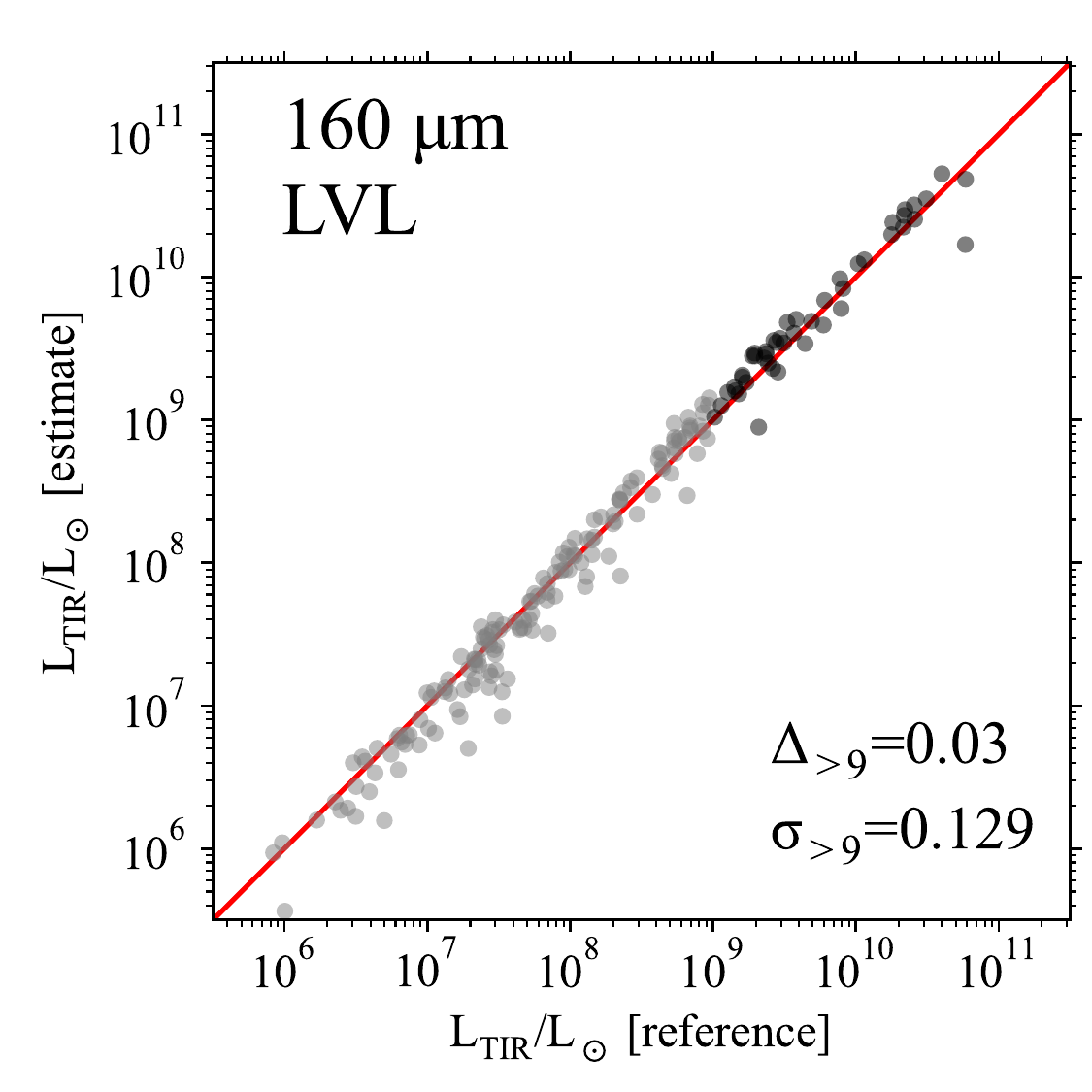}
  \includegraphics[width=0.32\textwidth]{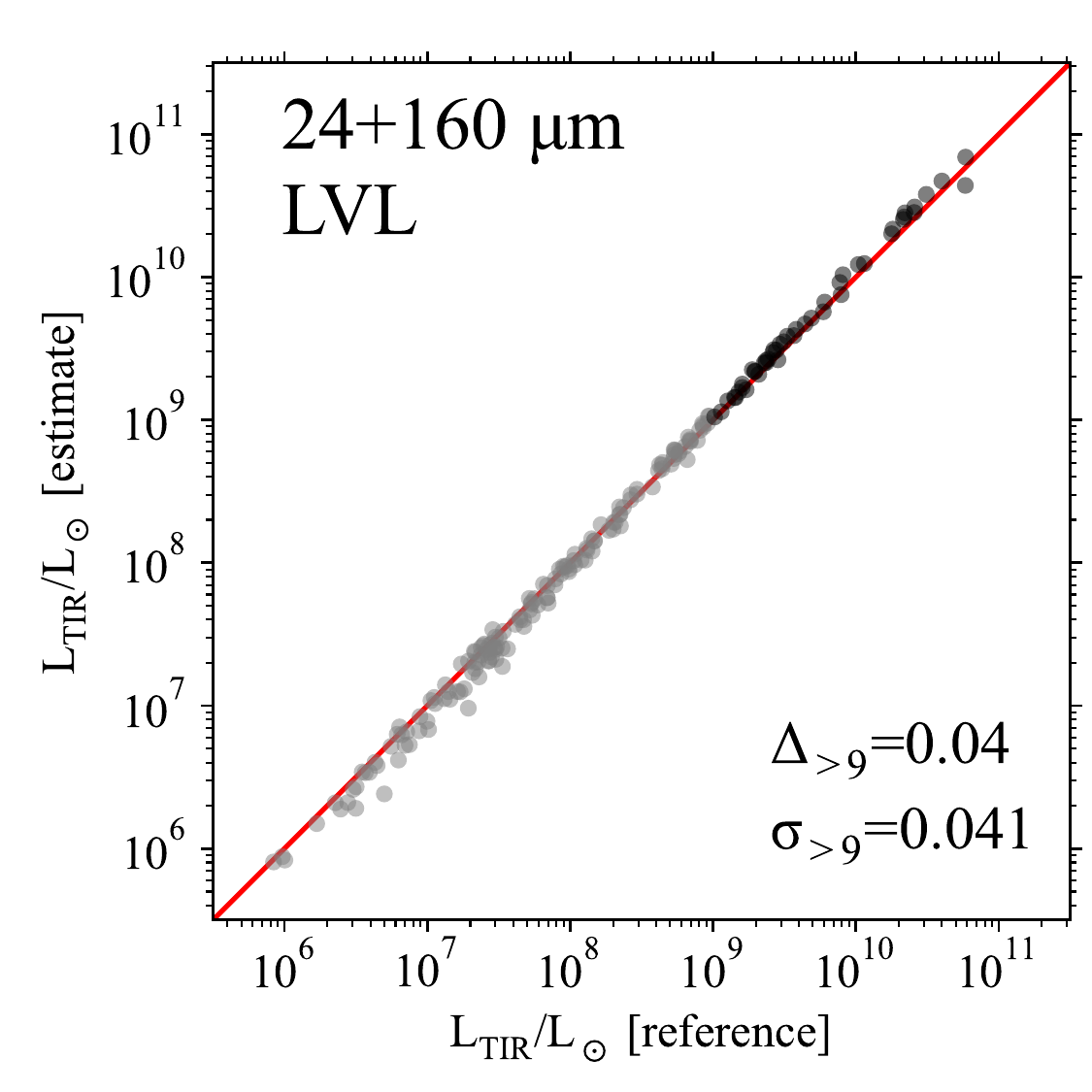}
  \includegraphics[width=0.32\textwidth]{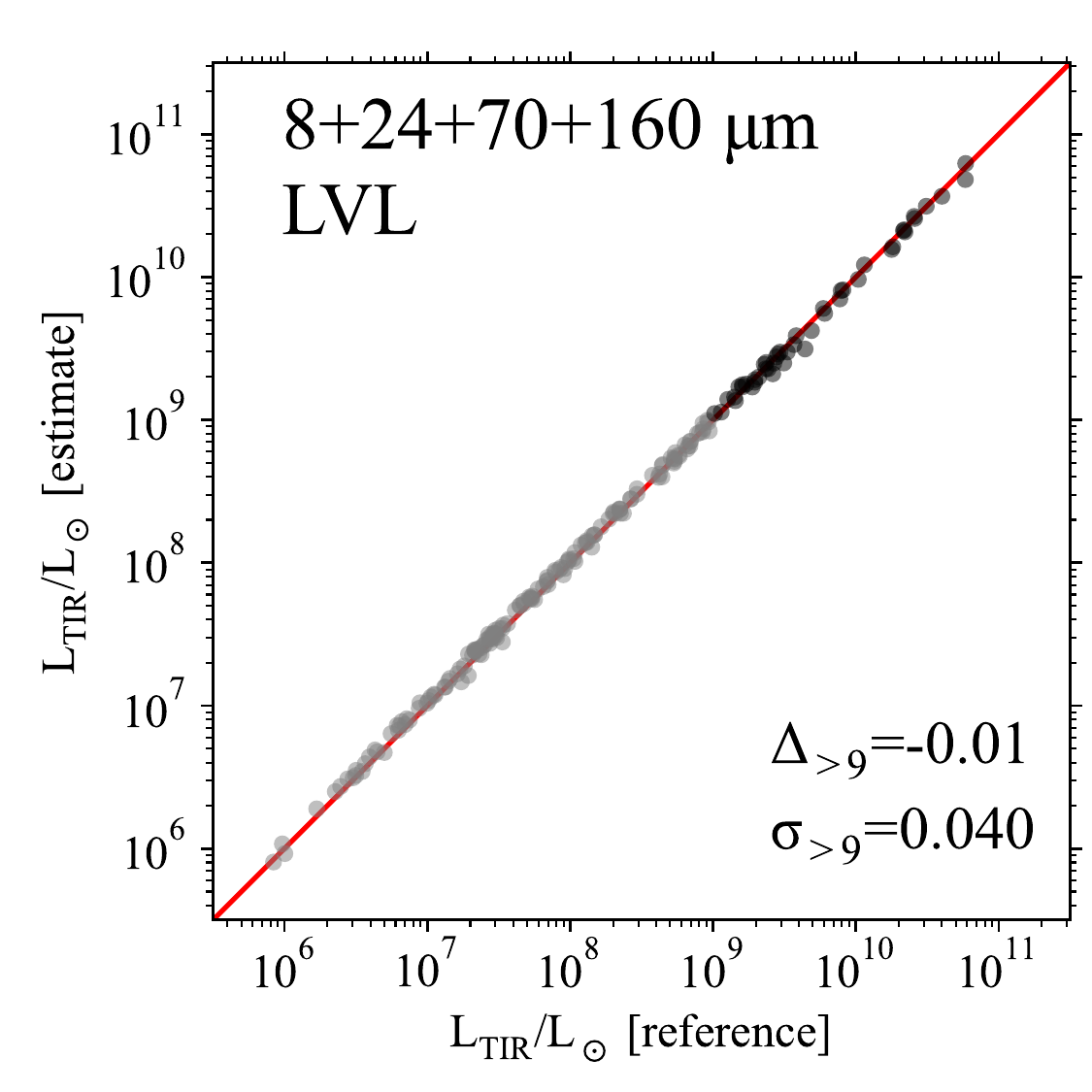}
 \caption{The accuracy and the precision of the recovery of the total IR luminosities for galaxies from the Local Volume Legacy (LVL) survey using our relations. Gray points represent galaxies that fall below the $L_{TIR}$ range of our H-ATLAS sample from which the relations were derived. Values in panels indicate the mean offset and standard deviation of the residuals for galaxies with $\log L_{TIR}/L_odot>9$. Estimates based on single bands take into account the sSFR dependence of the relations. The red line represents the 1:1 relation. Reference $L_{TIR}$ values come from using all IR bands and \citet{draine2014a} models. \label{fig:lvl}}
\end{figure*}

The performance of our relations for LVL is presented in Figure \ref{fig:lvl}.
For the luminosity range for which the relations are constructed ($L_{TIR}>9$) all our single-band estimates of $L_{TIR}$ are in excellent agreement with the reference values, with offsets no greater than 0.04 dex, and a scatter smaller than 0.13 dex. If we were to use relations without the sSFR dependence, the offset would in some bands exceed 0.1 dex. In the low-luminosity regime, the single band estimates are reasonably good, with almost no offset at 8~$\mu$m and 160~$\mu$m. Moving on to two-band estimates, and in particular the estimate based on 24~$\mu$m plus 160~$\mu$m, we see a remarkable improvement in the scatter at all luminosities. Finally, the four-band estimate follows the reference values almost perfectly. Apparently, the fact that our multiple-band relations are not sSFR dependent is accounted for by the presence of color terms. From this analysis we conclude that our relations are generally applicable even down to $\log L_{TIR}/L_\odot=6$, though some caution in the low-luminosity regime may be warranted for single-band estimates based on bands in the intermediate wavelength range. 

\subsubsection{Dusty local LIRGs}
We evaluate the applicability of the new relations with respect to the galaxies from the Great Observatories All-Sky LIRG Survey \citep[GOALS,][]{armus2009a}, which contains 181 LIRGs ($11<\log L_{TIR}/L_\odot<12$) and 21 ULIRGs ($12<\log L_{TIR}/L_\odot<12.4$). Since the majority of low-redshift U/LIRGs are merger driven and are much dustier than normal galaxies with similar $L_{TIR}$, we are testing potentially different IR SED shapes from any galaxy in our sample. 

First, we remove 64 galaxies (32\% of the sample) that are associated with an AGN in the SIMBAD database. The removal of AGN also removes essentially all ULIRGs in this sample, leaving galaxies that, at least in terms of $L_{TIR}$ fall within the range covered by our H-ATLAS sample. Furthermore, in order to use sSFR-dependent relations for single-band estimates, we require stellar masses, which we take from \cite{u2012a}. This leaves a final sample of 39 galaxies. For our analysis we use IRAS fluxes in the 12, 25, and 100~$\mu$m bands taken from the IRAS Revised Bright Galaxy Sample \citep[RBGS,][]{sanders2003a} and PACS/SPIRE \textit{Herschel} photometry from \cite{chu2017a}. For galaxies where \textit{Herschel} resolves the IRAS source into two components, we sum the component fluxes. Distances are taken from \cite{armus2009a}, which also provides luminosities derived using the \cite{sanders1996a} formula, which was constructed from single-temperature dust emission models and utilizes all four IRAS bands (12, 25, 60, and 100~$\mu$m). We use \textsc{cigale} with the \cite{draine2014a} models, with the parameters chosen the same way as in  Sect.~\ref{sssec:dustmodeling}, to derive new $L_{TIR}$ based on all of IRAS and \textit{Herschel} photometry (9 bands in total). Our own estimates of $L_{TIR}$ and those from \cite{armus2009a} have a mutual scatter of only 0.02 dex and no systematic bias. To evaluate the relations derived in this work we continue by using our $L_{TIR}$ values.

\begin{figure*}[!htbp]
  \centering
  \includegraphics[width=.24\textwidth]{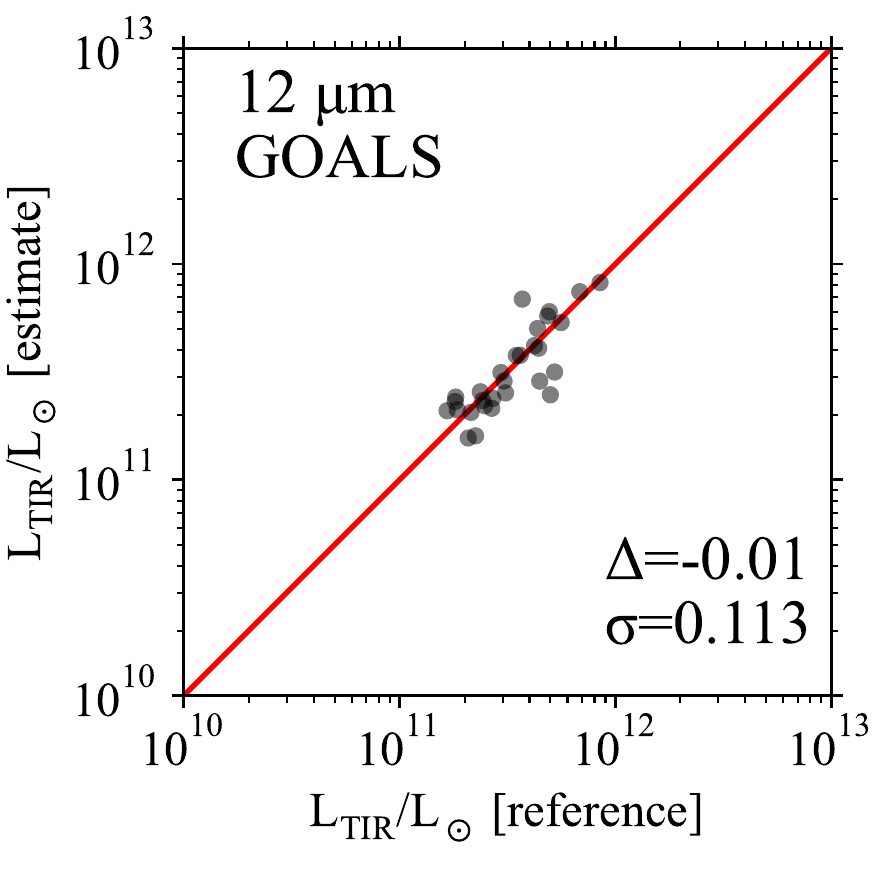}
  \includegraphics[width=.24\textwidth]{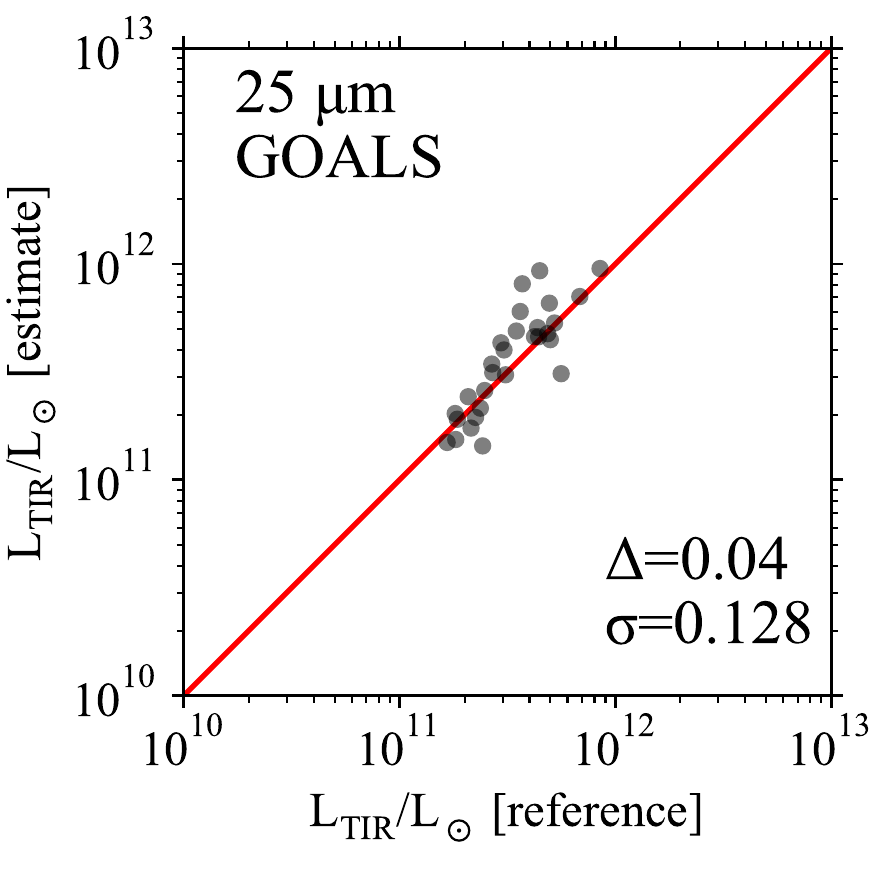}
  \includegraphics[width=.24\textwidth]{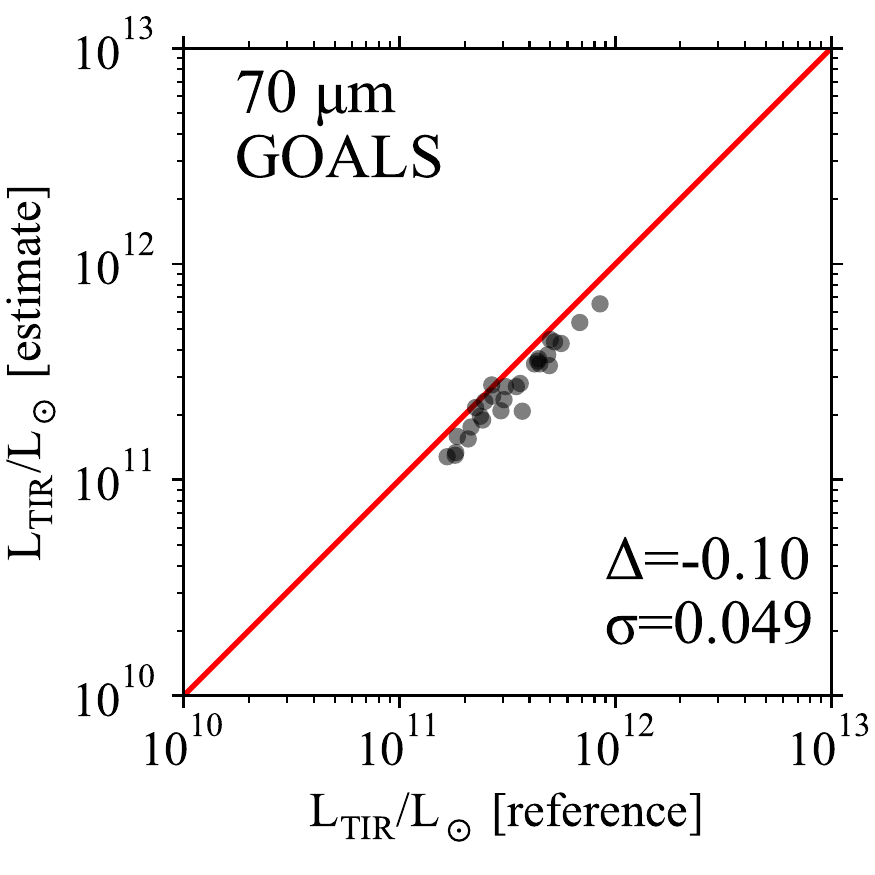}
  \includegraphics[width=.24\textwidth]{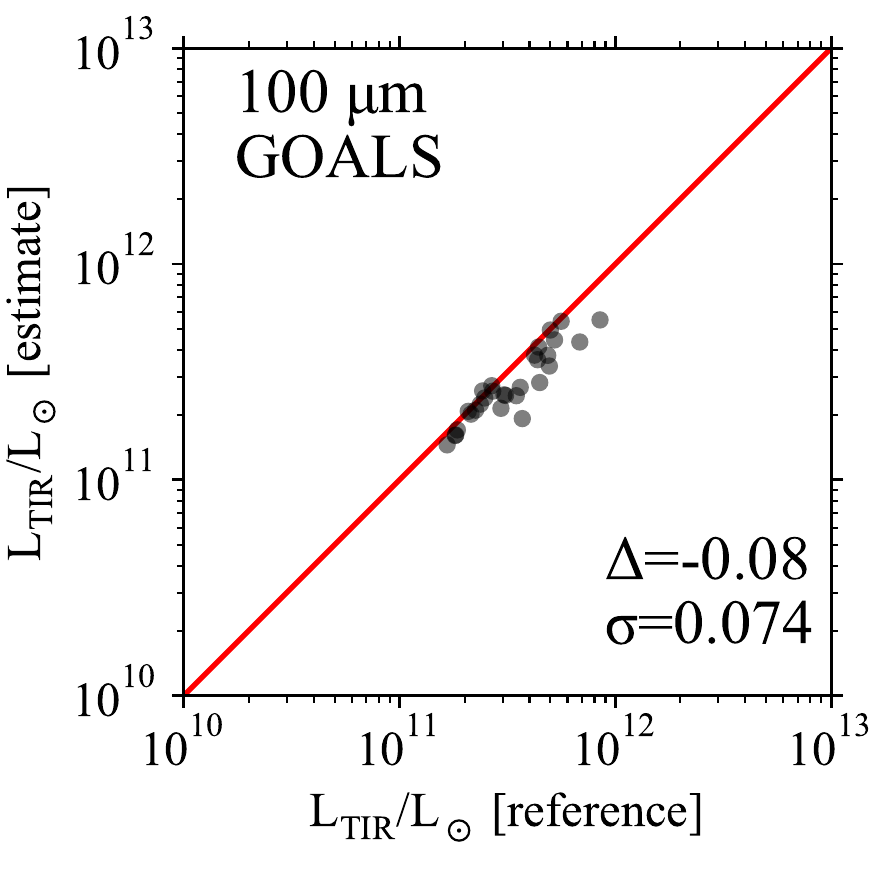}\\
  \includegraphics[width=.24\textwidth]{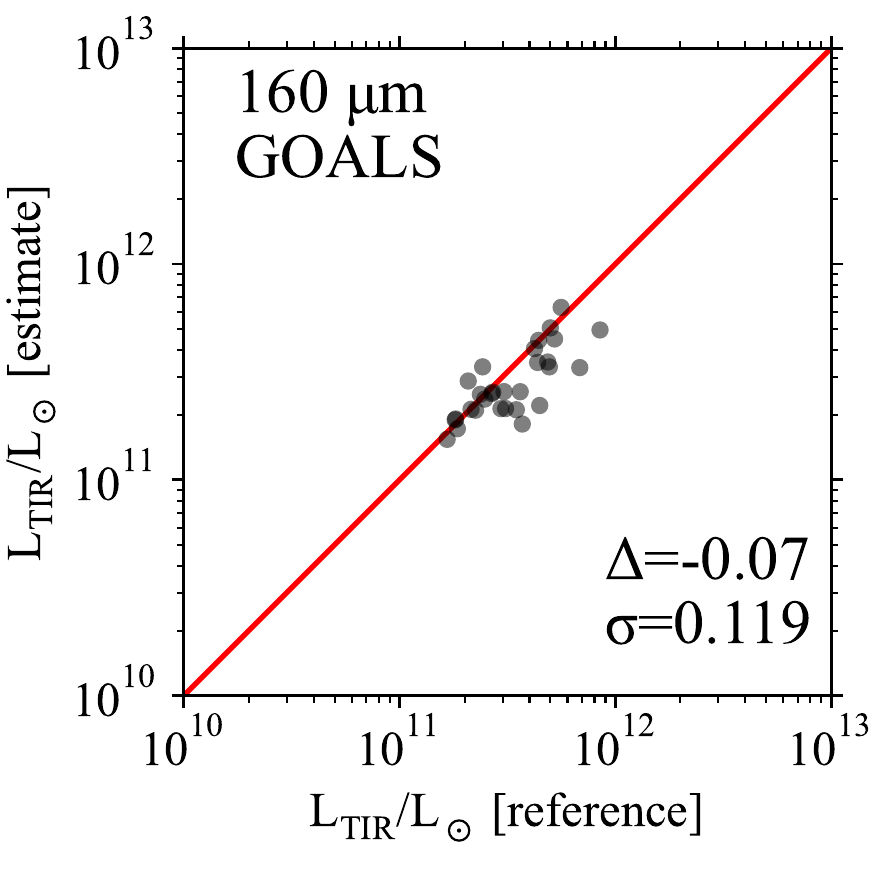}
  \includegraphics[width=.24\textwidth]{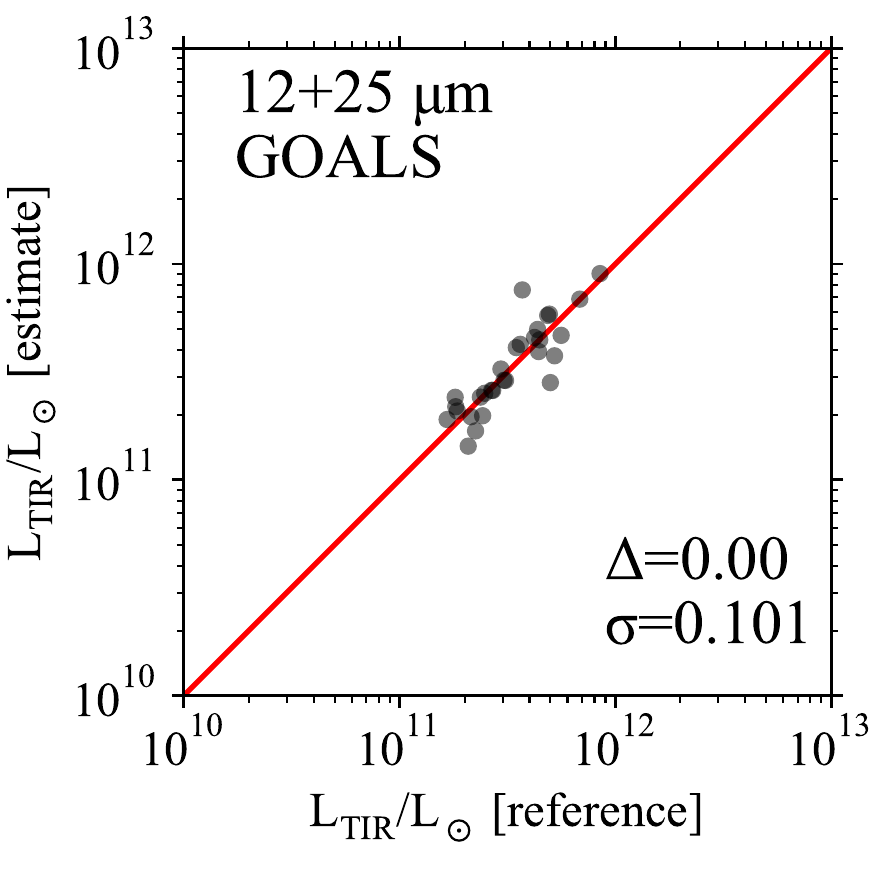}
  \includegraphics[width=.24\textwidth]{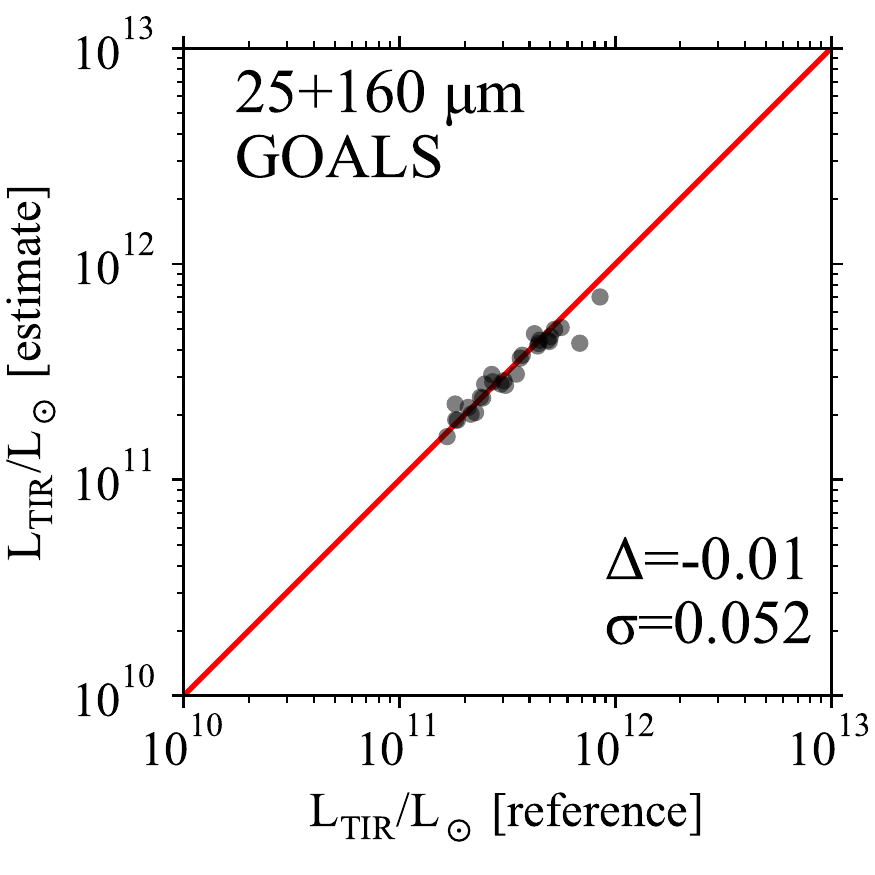}
  \includegraphics[width=.24\textwidth]{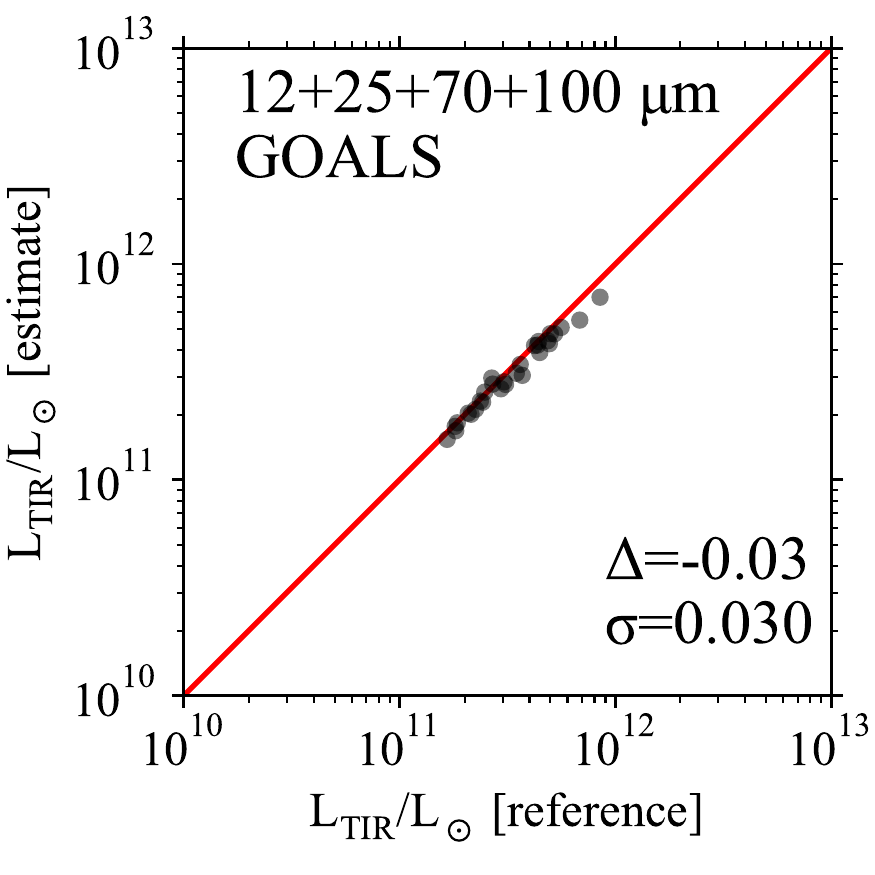}
 \caption{The accuracy and the precision of the recovery of the total IR luminosities for galaxies from the Great Observatories All-Sky LIRG Survey (GOALS) using our relations. Values in panels indicate the mean offset and standard deviation of the residuals. Estimates based on single bands take into account the sSFR dependence of the relations. Red line represents 1:1 relation. Reference $L_{TIR}$ values come from using all IR bands and \citet{draine2014a} models.\label{fig:goals}}
\end{figure*}

Since we have not calculated our relations for IRAS bands, we use WISE W3 and \textit{Spitzer} MIPS 24 $\mu$m bandpasses instead. The performance of our relations for GOALS LIRGs is presented in Figure \ref{fig:goals}.
We produce single-band estimates based on 12, 25, 70, 100, and 160~$\mu$m (100~$\mu$m is from \textit{Herschel}) and using sSFR-dependent relations. The systematics exist for PACS bands, but are relatively small (up to 0.1 dex). We believe these systematics emerge because nearby LIRGs, selected from shallow IRAS survey, would be biased towards higher IR luminosities for their sSFR and dust mass compared to a more typical population of galaxies with such dust mass and sSFR. In terms of the precision, the 70~$\mu$m has the smallest scatter, 0.05~dex, but there is a slight offset. For the estimate based on the 8~$\mu$m emission, the scatter is reasonably small (0.12 dex) and there is no bias.

We tested various two-band estimates, and, as expected, they provide certain improvements. In particular, estimates based on MIR alone (12 and 25~$\mu$m) and on the combination of the 25 and 160~$\mu$m fluxes (both shown in Fig.~\ref{fig:goals}) are both unbiased. Finally, the four-band estimate comprising of MIR IRAS bands and two two shorter PACS bands achieves a scatter of just 0.03 dex with respect to the reference luminosities. 

In conclusion, the analysis in this section provides reassurance that the application of single-band relations to JWST observations of higher-redshift galaxies, even if sub-millimeter selected (i.e., analogous to nearby LIRGs), should yield unbiased estimates. In general, however, high-redshift luminous galaxies (assuming no AGN is present) are more likely to resemble the high-luminosity galaxies in H-ATLAS, so our relations should be even more appropriate for them.

\section{Summary and conclusion\label{sec:conclusion}}
We have measured the physical properties of 2584 low-redshift star-forming galaxies by modeling their emission from the FUV to 500~$\mu$m with the spectral energy distribution modeling code \textsc{cigale} \citep{boquien2019a}, using in particular the stellar populations from \cite{bruzual2003a}, flexible dust attenuation curves, and the dust emission models from \cite{draine2007a,draine2014a}. Our sample spans a wide range of stellar masses and sSFR, and excludes AGN. The modeling facilitated a detailed study of the dependence of dust emission spectra on different physical properties ($M_{star}$, total SFR, sSFR, $L_{TIR}$, $M_{dust}$, and the oxygen abundance), from which we conclude the following:

\begin{itemize}
\item The overall shape of dust emission spectra varies strongly with $L_{TIR}$ (or, alternatively, with SFR), but also on sSFR, confirming the findings of other studies.
\item The dependence of the shape of the dust emission spectra on sSFR is independent from the $L_{TIR}$ dependence, i.e., sSFR is a second parameter.
\item Monochromatic IR luminosities best constrain $L_{TIR}$ at 90~$\mu$m rest frame, with the minimum dispersion of only $\sim$0.05 dex.
\item The current SFR is better constrained by the total IR luminosity (0.11 dex) than by any monochromatic luminosity, though the luminosity circa 90~$\mu$m rest-frame comes close in precision (0.12 dex).
\item Precise estimates of $L_{TIR}$ are possible with only a single band. In particular, single-band photometry in two IR regions provides constraints on $L_{TIR}$ that are better than 0.1 dex: 12--17 $\mu$m and 55--130 $\mu$m. Performance is significantly worse between 20 and 40 $\mu$m and beyond 200  $\mu$m (all wavelengths are rest-frame.) Thus the prospects for JWST MIRI are excellent.
\item Monochromatic IR luminosities constrain the total (obscured plus unobscured) SFR less well (minimum dispersion of 0.12 dex) than $L_{TIR}$ because of the diversity of stellar populations and attenuation curves at a given luminosity. The knowledge of the population age (sSFR) reduces this uncertainty to 0.09 dex. These precisions are in either case remarkable.
\item Dust mass can be constrained reasonably well (dispersion $<0.2$ dex) only at rest-frame $\lambda>200$~$\mu$m, with longer wavelengths giving progressively more accurate estimates ($\sim 0.1$ dex at 1 mm).
\end{itemize}

Informed by the above results, we constructed single-parameter IR templates parameterized on $L_{TIR}$ and (separately) on SFR, as well as the functional relations that allow the determination of these two properties using between 1 and up to 4 commonly used bandpasses from \textit{Spitzer}, WISE, \textit{Herschel}, and, in the future, JWST. These templates and relations can be used in a large number of circumstances, in particular when there is only a limited coverage of the dust emission that precludes the fitting of theoretical dust models that contain many free parameters, such as those of \cite{draine2007a} or \cite{dacunha2008a}. Furthermore, the dependence of the dust emission spectra on sSFR at fixed $L_{TIR}$ motivated the development of two-parameter dust templates and relations. Two-parameter templates increase the reliability of estimation of dust parameters (including $L_{TIR}$)  for galaxies with a wide range of star formation activity, from relatively quiescent local galaxies, to galaxies that, with their high sSFR, resemble typical star-forming populations at higher redshifts. Based on this we additionally conclude the following:

\begin{itemize}
\item Previously published dust emission templates based on samples selected from shallow FIR surveys generally have bluer peak than ours, consistent with our sample being on average more representative of normal star forming galaxies. Our two-parameter templates do, however, become progressively bluer for higher sSFR, as expected.
\item Our relations provide improvements over some literature relations at MIR wavelengths, in particular accounting for the effect of variations of the dust temperature.
\item Our relations used with the individual single bands in the MIR and FIR yield fairly unbiased estimates of $L_{TIR}$ (within 0.1 dex) for a diverse population of galaxies, including local LIRGs and dusty galaxies in the green valley.
\item Utilizing multiple bands, in particular from different wavelength ranges, considerably reduces the scatter in estimating $L_{TIR}$. Relations involving multiple bands were tested to be unbiased even down to $\log L_{TIR}/L_\odot=6$.
\end{itemize}

Finally, we make available software tools to generate $L_{TIR}$-dependent and $L_{TIR}$+sSFR-dependent spectra at any value of these parameters, eliminating the need for the traditional discrete templates, as well as the tools to derive $L_{TIR}$ and SFR from a wide selection of bands at $0<z<4$. We also provide explicit formulae to calculate $L_{TIR}$ and the total SFR directly from JWST 21 $\mu$m (F2100W) fluxes and redshifts. 

\begin{acknowledgements}
We thank the anonymous referee for helpful comments that have contributed to improving and clarifying the manuscript.

We also thank Dan Smith for kindly providing us his updated galaxy templates, Steve Maddox and Steve Eales for their suggestions regarding PACS photometry, and the entire H-ATLAS team for their wonderful survey.

M.B. was partially supported by FONDECYT regular grants 1170618 and 1211000, and the work of S.S.\ was partially supported by NASA award 80NSSC20K0440. 

This research made use of Astropy,\footnote{\url{http://www.astropy.org}} a community-developed core Python package for Astronomy \citep{astropy2013a, astropy2018a}.
\end{acknowledgements}

\bibliographystyle{aa}
\bibliography{aa}

\appendix
\section{Redshift-dependent relations for estimating $L_{TIR}$ and SFR from a single IR band\label{sec:appen_z}}
\begin{figure*}[!htbp]
 \centering
 \includegraphics[width=.33\textwidth]{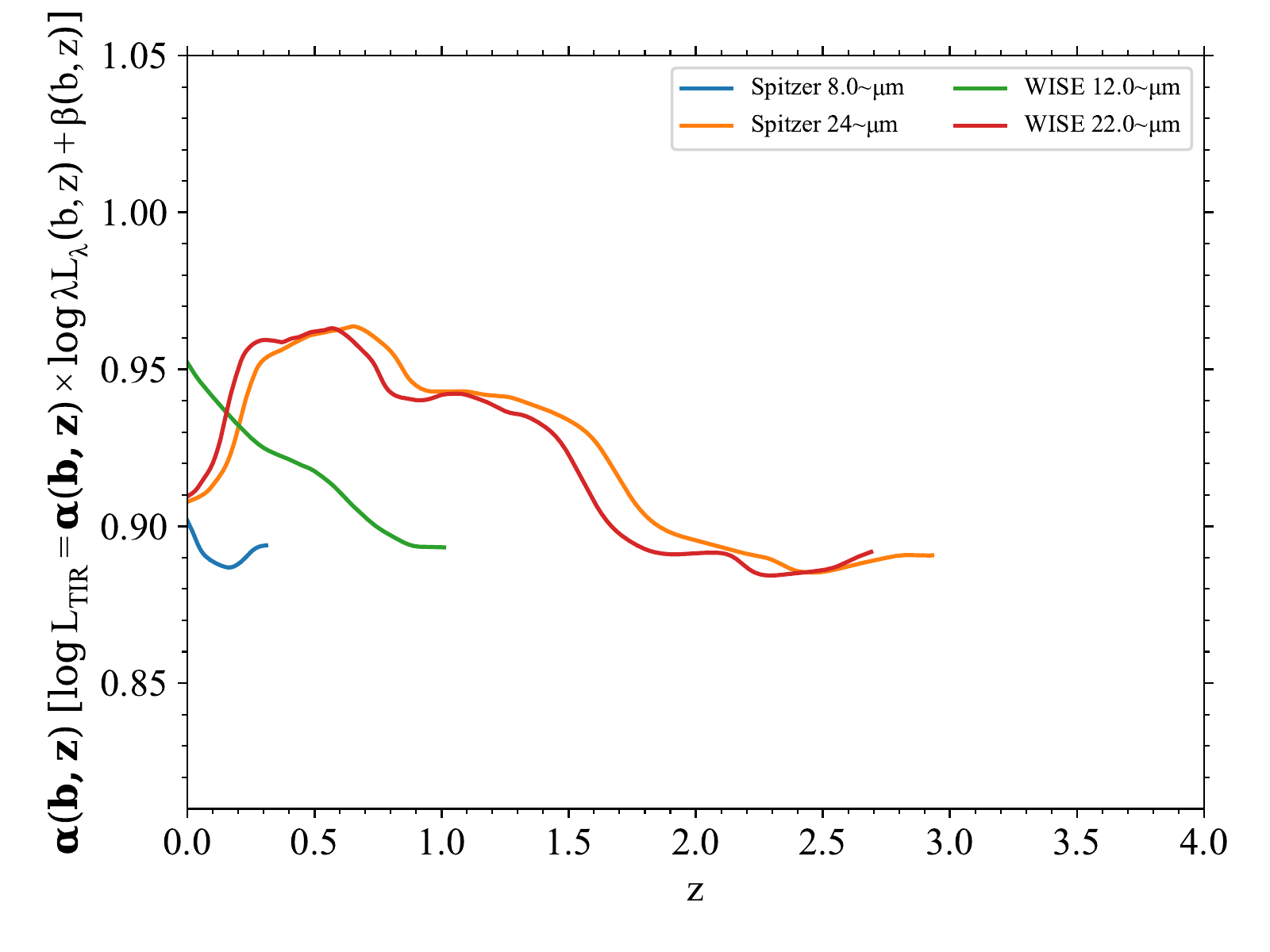}
 \includegraphics[width=.33\textwidth]{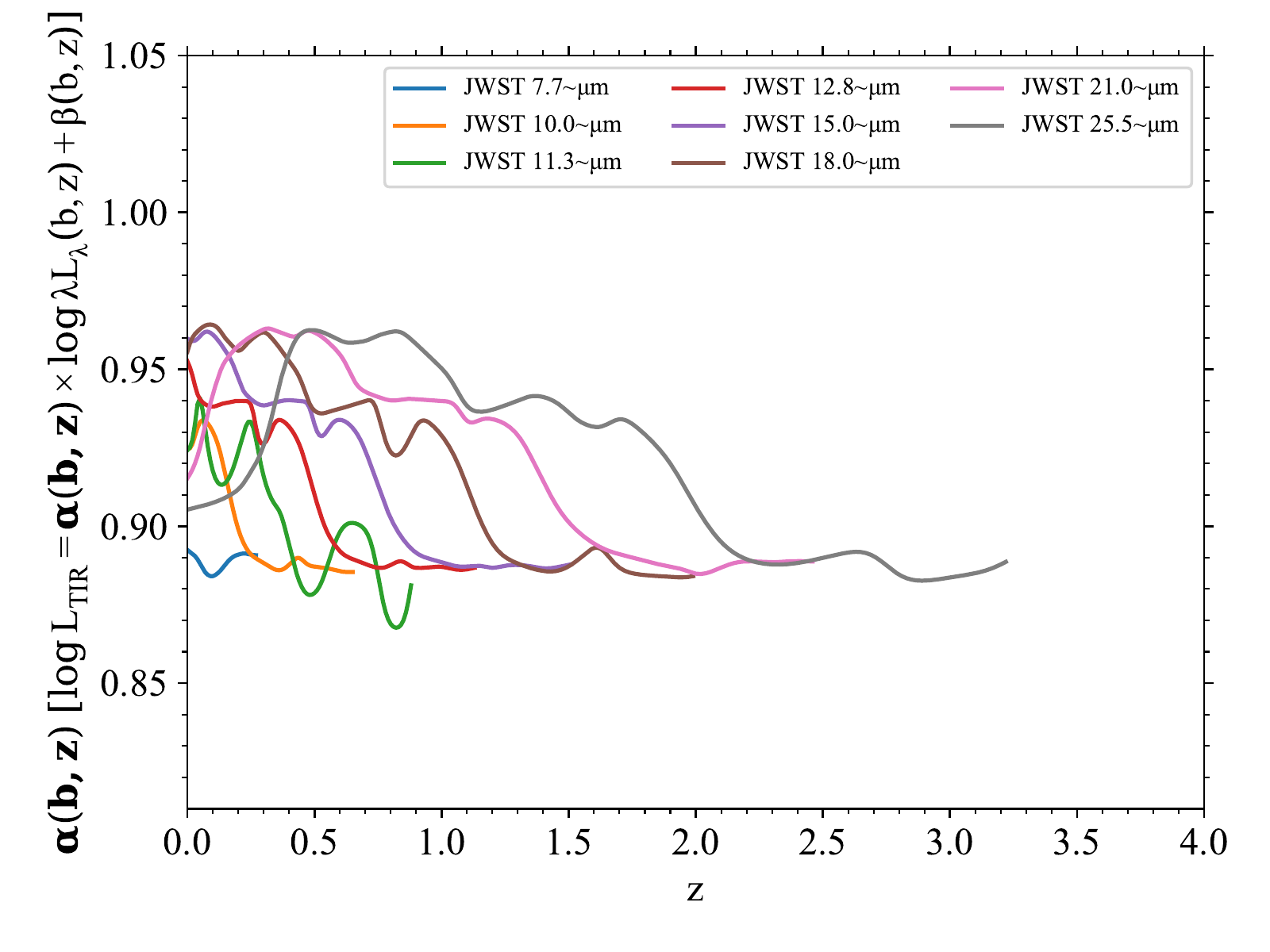}
 \includegraphics[width=.33\textwidth]{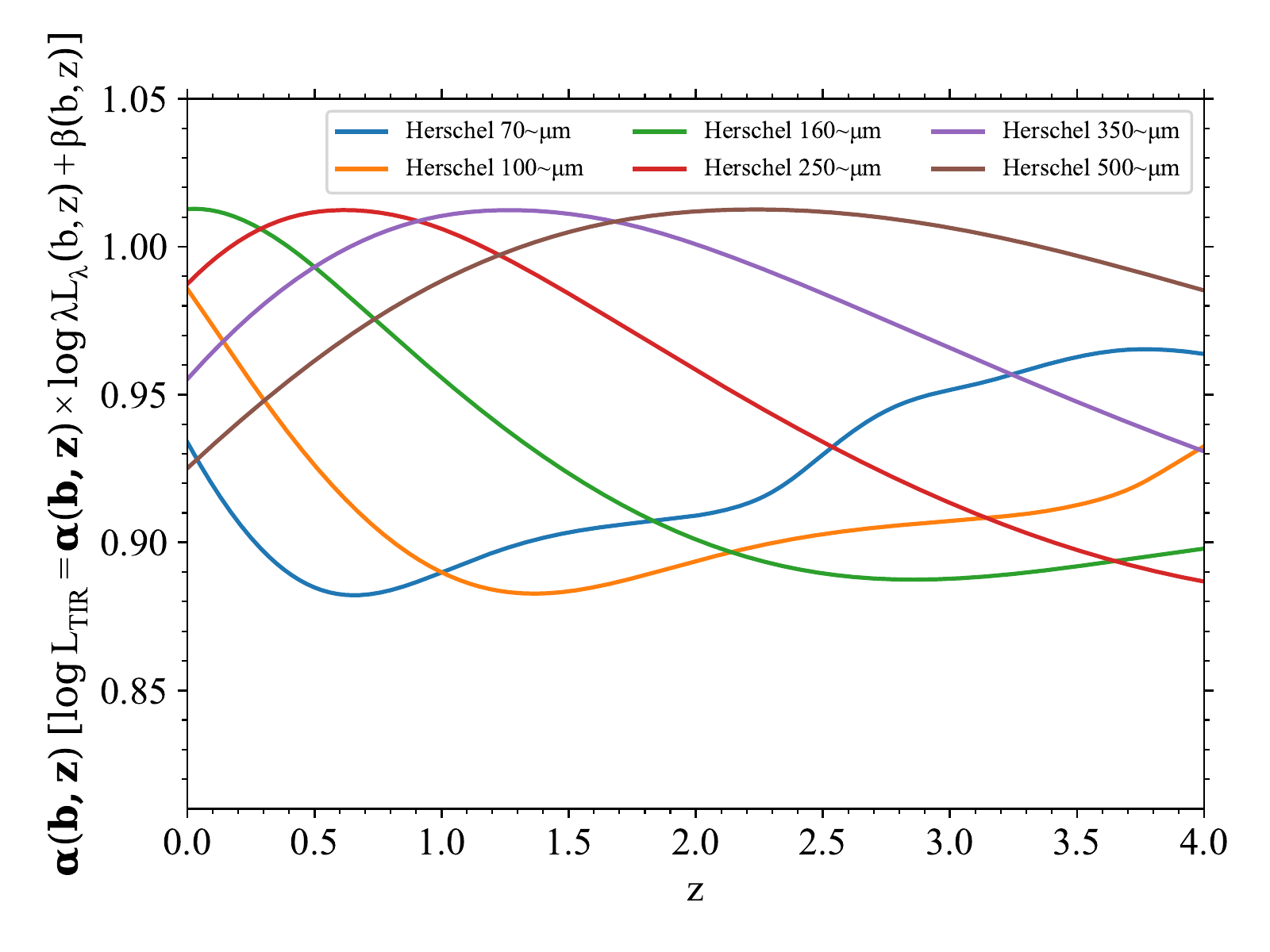}\\
 \includegraphics[width=.33\textwidth]{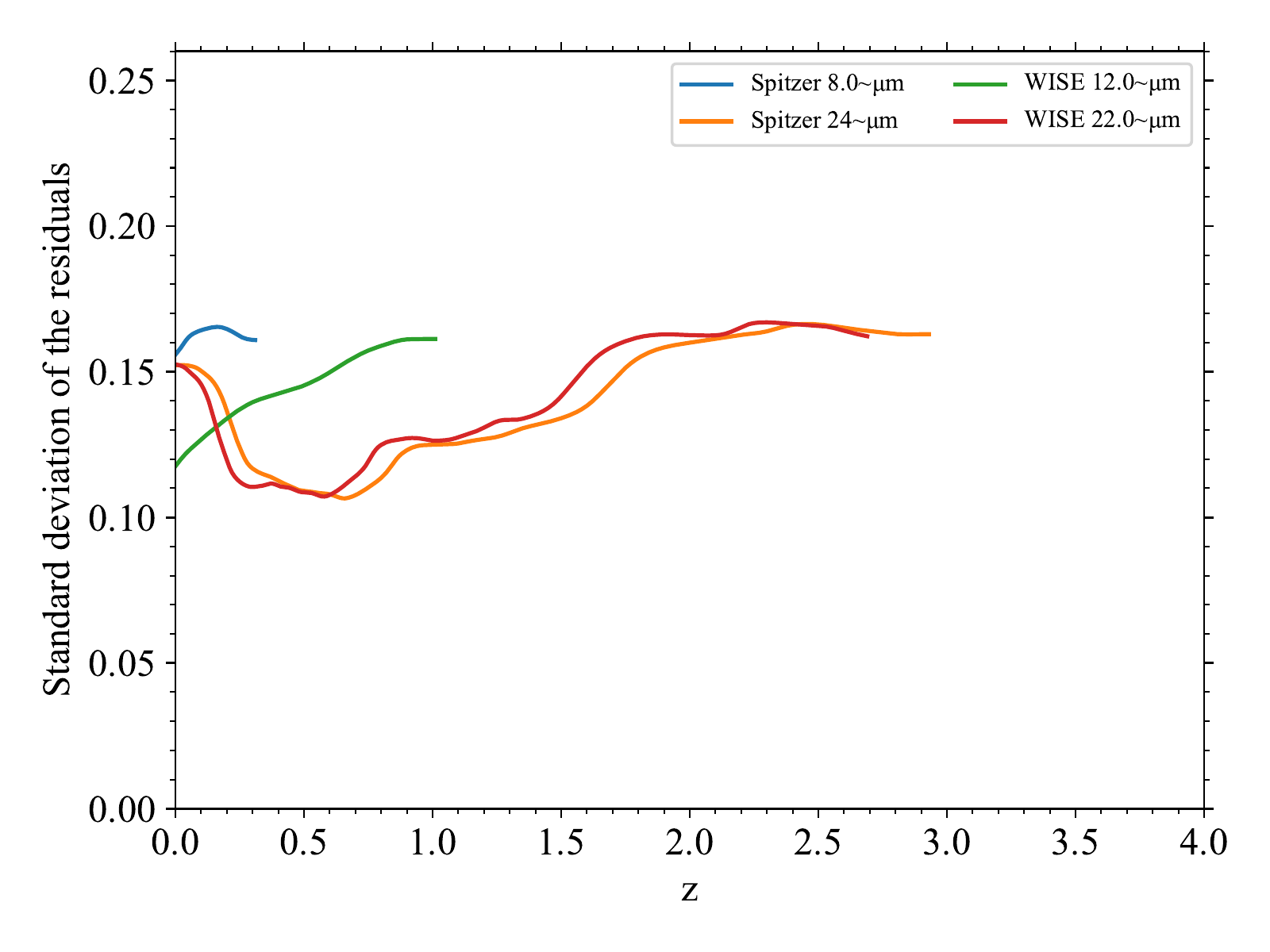}
 \includegraphics[width=.33\textwidth]{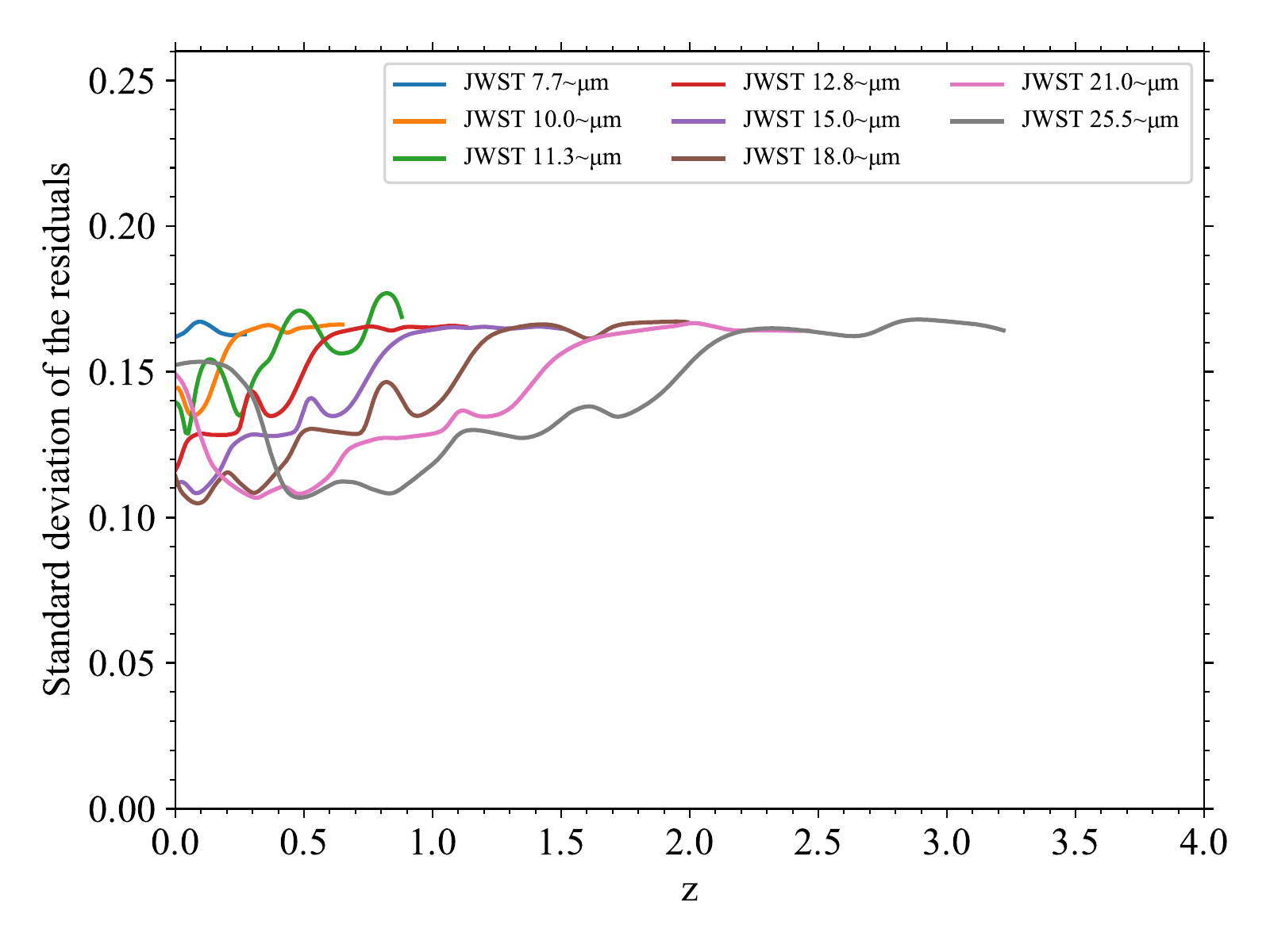}
 \includegraphics[width=.33\textwidth]{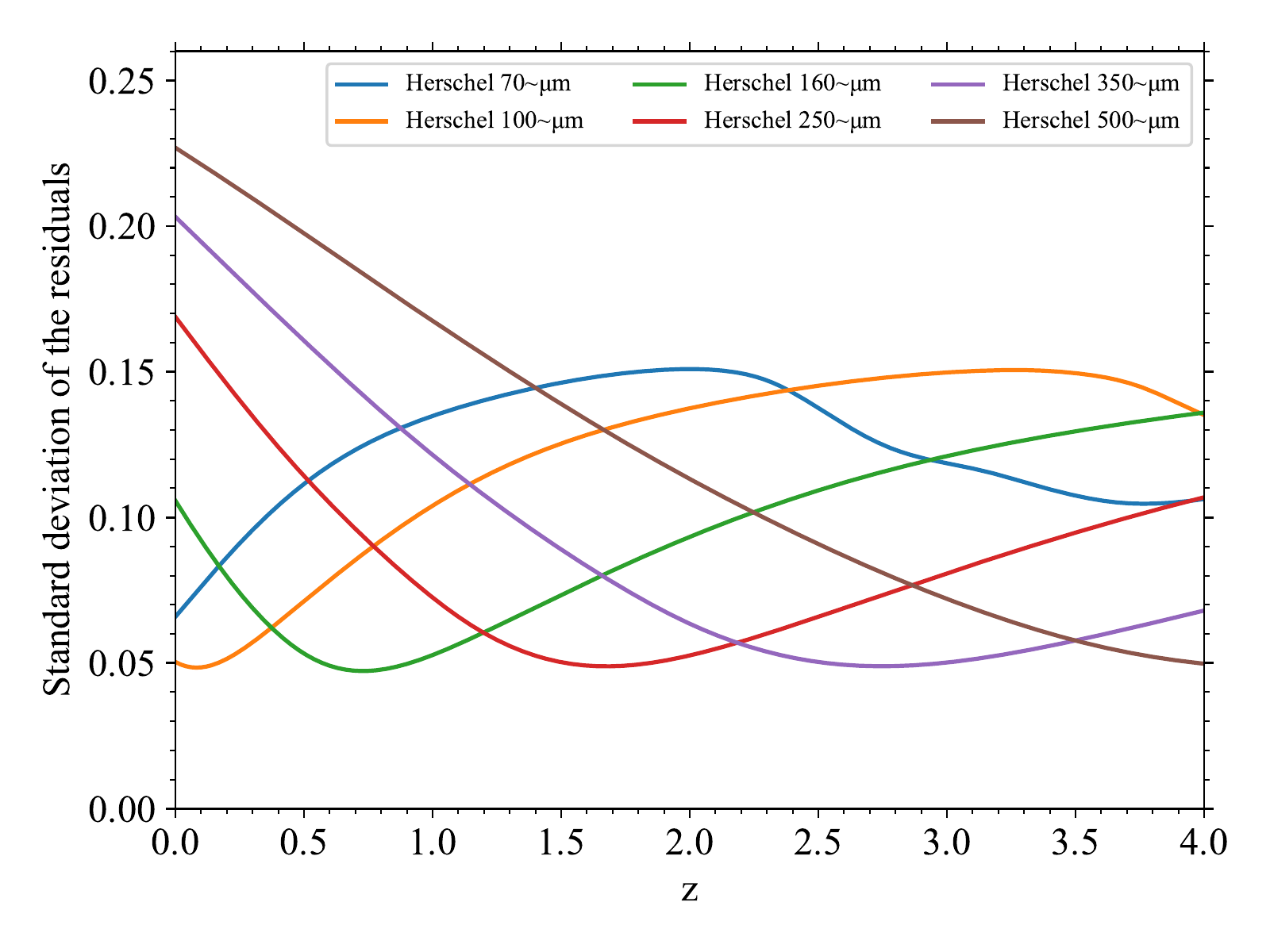}\\
 \caption{Top row. Evolution of $m\left(b, z\right)$ from Eq.~\ref{eqn:fit-LTIR-SFR-z} as a function of the redshift to estimate the TIR luminosity from the main WISE, \textit{Spitzer}, \textit{Herschel}, and JWST bands. Bottom row. Standard deviation of the residual of the TIR luminosity estimated from Eq.~\ref{eqn:fit-LTIR-SFR-z}. Fewer bands are available at higher redshifts as the computation is only carried out for $\lambda / (1+z)>6$~$\mu$m, with $\lambda$ the pivot wavelength of the filter.\label{fig:alpha-z-LTIR}}
\end{figure*}

\begin{figure*}[!htbp]
 \centering
 \includegraphics[width=.33\textwidth]{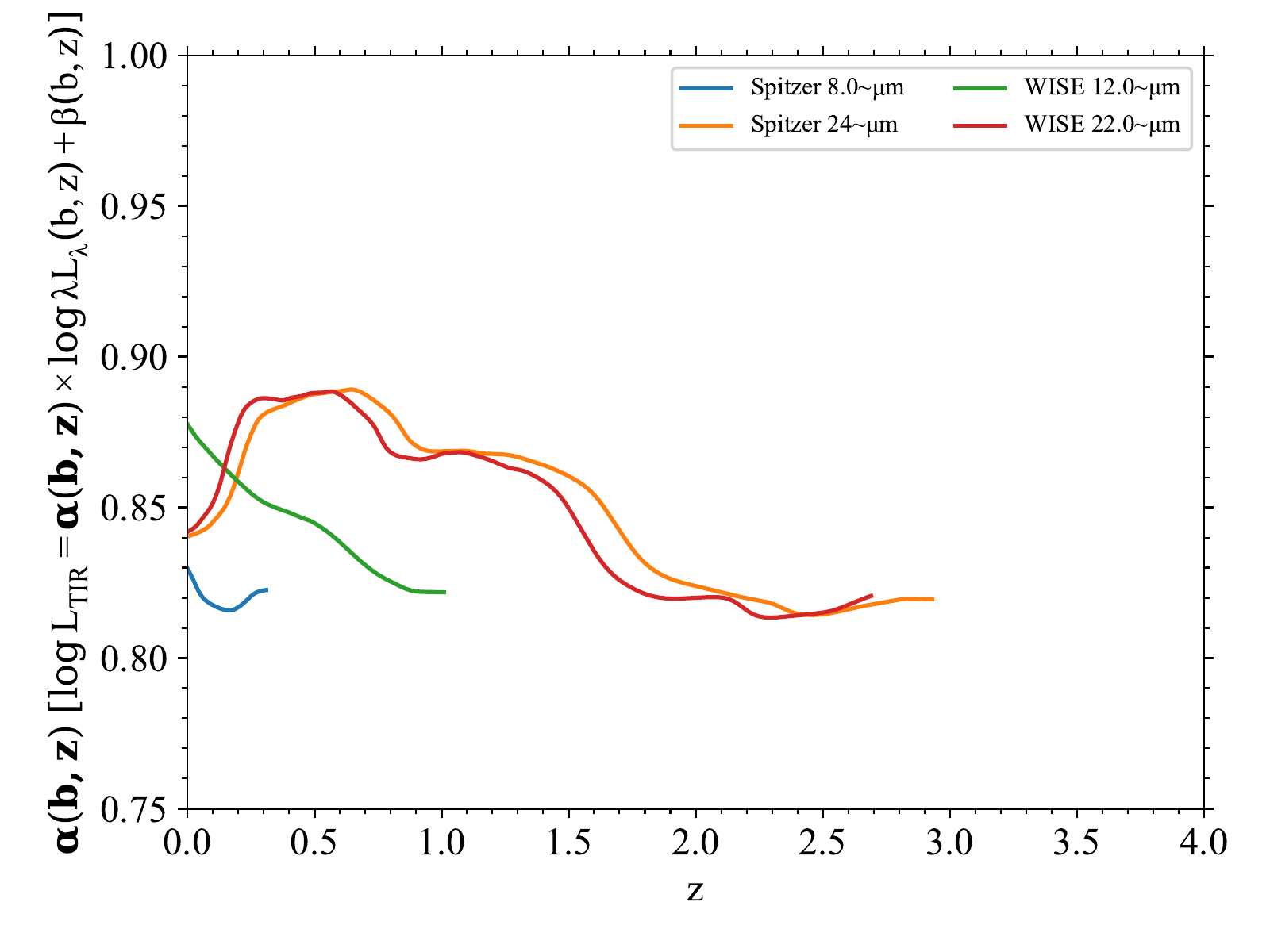}
 \includegraphics[width=.33\textwidth]{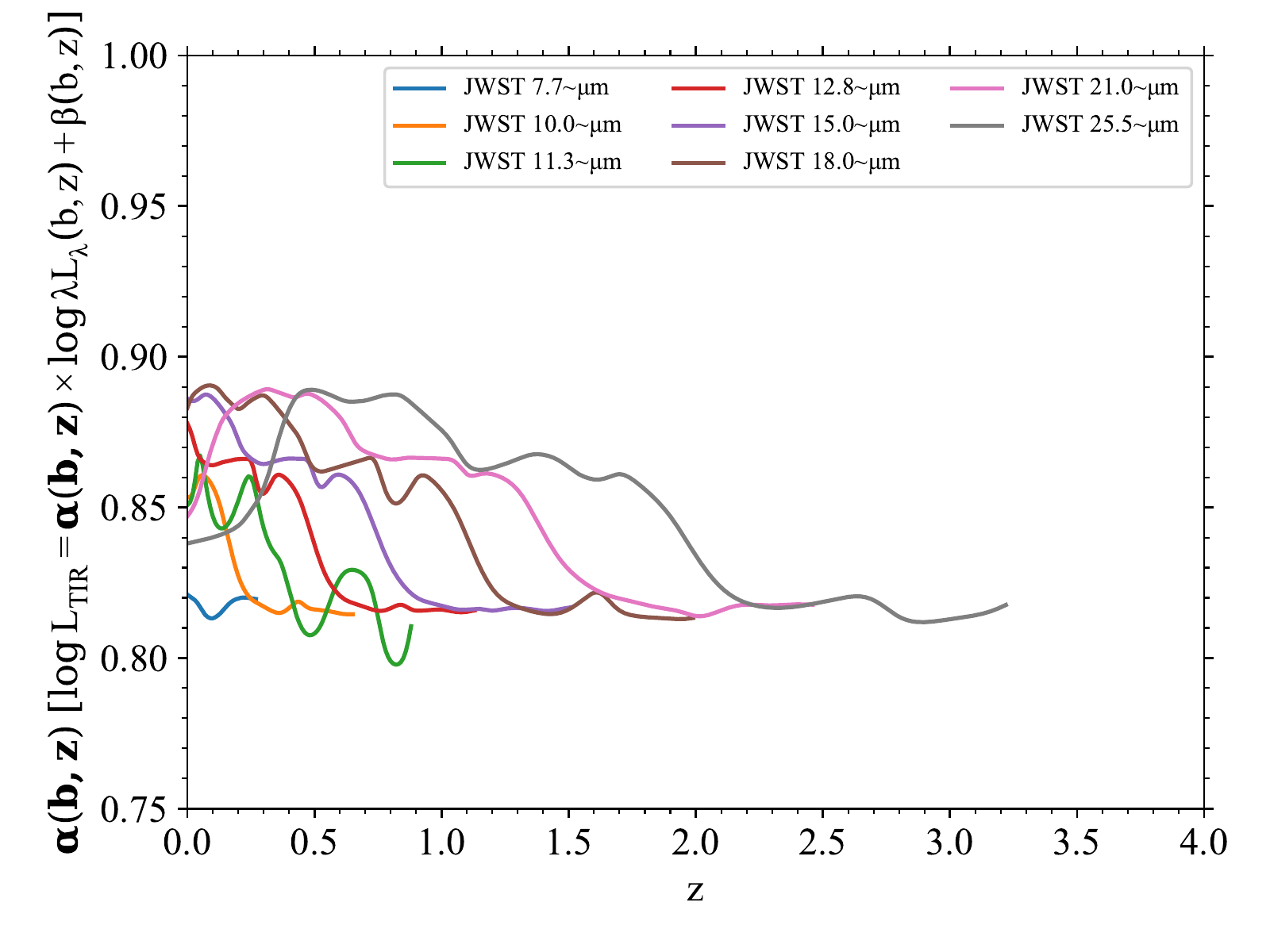}
 \includegraphics[width=.33\textwidth]{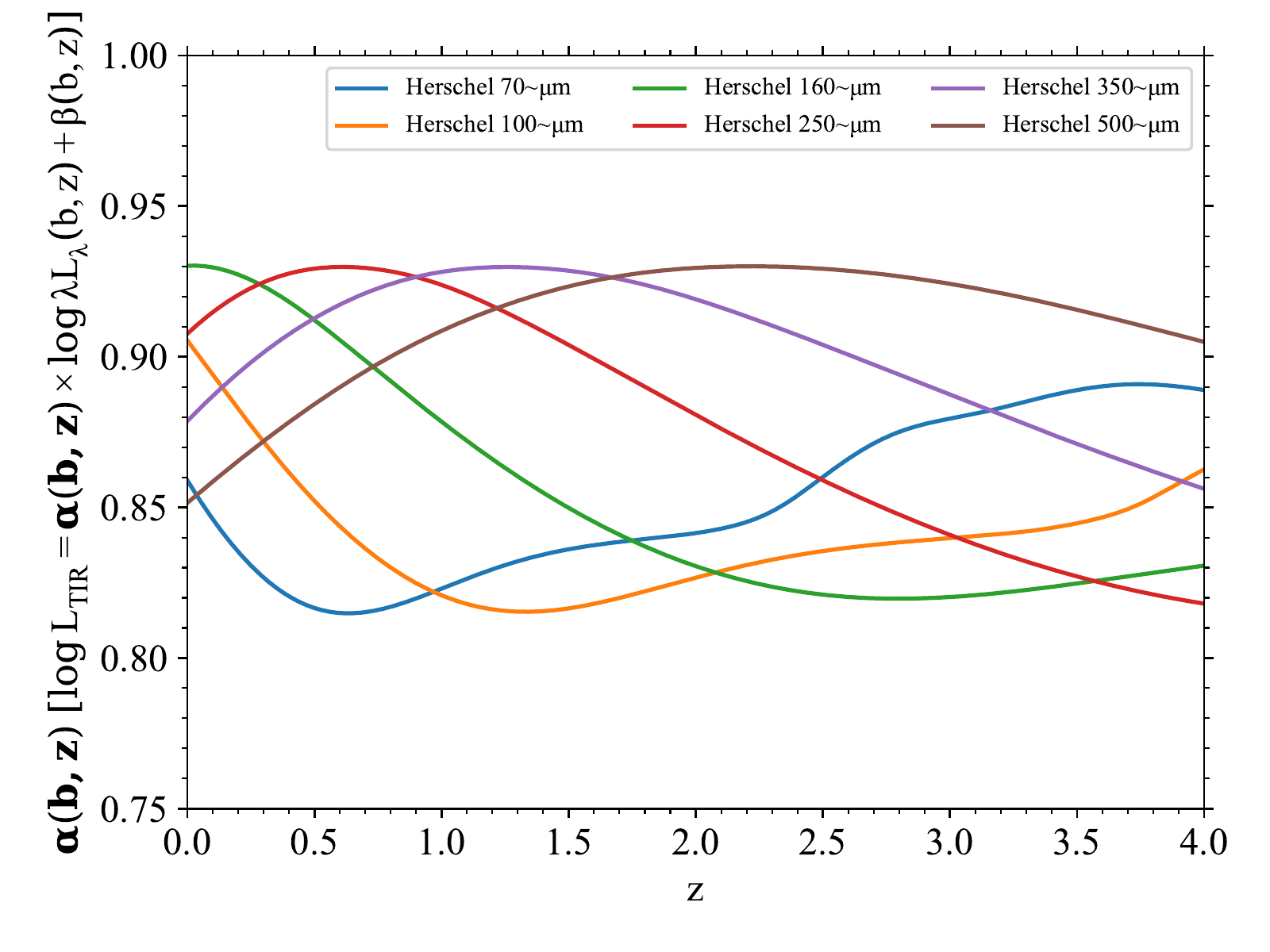}\\
 \includegraphics[width=.33\textwidth]{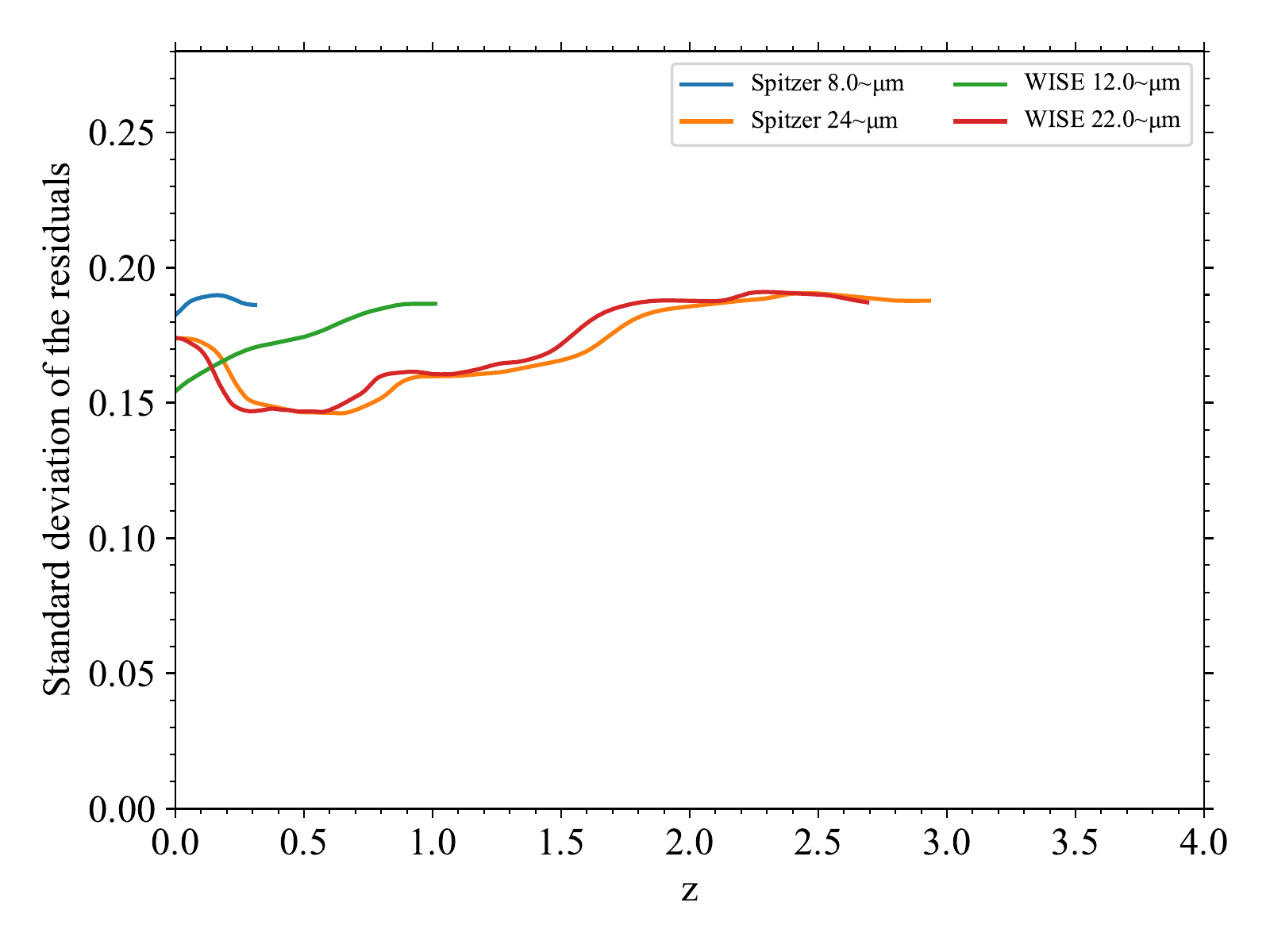}
 \includegraphics[width=.33\textwidth]{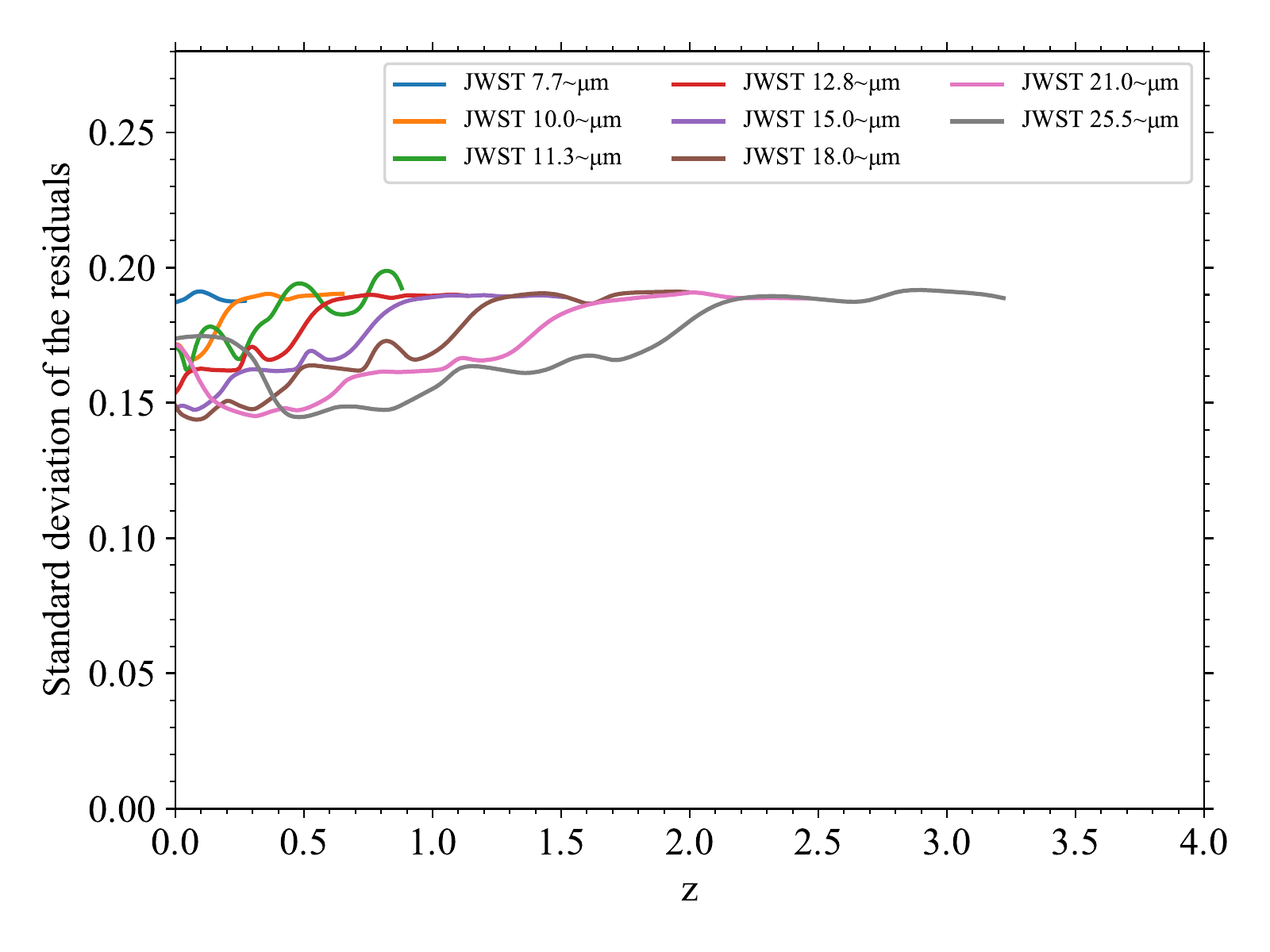}
 \includegraphics[width=.33\textwidth]{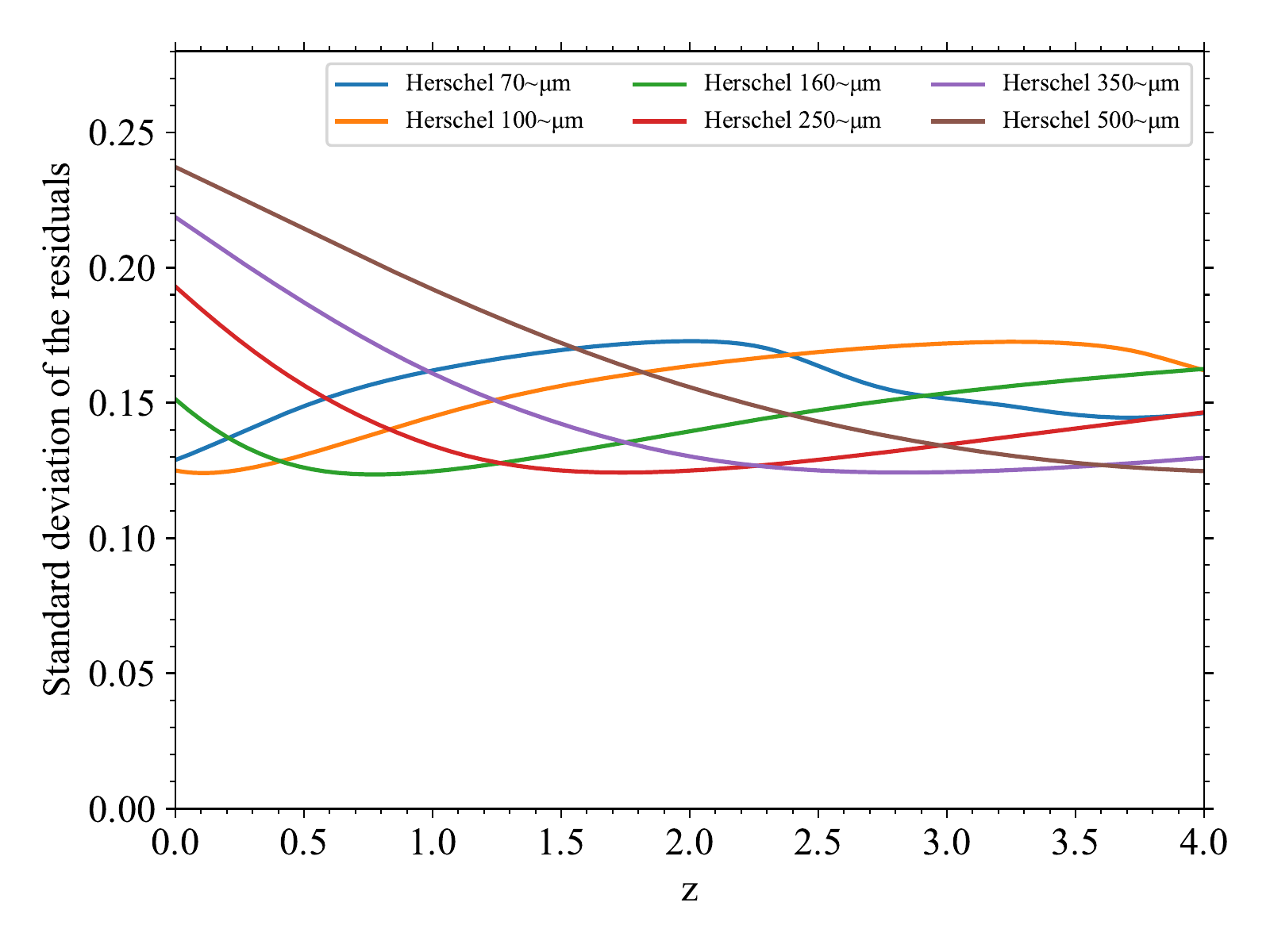}\\
 \caption{Same as Fig.~\ref{fig:alpha-z-LTIR} but for the SFR.\label{fig:alpha-z-SFR}}
\end{figure*}

In Sect.~\ref{ssec:single}, we examined the case of galaxies in the local universe. However with increasing redshifts, the K-correction becomes a major factor and a significant source of uncertainty, in particular given the complex shape of the dust emission. At the same time, more distant galaxies tend to have fewer bands available, increasing the need to have efficient single-band estimators of $L_{TIR}$ and the SFR. To fill this need, we have computed the coefficients for redshift-dependent relations, adding the redshift as a variable to Eq.~\ref{eqn:fit-LTIR-SFR}:

\begin{equation}
  \log p = m\left(b, z\right)\times\log \lambda L_\lambda\left(b\right) + n\left(b, z\right)\label{eqn:fit-LTIR-SFR-z},
\end{equation}
with $m\left(b,z\right)$ and $n\left(b,z\right)$ the coefficients obtained from the fit. The other terms are defined similarly as in Eq.~\ref{eqn:fit-LTIR-SFR}. We have built a grid of estimators up to $z=4$ by fitting the redshifted best-fit models with this equation. We show the evolution of $m\left(b,z\right)$ in Fig.~\ref{fig:alpha-z-LTIR} and \ref{fig:alpha-z-SFR} for the main \textit{Spitzer}, WISE, \textit{Herschel}, and JWST bands.
Naturally, the values at $z=0$ correspond to what was presented in Table~\ref{table:estimators-LTIR-SFR}. We observe that for the most part the curves appear to be qualitatively similar, but offset from one another with respect to the redshift, which is expected as different filters probe the same rest-frame wavelength at different redshifts and the derived estimators depend on the rest-frame wavelength. The exact filter shape only has a limited effect, as can be seen from the evolution of $m\left(b,z\right)$ as well as for the standard deviation of the residuals. Overall we find that at all redshifts the FIR bands sampling the peak of the emission show a better performance, with always a band simultaneously with $m\sim 1$ and a low scatter in the residuals. The performance of MIR bands tends to be significantly worse. The bands in the 20~$\mu$m to 25~$\mu$m range appear to work best at $z\sim 0.5$, with a progressive degradation with increasing redshift as they become heavily dominated by PAH bands. All MIR bands present a sub-linearity at all redshifts. The SFR estimators present a globally similar behavior.

\section{Construction of select previously published templates\label{sec:appen_temp}}
\subsection{\cite{chary2001a} templates}
\cite{chary2001a} use the following approach to construct their templates. They start with a collection of data from local galaxies observed at 6.7, 12, 15, 60 and 850~$\mu$m with ISOCAM, IRAS and SCUBA, as well as estimates of their total and FIR luminosities from IRAS bands (12, 25, 60 and 100 $\mu$m). From this they construct linear (or piecewise linear, in log) relations between various monochromatic and integrated IR luminosities. The samples are not necessarily the same ones in different relations, and vary in number from $\sim$ 50 to a few hundred. The galaxies generally have total IR luminosities that lie in the same range ($9<\log L_{TIR}/L_\odot<12$) as in our study, but probably with a greater share of starburst galaxies (i.e., galaxies with higher $L_{TIR}$ than typical for their stellar and dust masses). In order to get templates that are continuous in wavelength, \cite{chary2001a} used \cite{silva1998a} theoretical SED (but updated with real MIR spectra) that describe four nearby galaxies representative of different IR SEDs. These ``proto-templates'' were constructed by \cite{silva1998a}  by fitting a model that includes three grain species and three dust environments. \cite{chary2001a} then split the four proto-templates into two wavelength regimes at 20~$\mu$m and interpolated them over many luminosities. The final 105 luminosity-dependent templates were chosen among the interpolated proto-templates that best fit the observed relations described above. The two wavelength regimes were then joined, and the $L_{TIR}$ of each template was determined by integration. The templates span 0.1--1000 $\mu$m, i.e., they also include the (highly uncertain) stellar emission. The templates provided to the community span $8.4<\log L_{TIR}/L_\odot<13.5$, i.e., they have been extrapolated into the ULIRG range.

\subsection{\cite{dale2002a} and \cite{dale2014a} templates}
The templates of \cite{dale2002a} represent a refinement of the templates produced by \cite{dale2001a}, which in turn rely on models of \cite{desert1990a}. Specifically, \cite{dale2001a} start by producing their local (as opposed to galaxy-wide) model templates by combining theoretically predicted emission from large grains, emission from very small grains and an average PAH spectrum from actual observations. This PAH spectrum includes an empirically motivated damping factor that increases as the heating intensity ($U$) increases. A range of global model SED were obtained by summing up local SED according to the power-law exponent $\alpha_{\rm SF}$, which represents the varying mix of contributions of regions with different levels of activity (i.e., heating intensities, $U$). \cite{dale2002a} empirically refine the FIR portion of the \cite{dale2001a} templates ($\lambda>100~\mu$m) by allowing the emissivity to vary as a function of $U$ in a way that  minimizes residuals with respect to 850 $\mu$m observations. Unlike the templates of \cite{chary2001a}, the templates of \cite{dale2001a} are constrained by, but not directly fitted to the data. Nevertheless, they show that their models well reproduce the average SED of a sample of 69 normal galaxies with $8.2<\log L_{TIR}<12.0$ and spanning 6.7--100 $\mu$m. Relating the power-law exponent $\alpha_{\rm SF}$ via FIR color to the total IR luminosity allows  \cite{dale2002a} templates to be used with single-band measurements. Such a calibration by \cite{marcillac2006a}, based on IRAS Bright Galaxy Sample, associates 64 templates of \cite{dale2002a} with $8.3<\log L_{TIR}/L_\odot<14.3$. More recently, \cite{dale2014a} updated the PAH spectrum of the \cite{dale2002a} templates and allowed the addition of an AGN-heated component.

\subsection{\cite{rieke2009a} templates}
\cite{rieke2009a} construct what they refer to as {\em average} templates separately for galaxies above and below $\log L_{TIR}/L_\odot=11$, using somewhat different (but mostly empirical) methods for each group. For high-luminosity ($\log L_{TIR}/L_\odot>11$) templates, they first produce continuous SED for 11 individual galaxies (local LIRGs/ULIRGs), which we again refer to as proto-templates. These proto-templates are constructed from a combination of synthetic SED in the optical/near-IR, empirical spectra in the MIR (6-35~$\mu$m), and modified blackbodies fit to the photometry in the FIR ($>60~\mu$m). The proto-templates are then combined with different weights in order to match the empirical colors obtained from the relations between luminosities in different bands (8, 12, 24 and 60 $\mu$m) and $L_{TIR}$ of $\sim 70$ galaxies spanning $10<\log L_{TIR}/L_\odot<12.3$. To produce intermediate luminosity ($\log L_{TIR}/L_\odot<11$) templates, \cite{rieke2009a} rely on the combination of \cite{dale2002a} models for the FIR and empirical IRS templates for the MIR. \cite{dale2002a} models are implemented at $>70~\mu$m through the correlation between $\alpha_{\rm SF}$ and  $L_{TIR}$, whereas the MIR templates ($<37~\mu$m) are associated to  $L_{TIR}$ via the 12-to-25~$\mu$m color. The two were combined by matching to the 70 and 160~$\mu$m photometry and to the 25-to-60~$\mu$m color. The final set consists of 14 luminosity-dependent templates with $9.75<\log L_{TIR}/L_\odot<13.0$ and covering a wavelength range from 4~$\mu$m to 30~cm.

\subsection{\cite{smith2012b} templates}
Binned SED (templates) of \cite{smith2012b} are constructed by median averaging individual best-fit spectra obtained with UV/optical/FIR SED fitting with MAGPHYS. Modeling of the IR SED emission in \cite{dacunha2008a}, and implemented in MAGPHYS SED fitting code, is mostly phenomenological. First, they model IR SED as a sum of the emissions from two sources: birth clouds and the diffuse ISM. Birth cloud emission consists of 3 components: PAH emission lines, hot MIR continuum of very small grains (peaking around 20 $\mu$m), and warm grain continuum (peaking around 50 $\mu$m). Emission of the diffuse interstellar medium (ISM) has these three components plus the cold grain continuum (peaking around 100 $\mu$m). The PAH spectrum is empirical and of fixed shape, whereas continuum components are modified black bodies of fixed temperature, except for warm birth cloud grains and cold ISM grains, for which the temperature is adjustable. Each component contributes to some degree to the  $L_{TIR}$ budget of birth clouds or the diffuse ISM. For the diffuse ISM the contributions of PAH, hot and warm dust are mutually fixed. In the end, the model of \cite{dacunha2008a} has six adjustable parameters: one controlling the relative contribution of the diffuse ISM to the total dust luminosity ($f_{\mu}$), two describing the diffuse ISM emission (the relative contribution and temperature of cold grains) and three describing the birth clouds emission (two relative contributions plus the temperature of warm grains). The $f_{\mu}$ parameter is correlated with the sSFR and plays a strong role in the overall shape of the SED in the FIR since it de facto determines the relative strengths of the warm and cold components.

\cite{smith2012b} SED are based on a 250 $\mu$m-selected sample from an initial portion of H-ATLAS. Some 20\% of their initial sample of $\sim1000$ was detected at 160 $\mu$m and 10\% at 100 $\mu$m. WISE data were not available at the time. Given that \cite{dacunha2008a} models contain several flexible components in the wavelength range not covered by the data, the shape of those parts of the SED is not directly constrained. In this work, we compare our results with an updated version of \cite{smith2012b} templates, based on significantly larger number of galaxies and with greater fraction of PACS detections, but still without the constraints in the MIR.

\section{Impact on the accuracy of $L_{TIR}$ and SFR when using multiple IR bands\label{sec:appen_multi}}
\begin{figure*}[!htbp]
 \centering
 \includegraphics[width=\columnwidth]{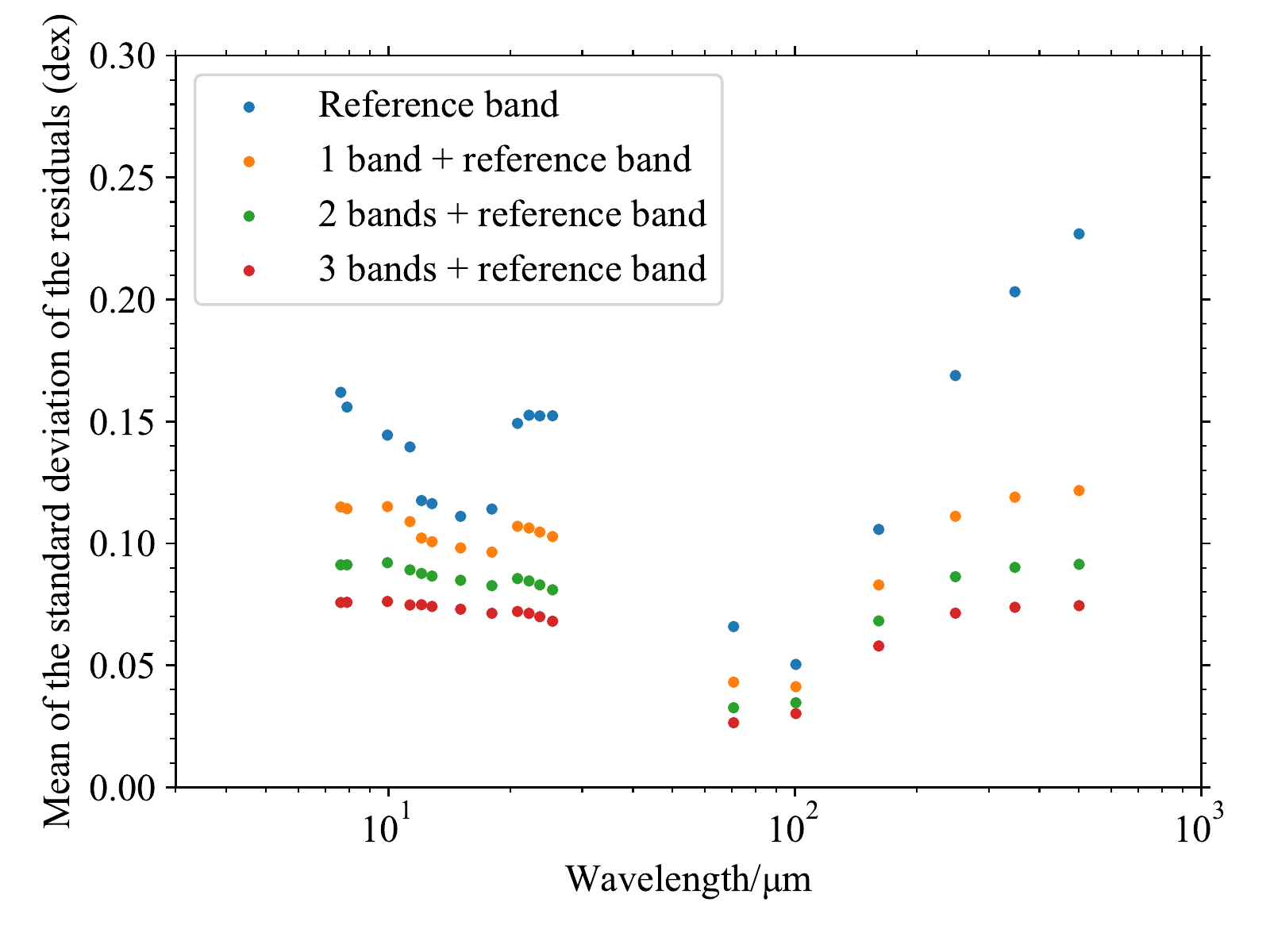}
 \includegraphics[width=\columnwidth]{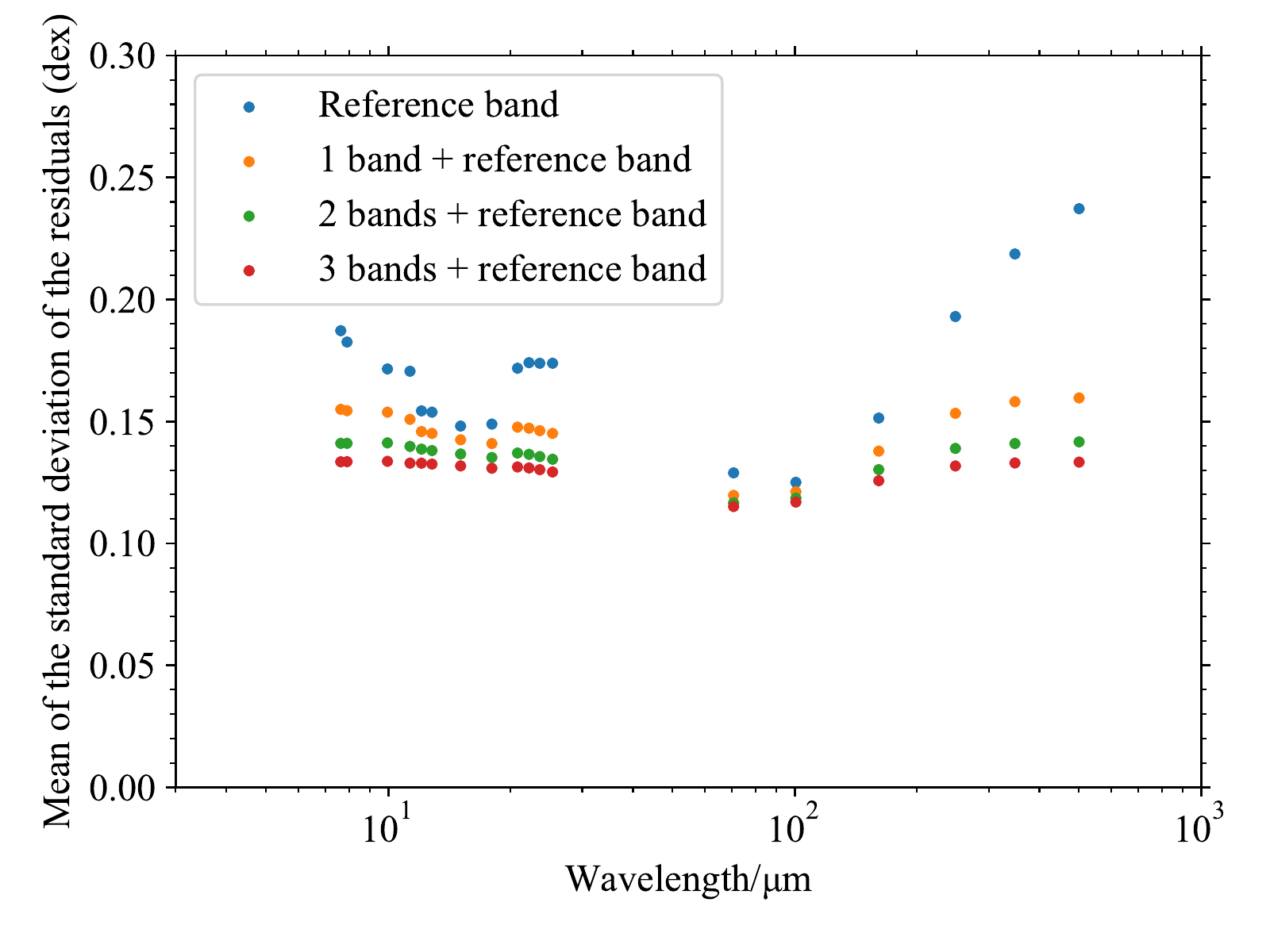}
 \caption{Mean standard deviation of the residuals of $L_{TIR}$ (left) or the SFR (right) using 1 (blue, 18 bands), 2 (orange, 153 band combinations), 3 (green, 816 band combinations), or 4 bands (red, 3060 band combinations). \label{fig:LTIR-SFR-mean-sigma-residuals}}
\end{figure*}

In order to quantify the effect of using an increasing number of bands, we have computed the mean standard deviation of the residuals for each band and when 1, 2, or 3 other bands are also used in addition to this reference band. For completeness, we use all possible combinations for each of the 18 reference bands when adding 1 (total of 153 combinations), 2 (816 combinations), and 3 (3060 combinations) bands. The results for $L_{TIR}$ and the SFR are presented in Fig.~\ref{fig:LTIR-SFR-mean-sigma-residuals}.
As expected, the standard deviation of the residuals diminishes with the increasing number of bands at all wavelengths. The effect is especially important for the determination of $L_{TIR}$ with sustained gains even with a larger number of bands. The gain is more moderate for the SFR beyond the addition of a second band, and the scatter always remains above 0.10~dex. This floor is set by the intrinsic scatter between $L_{TIR}$ and the SFR that is due to the variations in the SFH (stellar populations) and of the attenuation.

\begin{figure*}[!htbp]
 \centering
 \includegraphics[width=\columnwidth]{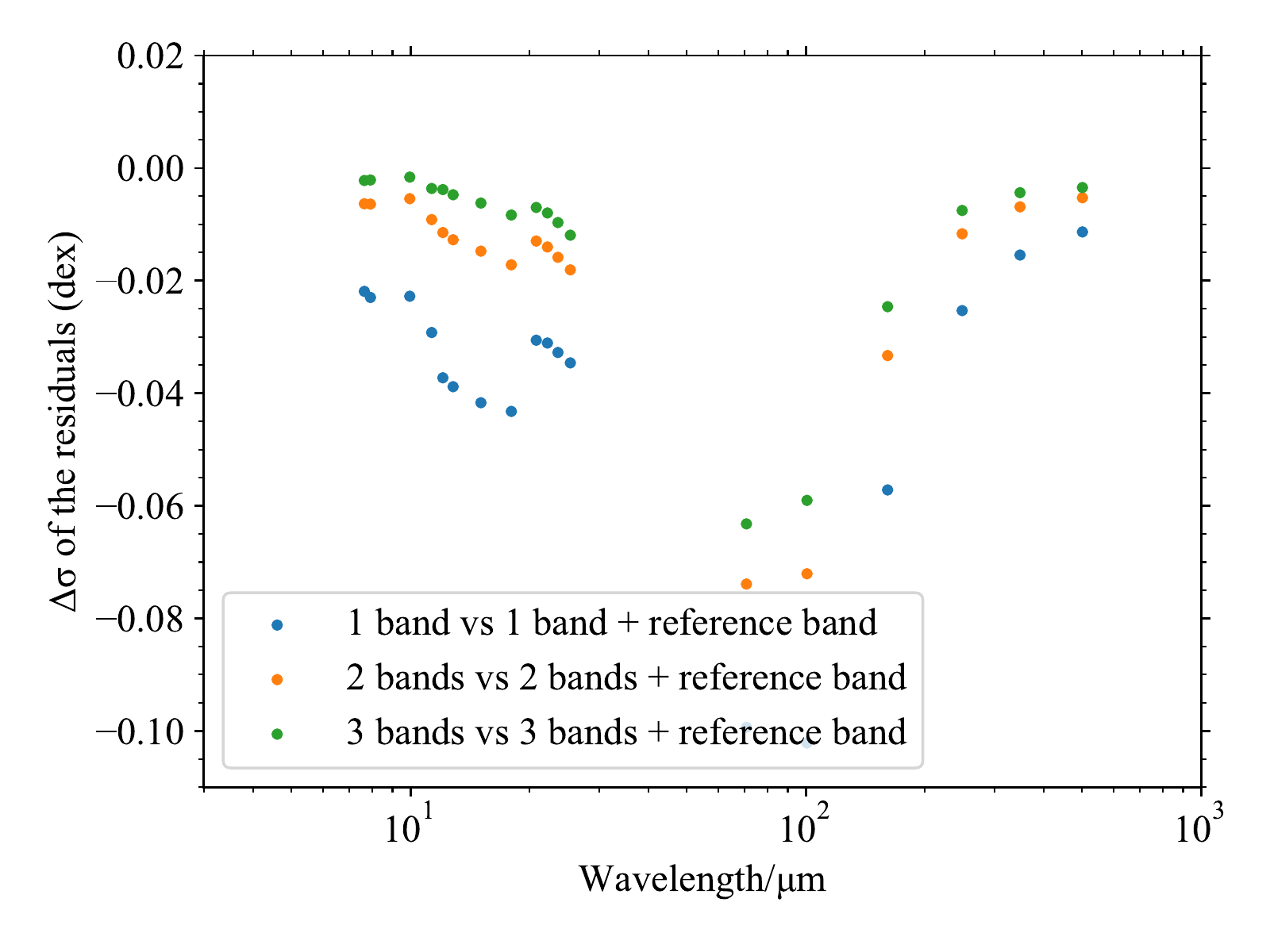}
 \includegraphics[width=\columnwidth]{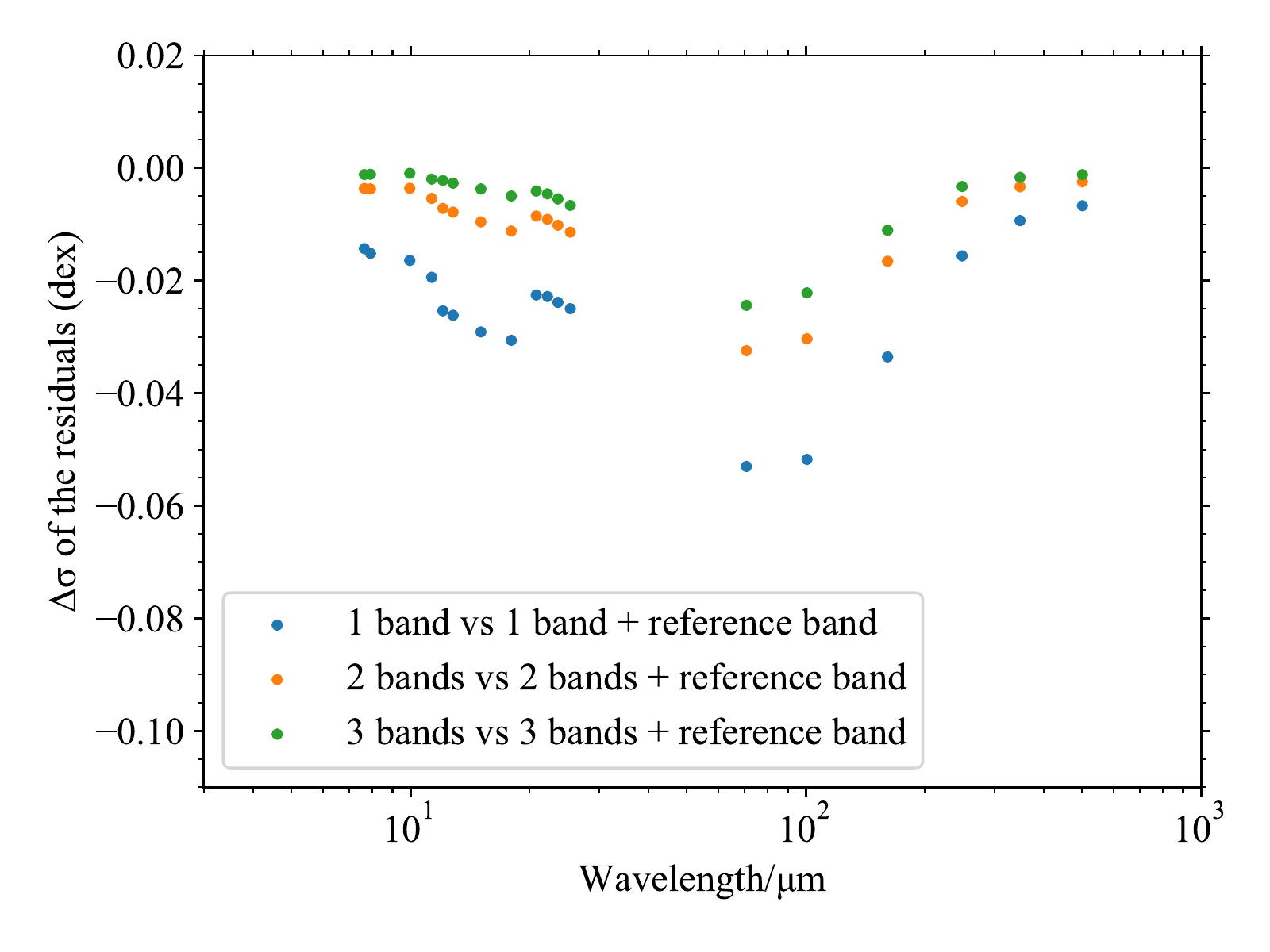}
 \caption{Reduction of the mean standard deviation of the residuals of $L_{TIR}$ (left) or the SFR (right) when adding a given band to a combination of 1 (blue, 153 band combinations), 2 (orange, 816 band combinations), or 3 (green, 3060 band combinations) bands at other wavelengths.\label{fig:LTIR-SFR-delta-sigma-residuals}}
\end{figure*}

This general overview, however, provides limited information on the added value of a given band because it averages over all possible combinations. To investigate this, in Fig.~\ref{fig:LTIR-SFR-delta-sigma-residuals} we plot the reduction in the standard deviation of the residuals when adding one band to a combination with 1, 2, or 3 reference bands at other wavelengths.
Unsurprisingly some bands have a strong impact on the reduction of the standard deviation. In the case of $L_{TIR}$, the \textit{Herschel} 70~$\mu$m and 100~$\mu$m bands generate the strongest improvement, reducing the standard deviation of the residuals by 0.1~dex, on average, when used as a second band. Even as a third or fourth band, they still bring on average an improvement better than 0.05~dex. This is a stronger added value than for the MIR bands that only bring a more limited improvement that is on average always under 0.05~dex, even as a second band. With the increasing number of MIR bands their value strongly decreases. However their weaker performance should be interpreted with caution. Because of the large number of MIR filters, there are numerous combinations of only MIR bands. As mentioned earlier, bands that are close in wavelength tend to bring very similar information. For instance, if we take the \textit{Herschel} 250~$\mu$m band as a reference, the standard deviation of the residuals is 0.17~dex. The addition of the JWST 18~$\mu$m band reduces it by \textit{0.08}~dex compared to a reduction of 0.12~dex when we rather include the \textit{Herschel} 70~$\mu$m band. For the SFR, the improvement brought by adding more bands is limited. Even in the most favorable of cases, the improvement is less than 0.05~dex and the marginal improvement readily drops below 0.02~dex when adding a third band. 

Overall for evaluating $L_{TIR}$ at $z\sim0$ the best band combinations are:
\begin{itemize}

 \item One band: \textit{Herschel} 100~$\mu$m ($\sigma=0.0504$~dex).
 \item Two bands: JWST 25.5~$\mu$m and \textit{Herschel} 100~$\mu$m ($\sigma=0.0231$~dex).
 \item Three bands: JWST 18~$\mu$m, and \textit{Herschel} 70~$\mu$m and 500~$\mu$m ($\sigma=0.0151$~dex).
 \item Four bands: JWST 12.8~$\mu$m and 25.5~$\mu$m, and \textit{Herschel} 70~$\mu$m and 500~$\mu$m ($\sigma=0.0142$~dex).
\end{itemize}
For the SFR the best band combinations are:
\begin{itemize}
 \item One band: \textit{Herschel} 100~$\mu$m ($\sigma=0.1250$~dex).
 \item Two bands: JWST 25.5~$\mu$m and \textit{Herschel} 100~$\mu$m ($\sigma=0.1148$~dex).
 \item three bands: JWST 18~$\mu$m, and \textit{Herschel} 70~$\mu$m and 500~$\mu$m ($\sigma=0.1126$~dex).
 \item Four bands: JWST 18~$\mu$m and 25.5~$\mu$m and \textit{Herschel} 70~$\mu$m and 500~$\mu$m ($\sigma=0.1124$~dex).
\end{itemize}
This shows that the estimation of $L_{TIR}$ strongly benefits from the combination of various bands that sample different regions from the spectrum, in order of decreasing priority, the peak of the emission by the warm dust around 100~$\mu$m, then the hot dust emission in the MIR, and finally the cold dust in the sub-millimeter range. In effect these relations embed the non linear variations of the shape of the dust emission spectrum with $L_{TIR}$. For the SFR a single band already provides results close to the best performance, probably because of the intrinsic scatter between $L_{TIR}$ and the SFR as mentioned earlier. Note that the assessments in this section do not take into account different levels of quality of data from different facilities and in different bands, and assumes all data having zero error. 

\section{Degeneracies and information contributed by individual bands\label{sec:appen_deg}}
\begin{figure}[!htbp]
 \centering
 \includegraphics[width=\columnwidth]{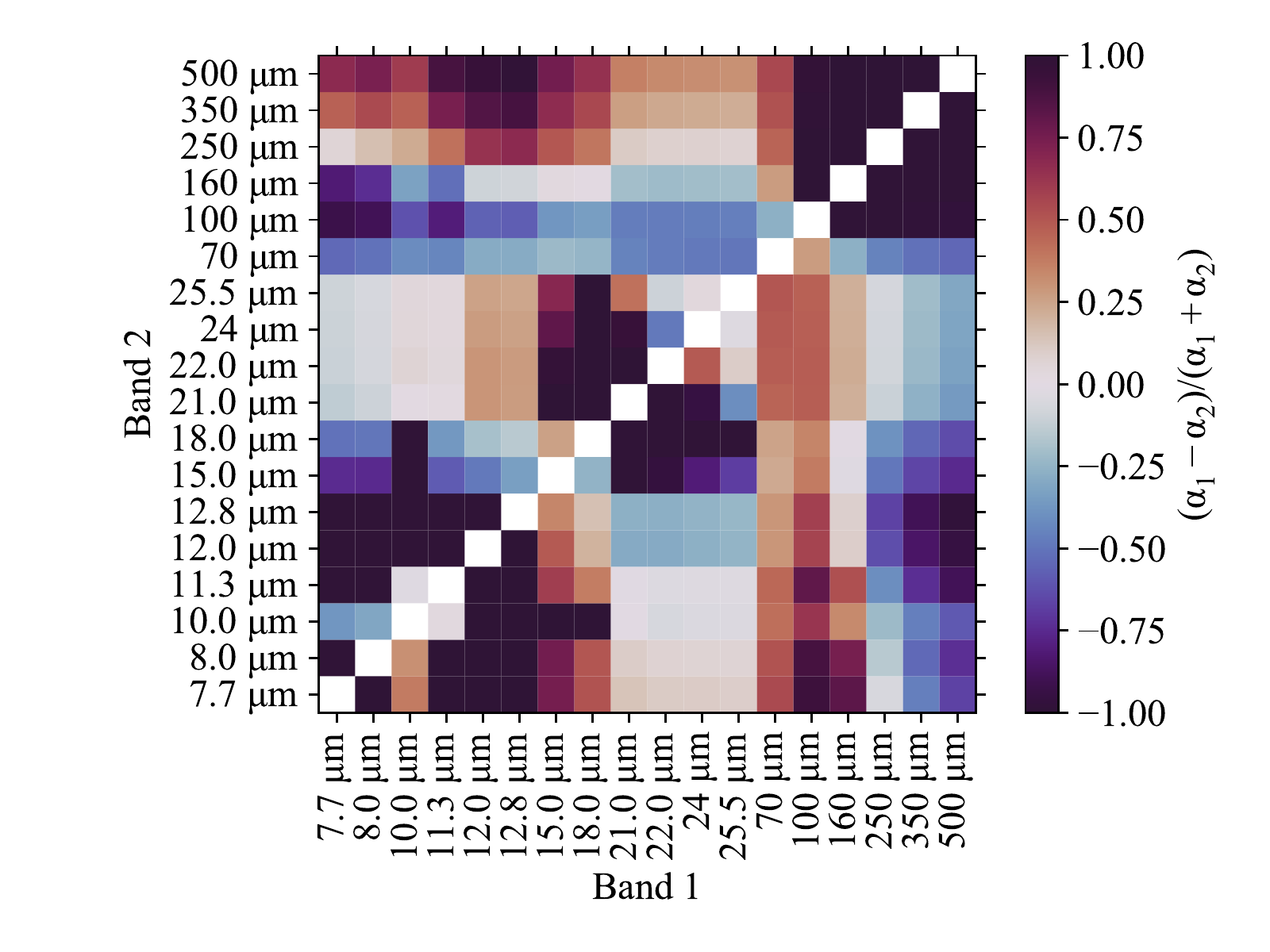}
 \caption{Relative difference between the $m_i$ coefficients for estimators using two bands. The lighter shades indicate that they are fairly commensurate with one another whereas the darker shades indicate that one is much larger than the other. This happens in particular when one of the two bands does not bring additional information. This is especially common for bands that are close in wavelengths while probing similar physical components, such as the PAH or the warm dust in the MIR or the Rayleigh-Jeans tail in the FIR.\label{fig:alpha-matrix}}
\end{figure}

The analysis in Appendix \ref{sec:appen_multi} suggests that when different bands bring similar information, it is common that all the weight is given to a single band, probably the one that is found to be slightly more informative. This means that the fitting process to determine $m_i$ and $n$ effectively eliminates some bands that do not bring additional information by assigning them $m_i \simeq 0$. In order to quantify this, we show in Fig.~\ref{fig:alpha-matrix} a matrix giving the relative difference between the $m_i$ coefficients of estimators using two bands for estimating $L_{TIR}$ (the figure is essentially identical for the SFR estimators).
It is readily apparent that when two bands probe a similar physical component (e.g., the PAH or the warm dust in the MIR, or the Rayleigh-Jeans tail in the FIR), one of these bands dominates, which effectively reduces these estimators to single-band estimators. For instance, if we consider the JWST 7.7~$\mu$m and \textit{Spitzer} 8.0~$\mu$m bands, $m_{7.7}<10^{-12}$ and $m_{8.0}=0.9018$, which corresponds exactly to the coefficient for the monochromatic estimator for the \textit{Spitzer} 8.0~$\mu$m band. By itself, this provides useful knowledge on which bands to preferentially acquire for estimating $L_{TIR}$ or the SFR when a broad sampling of the dust emission spectrum is not possible.

\end{document}